\documentclass[longauth]{aa}  
\usepackage{graphicx}
\usepackage{txfonts}
\usepackage[]{hyperref}
\usepackage{xcolor}
\usepackage[version=4]{mhchem}

\def\deg{^\circ}
\def\kms{\mathrm{km\,s}^{-1}}
\def\e#1{\times 10^{#1}}

\def\msol{\mathrm{M}_\odot}
\def\lsol{L_\odot}
\def\up#1{$^{#1}$}

\def\h2{$\mathrm{H}_2$}
\def\so2{$\mathrm{SO}_2$}

\def\spy{\;\msol~\mathrm{ yr}^{-1}}
\def\um{\;\mu\mathrm{m}}
\begin{document}

   \title{ATOMIUM: Halide molecules around the S-type AGB star W~Aquilae}

   \author{T. Danilovich
          \inst{1}\fnmsep\thanks{Senior Postdoctoral Fellow of the Fund for Scientific Research (FWO), Flanders, Belgium}
		\and
          M. Van de Sande\inst{1}
			\and
			J.~M.~C. Plane\inst{2}
			\and
			T. J. Millar\inst{3}
			\and
			P. Royer\inst{1}
			\and
			M. A. Amor\inst{4}
			\and
			K. Hammami\inst{4}
			\and 
          L.~Decock\inst{1}
			\and
			C. A. Gottlieb\inst{5}
			\and
			L. Decin\inst{1,2}
             \and
          	A. M. S. Richards\inst{6}
			\and
			E. De Beck\inst{7}
			\and
			A. Baudry\inst{8}
			\and
			J. Bolte\inst{1}
			\and
			E. Cannon\inst{1}
			\and
			F.~De~Ceuster\inst{20,1}
			\and
			A.~de~Koter\inst{9,1}
			\and
			S.~Etoka\inst{6}
			\and
			D. Gobrecht\inst{1}
			\and
			M. Gray\inst{6,10}
			\and
			F. Herpin\inst{8}
			\and
			W. Homan\inst{11,1}
			\and
			M. Jeste\inst{12}
			\and
			P.~Kervella\inst{13}
			\and
			T. Khouri\inst{7}
			\and
			E.~Lagadec\inst{14}
			\and
			S.~Maes\inst{1}
			\and
			J.~Malfait\inst{1}
			\and
			I.~McDonald\inst{6,22}
			\and
			K. M. Menten\inst{12}
			\and
			M. Montarg\`es\inst{13}
			\and
			H.~S.~P.~M\"uller\inst{15}
			\and
			B.~Pimpanuwat\inst{6,10}
			\and
			R. Sahai\inst{16}
			\and
			S.~H.~J.~Wallstr\"om\inst{1}
			\and
			L.~B.~F.~M.~Waters\inst{17,18}
			\and
			K. T. Wong\inst{19}
          \and
          J. Yates\inst{20}
          \and
          A.~Zijlstra\inst{6,21}
          }

  \institute{Department of Physics and Astronomy, Institute of Astronomy, KU Leuven, Celestijnenlaan 200D,  3001 Leuven, Belgium 
            \and
           University of Leeds, School of Chemistry, Leeds LS2 9JT, UK 
             \and
            Astrophysics Research Centre, School of Mathematics and Physics, Queen's University Belfast, University Road, Belfast BT7 1NN, UK 
            \and
            LSAMA, Department of Physics, Faculty of Sciences, University Tunis El Manar, Campus Universitaire, 1060 Tunis, Tunisia 
            \and
                        Harvard-Smithsonian Center for Astrophysics, 60 Garden Street, Cambridge, MA 02138, USA
            \and
            JBCA, Department Physics and Astronomy, University of Manchester, Manchester M13 9PL, UK
            \and
            Department of Space, Earth and Environment, Chalmers University of Technology, Onsala Space Observatory, 43992 Onsala, Sweden
             \and
             Universit\'e de Bordeaux, Laboratoire d'Astrophysique de Bordeaux, 33615 Pessac, France 
             \and
             University of Amsterdam, Anton Pannekoek Institute for Astronomy, 1090 GE Amsterdam, The Netherlands 
             \and
             National Astronomical Research Institute of Thailand, Chiangmai 50180, Thailand 
             \and
             Institut d'Astronomie et d'Astrophysique, Universit\'e Libre de Bruxelles (ULB), CP 226, 1060 Brussels, Belgium 
             \and
             Max-Planck-Institut f\"ur Radioastronomie, 53121 Bonn, Germany 
             \and
             LESIA, Observatoire de Paris, Universit\'e PSL, CNRS, Sorbonne Universit\'e, Universit\'e de Paris, 5 place Jules Janssen, 92195 Meudon, France 
             \and
             Universit\'e C\^ote d'Azur, Laboratoire Lagrange, Observatoire de la C\^ote d'Azur, F-06304 Nice Cedex 4, France 
             \and
             Universit\"at zu K\"oln, I. Physikalisches Institut, 50937 K\"oln, Germany 
             \and
             California Institute of Technology, Jet Propulsion Laboratory, Pasadena CA 91109, USA 
             \and
             SRON Netherlands Institute for Space Research, NL-3584 CA
Utrecht, The Netherlands 
			\and
			Radboud University, Institute for Mathematics, Astrophysics and Particle Physics (IMAPP), Nijmegen, The Netherlands 
			\and
			Institut de Radioastronomie Millim\'etrique, 300 rue de la Piscine, 38406 Saint Martin d'H\'eres, France 
			\and
			University College London, Department of Physics and Astronomy, London WC1E 6BT, United Kingdom 
			\and
			University of Hong Kong, Laboratory for Space Research, Pokfulam, Hong Kong 
			\and
			School of Physical Sciences, The Open University, Walton Hall, Milton Keynes, MK7 6AA, UK 
             \\
             }

   \date{Received ; accepted }

 
  \abstract
   {S-type asymptotic giant branch (AGB) stars are thought to be intermediates in the evolution of oxygen- to carbon-rich AGB stars. The chemical compositions of their circumstellar envelopes are also intermediate, but have not been studied in as much detail as their carbon- and oxygen-rich counterparts. W~Aql is a nearby S-type star, with well known circumstellar parameters, making it an ideal object for in-depth study of less common molecules.}
   {We aim to determine the abundances of AlCl and AlF from rotational lines, which have been observed for the first time towards an S-type AGB star. In combination with models based on PACS\thanks{{\it Herschel} is an ESA space observatory with science instruments provided by European-led Principal Investigator consortia and with important participation from NASA.} observations, we aim to update our chemical kinetics network based on these results.}
   {We analyse ALMA observations towards W~Aql of AlCl in the ground and first two vibrationally excited states and AlF in the ground vibrational state. Using radiative transfer models, we determine the abundances and spatial abundance distributions of Al$^{35}$Cl, Al$^{37}$Cl, and AlF. We also model HCl and HF emission and compare these models to PACS spectra to constrain the abundances of these species.}
   {AlCl is found in clumps very close to the star, with emission confined within  0\farcs1 of the star. AlF emission is more extended, with faint emission extending 0\farcs2 to 0\farcs6 from the continuum peak. We find peak abundances, relative to \h2, of $1.7\e{-7}$ for Al$^{35}$Cl, $7\e{-8}$ for Al$^{37}$Cl and $1\e{-7}$ for AlF. From the PACS spectra, we find abundances of $9.7\e{-8}$ and $\leq 10^{-8}$, relative to \h2, for HCl and HF, respectively.}
   {The AlF abundance exceeds the solar F abundance, indicating that fluorine synthesised in the AGB star has already been dredged up to the surface of the star and ejected into the circumstellar envelope. From our analysis of chemical reactions in the wind, we conclude that AlF may participate in the dust formation process, but we cannot fully explain the rapid depletion of AlCl seen in the wind.}


   \maketitle
%

\section{Introduction}

Stars on the Asymptotic Giant Branch (AGB) of the Hertzsprung-Russell diagram are an evolved form of low- and intermediate-mass stars with initial masses in the range $\sim 0.8$--$8~\msol$. The AGB evolutionary stage is characterised by intense mass loss, on the order of $\sim 10^{-8}$--$10^{-4}\spy$ \citep{Hofner2018}. The gas ejected by these stars forms molecules and dust in an expanding region surrounding the star, known as a circumstellar envelope (CSE). The CSEs of AGB stars are rich chemical laboratories and a large number of different molecular species have been detected towards various AGB stars \citep{Agundez2020}.

The chemical composition of the CSE is, to the first order, determined by the photospheric abundances of C and O of the star. AGB stars are oxygen-rich if C/O < 1 and carbon-rich if C/O > 1. Generally speaking, more oxygen-bearing molecules are found in the CSEs of oxygen-rich stars, while carbon-bearing molecules are more prevalent in the CSEs of carbon-rich stars. There is thought to be an evolutionary progression such that all stars are oxygen-rich when they transition to the AGB and then a subset of these gradually become carbon-rich as freshly nucleosynthesised carbon is dredged up from the core of the star to the surface, increasing the C/O ratio. S-type stars, with \mbox{C/O $\sim 1$}, are thought to be transition objects that arise during this evolutionary process \citep{Herwig2005}. The circumstellar chemistry of S-type stars has been generally found to be intermediate between carbon- and oxygen-rich chemistry \citep[see, for example,][]{Danilovich2014}.

Halogen-bearing molecules have not been extensively studied in the circumstellar envelopes of many AGB stars. 
{Chlorine has been found to be a tracer of metallicity \citep{Maas2016} and fluorine is thought to be produced in AGB stars and dredged up to the surface of the star \citep{Kobayashi2020}. By understanding the total abundance of Cl or F around AGB stars, we can better understand their metallicities or AGB-ages, respectively.}
To date, halide molecules have been studied in most detail towards the nearby, high mass-loss rate carbon star CW~Leo (IRC+10216), towards which AlCl, NaCl, KCl, AlF, HCl, and HF have been detected \citep{Cernicharo1987a,Cernicharo2010,Agundez2011,Agundez2012}. Aside from CW~Leo, halogen-bearing molecules have only been detected towards a handful of AGB stars, such as Al$^{35}$Cl tentatively seen towards the oxygen-rich stars IK~Tau and R~Dor \citep{Decin2017}, and NaCl seen towards IK~Tau \citep{Milam2007,Decin2018} and tentatively R~Dor \citep{De-Beck2018}. No spectrally resolved halogen-bearing species have been previously reported towards any S-type stars\footnote{Aside from a misidentification of NaCl towards W~Aql by \citet{De-Beck2020}, which will be discussed further in Sect. \ref{saltobs}.}, although spectrally unresolved infrared observations of HCl in the atmosphere of the S-type star R~And have been reported by \cite{Yamamura2000}.

\cite{Agundez2020} undertook an extensive study of molecular abundances in the inner regions of AGB CSEs under the assumption of thermochemical equilibrium, making several predictions for molecular abundances in the inner 10 stellar radii ($R_\star$). They predict AlF and AlCl to be the dominant F- and Cl-bearing molecules from $\sim 3R_\star$ outwards for S-type stars, with HF and HCl dominating the innermost regions ($\lesssim 3R_\star$). Hence, we should expect to see both aluminium and hydrogen halides towards S-type stars.

In recent years, \object{W Aql} has become the most-studied S-type AGB star, thanks to a combination of its proximity \citep[$374 \pm 22$~pc,][]{Gaia-2016,Gaia-EDR3,Lindegren2021}
moderately high mass-loss rate \citep[$3\e{-6}\spy$,][]{Ramstedt2017} and {equatorial position in the sky (declination $\sim-7\deg$)}. It has been observed by two instruments aboard \textsl{Herschel} \citep{Mayer2013,Danilovich2014} and  the Atacama Large Millimetre/sub-millimetre Array \citep[ALMA,][]{Ramstedt2017,Brunner2018}, as well as a variety of other telescopes \cite[see for example][]{De-Beck2020}, which helped constrain the conditions in its circumstellar envelope. In addition to (sub)millimetre observations, optical and infrared observations have provided information on the dust around this star \citep{Hony2009,Mayer2013,Ramstedt2011}, and its companion, which has been characterised as an F9 main sequence star \citep{Danilovich2015}, located at a projected distance of 0\farcs46 \citep{Ramstedt2011}, which corresponds to approximately 170 AU, at the distance given by Gaia.

In this study we focus on the rotational lines of halogen-bearing molecules towards W~Aql, especially AlCl and AlF which were observed with ALMA. These data are presented in Section \ref{almaobs}. {Additionally, we examine observations of HCl and HF obtained by \textsl{Herschel}/PACS (Section \ref{otherobs}). Radiative transfer modelling is performed for all four of these molecules (Section \ref{sec:modov}), with the results presented in Section \ref{sec:modresults}. We discuss our results in the context of the literature and use them to update our chemical kinetics model in Section \ref{sec:discussion}. Section \ref{conclusion} summarises our conclusions.}

\section{ALMA observations of AlCl and AlF}\label{almaobs}

As part of the ATOMIUM\footnote{\url{https://fys.kuleuven.be/ster/research-projects/aerosol/atomium/atomium}} programme (2018.1.00659.L, PI: L.~Decin), W~Aql was observed with three configurations of ALMA, which we will refer to as the compact (angular resolution $1\farcs11\times0\farcs88$ at 262.2 GHz and maximum recoverable scale (MRS) = 8\farcs9), mid (angular resolution $0\farcs374\times0\farcs250$ and MRS = 3\farcs9) and extended (angular resolution $0\farcs024\times0\farcs021$ and MRS = 0\farcs4) arrays \citep[see][ and \cite{Gottlieb2021}, for details]{Decin2020}. These three datasets were combined to produce more sensitive data cubes to allow us to examine the observations in more detail. When data from the different ALMA configurations are combined, different resolutions can be chosen to emphasise different aspects of the data. {We found the most useful combined data to have angular resolution of $32\times30$~mas for AlCl and $150\times130$~mas for AlF.} For this study, we aim to use the best dataset in each context, as will be discussed in detail below.

We detected Al$^{35}$Cl in the ground, first, and second excited vibrational states ($\varv=0,\,1,\,2$), Al\up{37}Cl in the ground and first excited vibrational states ($\varv=0,\,1$), and AlF in the ground vibrational state ($\varv=0$). 
We also use tentative or undetected lines of Al$^{35}$Cl in the third vibrationally excited state, Al\up{37}Cl in the second vibrationally excited state, and AlF in the first vibrationally excited state as upper limits when we perform our radiative transfer analysis of the data (see Sections \ref{sec:modov} and \ref{sec:modresults}).
These lines are listed with their frequencies, upper level energies and central velocities in Table \ref{tab:lines}. The central velocities were found by fitting Gaussian profiles to spectra extracted from the combined cubes. {Based on these central velocities, we find an average LSR velocity of $-23.1\pm0.9~\kms$, which is in good agreement with the values found by \cite{Danilovich2014} and \cite{De-Beck2020} from single dish observations of a variety of molecules.}

Angular sizes are given in Table \ref{tab:lines} for sufficiently bright lines. They have been measured by: examining zeroth moment
maps (of the velocity-integrated emission) of each line over the velocity range indicated in Table \ref{tab:lines}; and creating contours enclosing the flux at the $2\sigma$ level.
{The $2\sigma$ level was chosen since this gives a more accurate estimate of total extent, including weaker emission, than the 3$\sigma$ level. Isolated islands only detected at $2\sigma$ levels are not included since we require at least 3$\sigma$ certainty to consider emission to be detected.}
{This gave a table of $x$ and $y$ values of the coordinates enclosing the flux, centred on the star, which could be transformed to polar coordinates $r$, $\theta$. To reduce random noise affecting the contour, we binned and averaged the values with at least 10 samples per bin, corresponding to angular ranges of $\geq 60\deg$.}
We then measured the longest ($R_\mathrm{max}$) and shortest ($R_\mathrm{min}$) radial distances from the continuum peak to the binned contour.
To give an indication of the regularity or irregularity of the shape of the emission, {we also note the angle ($\theta_\mathrm{R}$) between the radii of the nearest and farthest angle of the contour.  If these are orthogonal the radii can correspond to semi-major and semi-minor axes, suggesting a more regular distribution, but if the angle is very different from $90\deg$, then the distribution is asymmetric.}
For AlF the measurement was done for data that had been combined with a taper of 0\farcs2, giving a lower resolution image but avoiding the irregularities seen in Fig. \ref{alfvsalcl}. {The uncertainty in the radii is $\sim 3$~mas for AlCl and $\sim 18$~mas for AlF.}

We characterise the 1D spectra extracted from ALMA cubes by an aperture size, which is the size of a circular region over which the spectrum has been extracted, and is always centred on the continuum peak. Since the AlF ($7\to6$) $\varv=0$ line is relatively weak, we average channels to give a lower velocity resolution of $\sim2.5~\kms$ (compared with $\sim1.1~\kms$ for AlCl) and a higher signal-to-noise ratio. For our modelling, we also use {what we refer to as} azimuthally averaged radial profiles of the ALMA lines.
{These are extracted from the zeroth moment maps by obtaining the average flux in concentric annuli, plus the flux in the central circular region, centred on the continuum peak.
These radial profiles} allow us to more easily compare ALMA data with our spherically symmetric models (see Sect. \ref{sec:modov} for details of modelling).

\begin{table*}
\caption{Properties of lines of AlCl and AlF covered by ATOMIUM towards W Aql}             
\label{tab:lines}      
\centering                          
\begin{tabular}{l c c c r |c c c |c c cc}        
\hline\hline                 
Molecule	&	\multicolumn{2}{c}{Transition}			&	Frequency	& $E_\mathrm{up}$ & \multicolumn{1}{c}{Aperture} & \multicolumn{1}{c}{$\upsilon_\mathrm{cent}$} & Int. flux\tablefootmark{a} & $R_\mathrm{max}$ & $R_\mathrm{min}$ & $\theta_\mathrm{R}$ & Vel. range\\
	&	$v$	&	$J'\to J$	&	[GHz]& [K]	& \multicolumn{1}{c}{[mas]} & \multicolumn{1}{c}{[$\kms$]} & [Jy $\kms$] & [mas] & [mas] & [$\deg$] & [$\kms$]\\
\hline
Al$^{35}$Cl	&	0	& $	18 \to 17	$ &	262.219	&	120	&	200	&	-23.8	&	0.415	&	93	&	39	&	100	& $-29.4, -16.0$\\
&&&&&80 &-23.7 & 0.279 &	...	&	...	&	...&	...	\\
Al$^{35}$Cl	&	1	& $	17 \to 16	$ &	246.037	&	794	&	80	&	-23.2	&	0.111	&	62	&	28	&	120	& $-14.5, -31.2$\\
Al$^{35}$Cl	&	1	& $	18 \to 17	$ &	260.491	&	807	&	80	&	-22.7	&	\phantom{$^\dagger$}0.138$^\dagger$ &	...	&	...	&	...	&	...\\
Al$^{35}$Cl	&	2	& $	16 \to 15	$ &	230.053	&	1463	&	80	&	-22.3	&	0.066	&	...	&	...	&	...	&	...\\
Al$^{35}$Cl	&	2	& $	17 \to 16	$ & \phantom{*}	244.414	* &	1475	&	80	&	-23.2	&	0.048	&	...	&	...	&	...	&	...\\
Al$^{35}$Cl	&	2	& $	18 \to 17	$ &	258.773	&	1488	&	80	&	-22.7	&	0.068	&	...	&	...	&	...	&	...\\
Al$^{35}$Cl	&	3	& $	15 \to 14	$ &	214.265	&	2117	&	80	&	...	&	ND	&	...	&	...	&	...	&	...\\
Al$^{35}$Cl	&	3	& $	16 \to 15	$ &	228.534	&	2128	&	80	&	...	&	ND	&	...	&	...	&	...	&	...\\
Al$^{37}$Cl	&	0	& $	16 \to 15	$ &	227.643	&	93	&	200	&	-23.3	&	0.209	&	63	&	23	&	200	& $-27.8, -18.7$\\
Al$^{37}$Cl	&	1	& $	18 \to 17	$ &	254.396	&	796	&	80	&	-23.7	&	0.087	&	...	&	...	&	...	&	...\\
Al$^{37}$Cl	&	1	& $	19 \to 18	$ & \phantom{*}	268.509	* &	809	&	80	&	-23.5	&	0.084	&	...	&	...	&	...	&	...\\
Al$^{37}$Cl	&	2	& $	18 \to 17	$ & \phantom{*}	252.738	* &	1020	&	80	&	-20.8	&	0.083	&	...	&	...	&	...	&	...\\
AlF	&	0	& $	7 \to 6	$ &	230.794	&	44	&	600	&	-24.3	&	0.635	&	769	&	284	&	91	& $-50.7, 5.2$\\
&&&&&300 &-23.8 & 0.383 &	...	&	...	&	...	&	...\\
AlF	&	1	& $	7 \to 6	$ &	228.717	 &	1185	&	200	&	...	&	ND	&	...	&	...	&	...	&	...\\
\hline                                   
\end{tabular}
\tablefoot{(\tablefootmark{a}) Integrated flux density. (*) indicates a tentative detection. ($^\dagger$) indicates an uncertain measurement due to a partial overlap with the wing of an adjacent line. Aperture gives the radius of the spectral extraction aperture; $R_\mathrm{min}$ and $R_\mathrm{max}$ give the minimum and maximum angular extents of the emission, measured from the continuum peak for a zeroth moment map {made by summing over the channels in the velocity range indicated}, and $\theta_\mathrm{R}$ is the angle between $R_\mathrm{min}$ and $R_\mathrm{max}$. {Angular extents are omitted for unresolved emission.} ND indicates a non-detection. \textit{References:} line data retrieved from CDMS \citep{Muller2001,Muller2005} with Al$^{35}$Cl and Al$^{37}$Cl frequencies from \cite{Wyse1972} and \cite{Hensel1993} and AlF frequencies from \cite{Wyse1970} and \cite{Hoeft1970}.}
\end{table*}

\subsection{AlCl}

\begin{figure*}
   \centering
   \includegraphics[width=\hsize]{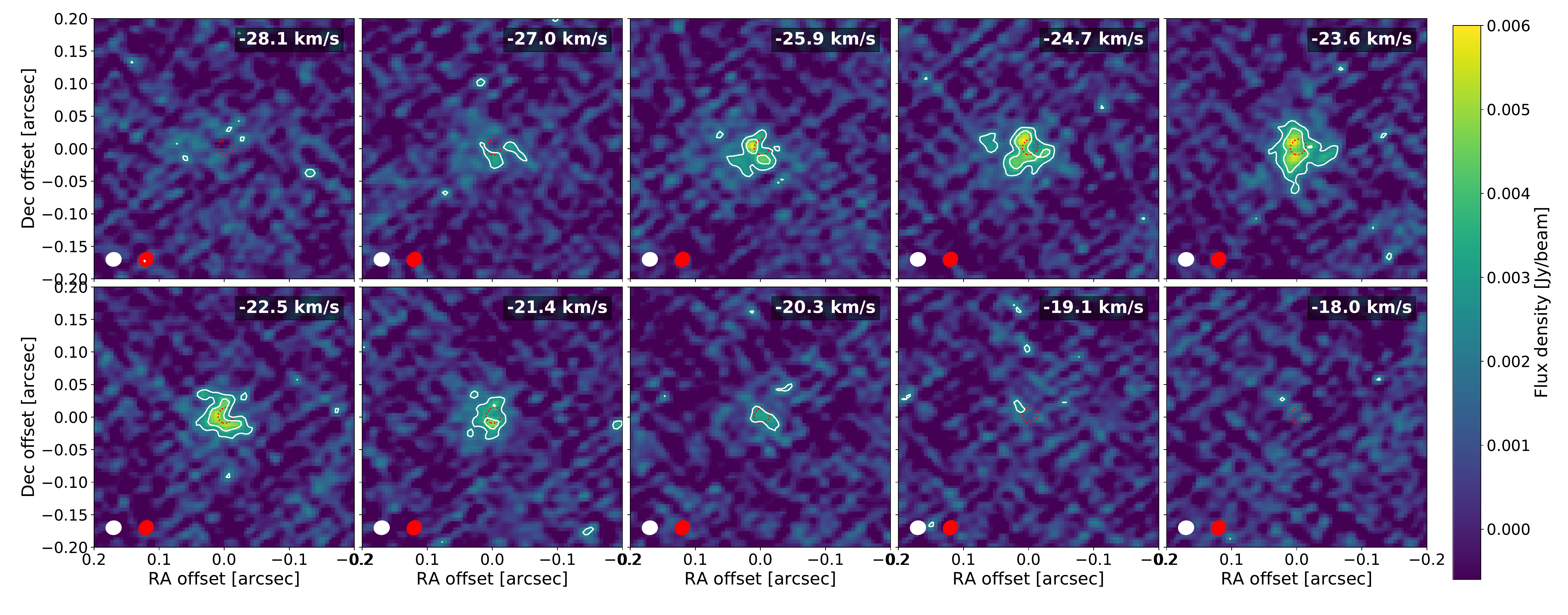}
      \caption{Channel maps of Al$^{35}$Cl ($18\to17$) observed towards W~Aql using ALMA in the extended array configuration. The dotted red contours enclose 50\% of the total continuum flux. The white contours indicate levels of 3 and 5$\sigma$. The ellipses in the bottom left of each panel indicate the synthetic beams for the molecular emission (\textit{white}) and continuum emission (\textit{red}).}
         \label{alclchanmaps}
   \end{figure*}

\begin{figure*}
   \centering
   \includegraphics[width=0.33\hsize]{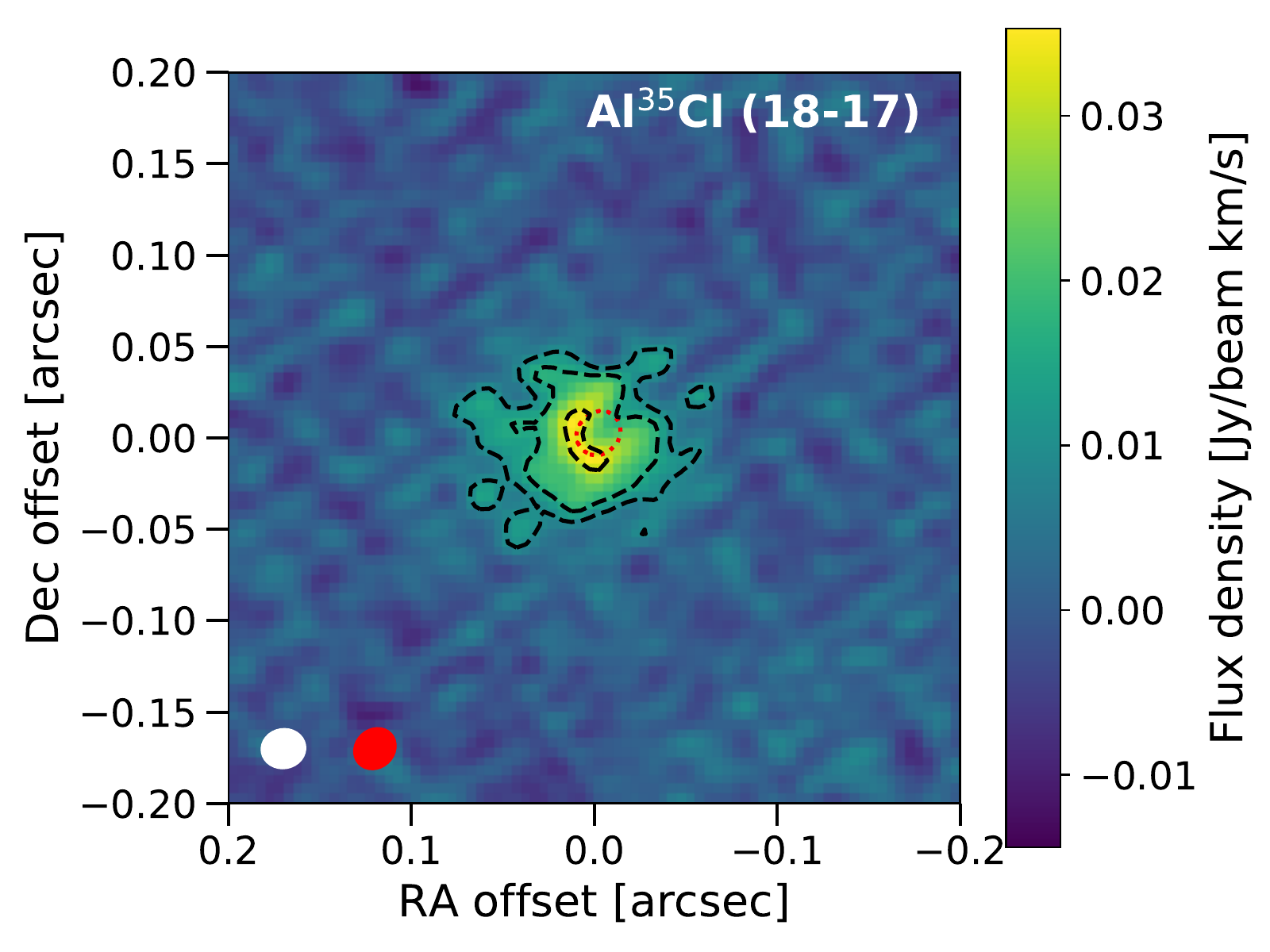}
   \includegraphics[width=0.33\hsize]{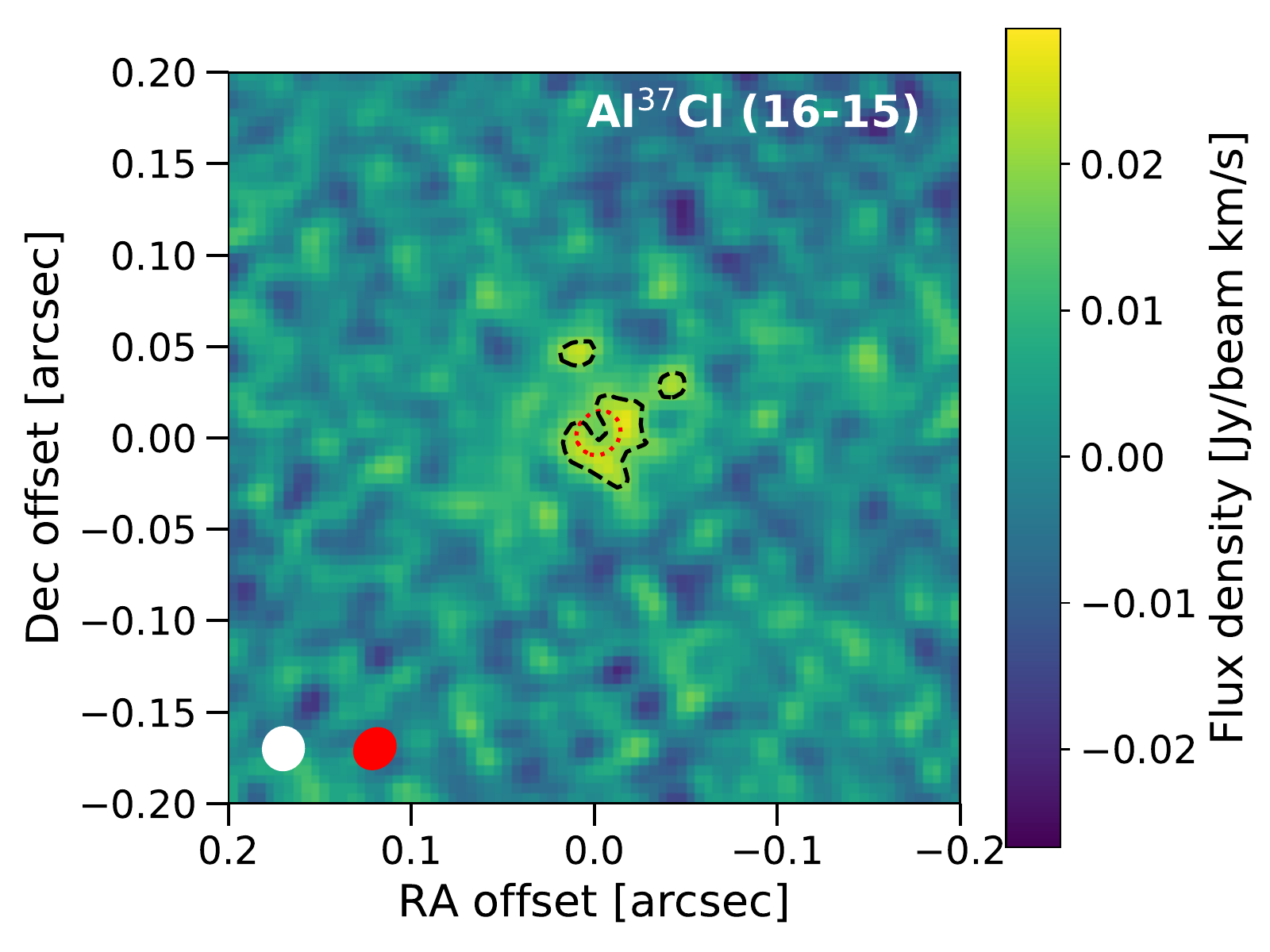}
      \includegraphics[width=0.33\hsize]{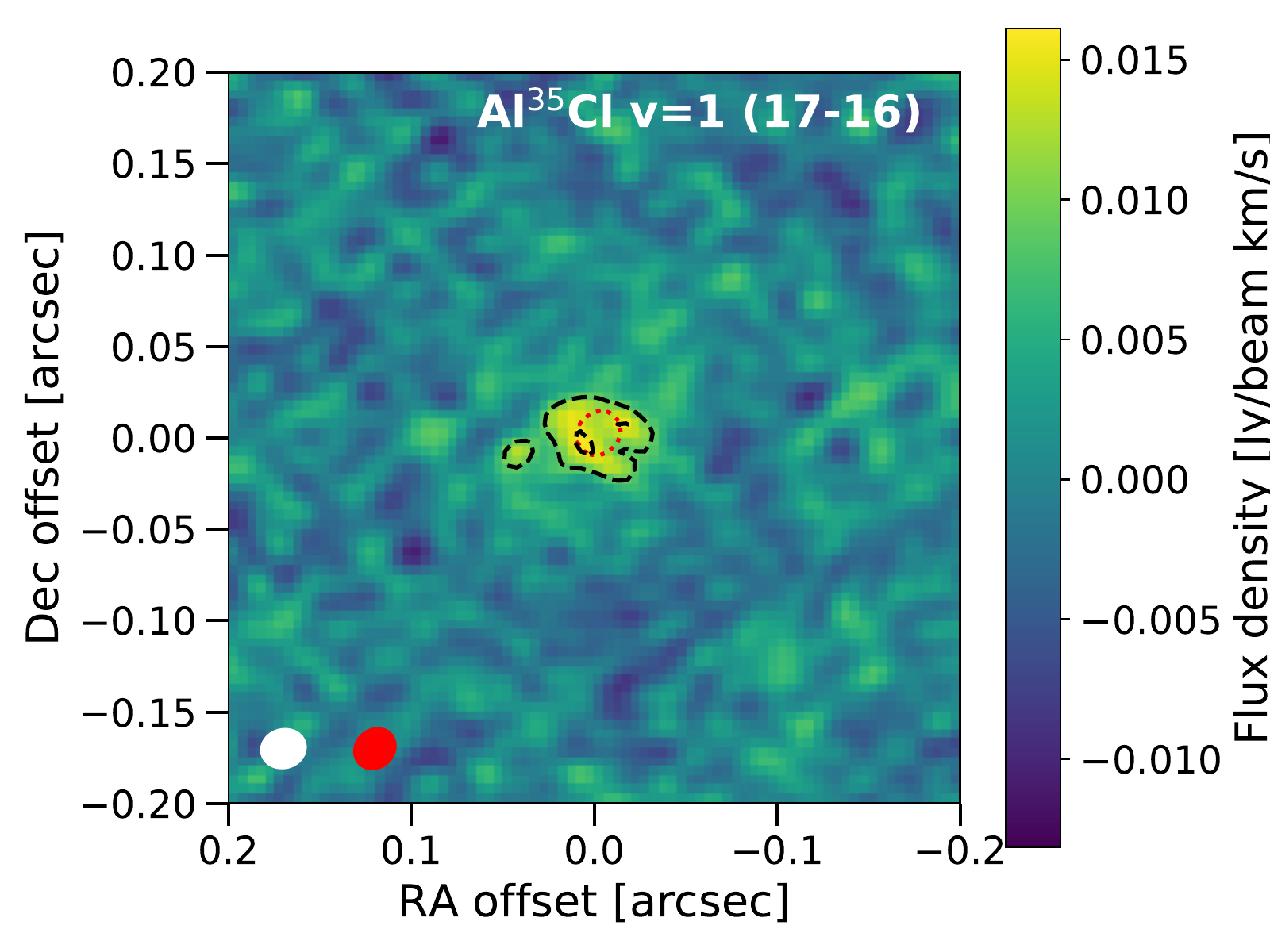}
      \caption{Zeroth moment maps of AlCl observed towards W~Aql using ALMA in the extended array configuration. Transitions are labelled in the top right of each panel. The black contours indicate levels of 3, 5, 10$\sigma$. The dotted red contour encloses 50\% the total continuum flux. The ellipses in the bottom left of each panel indicate the synthetic beams for the molecular emission (\textit{white}) and continuum emission (\textit{red}).}
         \label{alclmom0}
   \end{figure*}

The AlCl lines seen towards W~Aql are all fairly compact, including the $\varv=0$ lines. The channel maps for Al$^{35}$Cl ($18\to17$) in the ground vibrational state, observed with the ALMA extended array, are shown in Fig. \ref{alclchanmaps}. There it can be seen that the emission is resolved and clumpy, but seen within 0\farcs1 of the star. The Al$^{35}$Cl $\varv=1$ lines and the Al$^{37}$Cl ($16\to15$) line in the ground vibrational state are found in similar regions but are seen less clearly, due to their expected lower intensities. We show the zeroth moment maps of Al$^{35}$Cl ($18\to17$) and Al$^{37}$Cl ($16\to15$) in the ground vibrational state, and ($17\to16$) in the first vibrationally excited state in Fig. \ref{alclmom0}.
The zeroth moment maps were all constructed so as to include all of the line flux with no contamination from adjacent lines and with minimal dilution due to line-free noise-dominated channels. The zeroth moment map of Al$^{35}$Cl $\varv=0$ ($18\to17$) shows the emission is not centred on the star, with more emission seen to the southeast than the northwest. However the emission from the other transitions shown in Fig. \ref{alclmom0} is not consistently offset in the same direction. We conclude that the cause of this asymmetry is {either noise or} clumpy emission, rather than a specific directional bias for the formation of AlCl, {though the latter is not decisively ruled out}.
The spectral lines of AlCl are presented in conjunction with our models in Sect. \ref{sec:modresults}, {where we show all the lines extracted for an aperture of radius 0\farcs08 and include the Al$^{35}$Cl $\varv=0$ ($18\to17$) line extracted for a radius of 0\farcs2, to ensure all the flux is captured for the purposes of comparisons with our model.}

\subsection{AlF}

\begin{figure}
   \centering
   \includegraphics[width=\hsize]{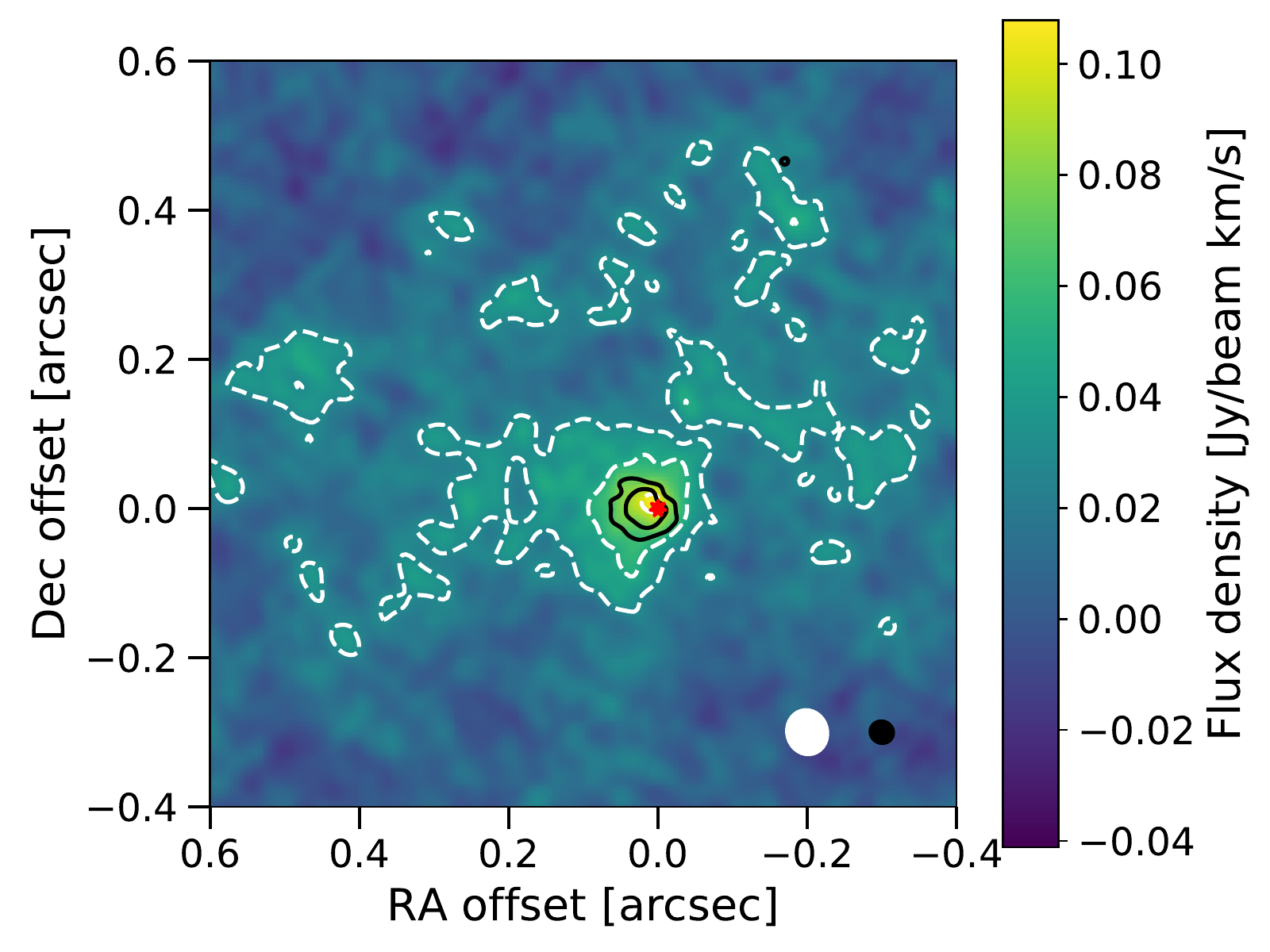}
      \caption{The zeroth moment map of the AlF ($7\to6$) line in the ground vibrational state (colours and white dashed contours) plotted with the zeroth moment map of the Al$^{35}$Cl ($18\to17$) line in the ground vibrational state (black solid contours). The contours are plotted at levels of 3, 5, 10$\sigma$ for AlF and $5,10\sigma$ for Al$^{35}$Cl. The red star indicates the position of the continuum peak. The white and black ellipses in the bottom right indicate the synthetic beam sizes for the AlF and Al$^{35}$Cl emission, respectively.}
         \label{alfvsalcl}
   \end{figure}

The ground state AlF emission is more extended than AlCl. It has a bright central region, within $\sim0\farcs15$ of the star, and a more extended region of faint, clumpy emission. The extended emission is seen most clearly in a zeroth moment map with a lower angular resolution of $150\times130$~mas (Fig. \ref{alfvsalcl}), which allows us to see the faint emission above the noise, with dashed white contours outlining the emission down to the $3\sigma$ level. As can be seen in Fig. \ref{alfvsalcl}, the extended diffuse emission is mostly found to the north of the stellar position. Relative to the stellar continuum peak, the diffuse emission extends to $\sim 0\farcs5$ north, $\sim 0\farcs6$ east, $\sim 0\farcs4$ west and only $\sim 0\farcs2$ south. 
We include contours of the Al$^{35}$Cl ($18\to17$) zeroth moment map (with an angular resolution of $32 \times 30$~mas) in Fig. \ref{alfvsalcl}. Although the offsets noted for AlF and Al$^{35}$Cl are not in the same direction, we find that the central peak of AlF emission corresponds reasonably well with the region of Al$^{35}$Cl emission.

\section{Observations of other halide molecules}\label{otherobs} 

\subsection{NaCl and KCl}\label{saltobs}

Several transitions of NaCl and KCl were covered in the ATOMIUM observations but not detected towards W~Aql. ATOMIUM detections of these two molecules were made towards the higher mass-loss rate oxygen-rich stars (and the red supergiant VX Sgr) and will be examined in more detail in a future work. We calculate the rms values as detection limits for NaCl and KCl towards W~Aql in Appendix \ref{saltnds}.

In their recent study, \cite{De-Beck2020} analysed a spectral survey of W~Aql conducted by the Atacama Pathfinder Experiment (APEX). They find emission from 13 species and their isotopologues, in addition to two unidentified lines (the latter falling outside of our observed frequency range). They do not detect AlCl or AlF but report tentative detections of NaCl, both of which fall within our observed frequency range at 247.2397 GHz for the ($19\to18$) line and 260.2231 GHz for the ($20\to19$) line.

The line at 247.2397~GHz falls on the edge of a band in our data, however a clear partial detection is visible, both in the spectrum and the channel maps. After correcting for the LSR velocity of $-23~\kms$ \citep{Danilovich2014}, we found that the 247.2397~GHz line did not coincide with the emission we detect. A more likely carrier of the emission seen near this band edge is the $^{30}$SiO ($6\to5$) line in the $\varv=4$ state at 247.244~GHz, owing to a better agreement with the systemic velocity of W~Aql {for the spatially compact emission of this line}. In the case of the possible NaCl line at 260.2231~GHz, emission is present at this frequency, however we identify it as the 260.2248~GHz line of H$^{13}$CN, with $J=3\to2$ in the $\nu_2=1$ excited bending vibrational level, which is in better agreement with the systemic velocity of W~Aql. The H$^{13}$CN line is also similar in brightness and appearance to the other {component of the $l$-type doublet} (with the same $J$ and $\nu_2$ values) seen at 258.9361~GHz, thereby confirming the assignment. {An additional line of NaCl, ($17\to16$) at 221.2601~GHz, was covered in our observations and by \cite{De-Beck2020}, but was not detected in either case.}

\subsection{HCl and HF}

{A search of the literature revealed that, aside from the metal halides discussed above, the only other published detections} of halogen-bearing molecules towards AGB stars are the hydrides HCl and HF towards the carbon star CW~Leo \citep{Cernicharo2010,Agundez2011}. The frequencies of both HCl and HF are well outside of the observing window of the ATOMIUM project. However, some transitions of HCl and HF were covered by the \textsl{Herschel}/PACS spectrograph \citep{Pilbratt2010,Poglitsch2010} and observed towards several AGB stars \citep{Groenewegen2011,Nicolaes2018}. Although spectrally resolved transitions of HCl and HF were observed by \cite{Agundez2011} towards CW~Leo with \textsl{Herschel}/HIFI \citep{de-Graauw2010}, all of these lines are outside of the frequency range at which W~Aql was observed with \textsl{Herschel}/HIFI \citep{Danilovich2014}.

We searched the PACS spectrum of W~Aql \citep{Nicolaes2018} and found evidence of H$^{35}$Cl and H$^{37}$Cl emission in the $J=3\to2$ and $J=4\to3$ lines. In both cases, there is a partial overlap of the H$^{35}$Cl and H$^{37}$Cl lines, since the emission is not spectrally resolved by PACS. The H$^{37}$Cl ($4\to3$) line is also known to be blended with H$^{13}$CN ($29\to28$) at $120.12~\um$ and SiO ($58\to57$) at $120.14~\um$. Without spectrally resolved observations available, we cannot disentangle the contributions of these lines to the H$^{37}$Cl ($4\to3$) line, but they are the most likely reason that the H$^{37}$Cl appears brighter than the {more abundant} H$^{35}$Cl.
We do not find clear detections of the $J=6\to5$ and $J=7\to6$ lines for either HCl isotopologue. However, we still compare the PACS spectrum in the region of these two lines to our model results (see Sect. \ref{sec:hclmod}), so that they can serve as upper limits. The details of these lines are listed in Table \ref{tab:PACSlines}. We also include the measured fluxes of the detected lines, which were calculated by fitting gaussians to the spectra.

We also searched the PACS spectrum for HF, for which three transitions fall within the observed PACS range: $J=2\to1$, $J=3\to2$, and $J=4\to3$ (see Table \ref{tab:PACSlines}). The lowest energy of these lines, at 121.70$\um$, is unfortunately blended with a bright ortho-\h2O line at 121.721$\um$. Higher spectral resolution observations would be needed to distinguish these two lines if the HF line is indeed present. We did not conclusively detect the two higher energy lines, {although there is a faint line ($\sim2\sigma$) that we tentatively attribute to HF ($4\to3$).} As for HCl, we include these lines in our model so that they might allow us to derive an upper limit for the abundance of HF.

\begin{table}
\caption{HCl and HF lines covered by \textsl{Herschel}/PACS towards W~Aql}             
\label{tab:PACSlines}      
\centering                          
\begin{tabular}{c c c r c}        
\hline\hline                 
	&			&	$\lambda$	& $E_\mathrm{up}$ & Flux \\
	& $J'\to J$	&	\multicolumn{1}{c}{[$\mu$m]}& [K]	& [ $10^{-17}$ W m$^{-2}$]\\
\hline
H$^{35}$Cl & $3\to2$ & 159.78 & 180 & $2.08\pm 1.00$\\
H$^{37}$Cl & $3\to2$ & 160.02 & 180 & ($0.77\pm 1.02$)\\
H$^{35}$Cl & $4\to3$ & 119.92 & 300 & $3.15\pm0.88$\\
H$^{37}$Cl & $4\to3$ & 120.10 & 300 & $4.23 \pm 1.27$\\
H$^{35}$Cl & $6\to5$ & \phantom{0}80.11 & 630 &ND\\
H$^{37}$Cl & $6\to5$ & \phantom{0}80.23& 629& ND\\
H$^{35}$Cl & $7\to6$ & \phantom{0}68.76& 840 &ND\\
H$^{37}$Cl & $7\to6$ & \phantom{0}68.86 & 838 & ND\\
\hline
HF & $2\to 1$ & 121.70 & 178 & \h2O blend\\
HF & $3\to 2$ & \phantom{0}81.22 & 355 & ND\\
HF & $4\to 3$ & \phantom{0}61.00 & 591 & ND\\
\hline                                   
\end{tabular}
\tablefoot{ND denotes a non-detection, parentheses indicate a tentative detection. Wavelengths and energies taken from \cite{Nolt1987} via the JPL Molecular Spectroscopy Database \citep{Pickett1998}.}
\end{table}

\section{Modelling overview}\label{sec:modov}

\subsection{Radiative transfer model}\label{sec:rtmodov}

\begin{figure}
   \centering
   \includegraphics[width=\hsize]{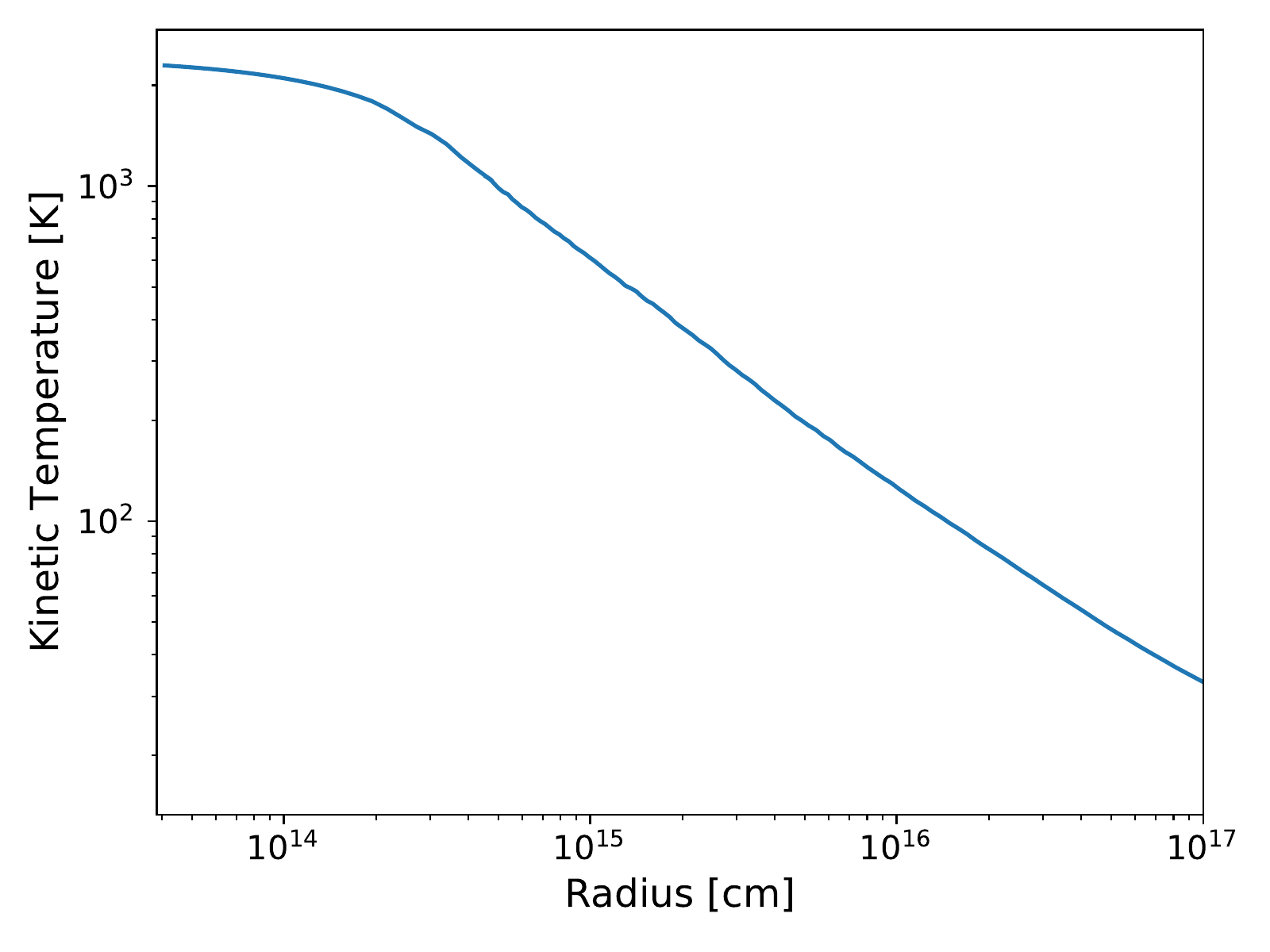}
      \caption{The radial gas kinetic temperature profile used in our modelling of W Aql.}
         \label{kintemp}
   \end{figure}

{Radiative transfer modelling is performed to determine the abundances and abundance distributions of our observed halide molecules.}
For the radiative transfer modelling of AlCl and AlF, we use a spherically symmetric model and the accelerated lambda iteration method \citep[ALI,][]{Rybicki1991}, previously used to model various molecules in the circumstellar envelope of W~Aql \citep{Danilovich2014,Ramstedt2017,Brunner2018}. For this study, we use the circumstellar parameters derived by \cite{Danilovich2014}, including the assumption of silicate dust, and include the adjusted mass-loss rate found by \cite{Ramstedt2017}. {The same velocity profile\footnote{We note our use of the expansion velocity implemented in previous radiative transfer models of W~Aql, $\upsilon_\infty=16.5~\kms$ \citep{Danilovich2014,Ramstedt2017,Brunner2018}, which is smaller than the maximum velocity derived by \cite{Gottlieb2021}, of $\upsilon_\infty=27.1~\kms$, partly due to low-intensity, high-velocity wings. As discussed in \cite{Danilovich2014}, larger expansion velocities can be found for W~Aql due to an auxiliary feature seen in the blue part of many molecular line profiles. This is routinely excluded from 1D radiative transfer modelling, an especially valid approach here, since we do not detect this feature or any other high velocity components in the AlCl or AlF emission.} is used here as in \cite{Danilovich2014}:
\begin{equation}\label{eq:vel}
\upsilon(r) = \upsilon_0 + (\upsilon_\infty - \upsilon_0)\left( 1 - \frac{R_\mathrm{in}}{r} \right)^\beta
\end{equation}
with the parameters listed} in Table \ref{tab:stellarparams} with the other circumstellar and stellar parameters. {The gas kinetic temperature was derived in \cite{Danilovich2014} and \cite{Ramstedt2017} from CO radiative transfer modelling \citep[see][for a discussion of the heating and cooling terms]{Schoier2001,Danilovich2014}. As plotted in Fig. \ref{kintemp} we extend the kinetic temperature profile inwards such that the stellar effective temperature is not exceeded at the stellar surface.}

{For consistency with previous models, we use the distance for W~Aql obtained by \cite{Danilovich2014} of 395~pc, which was found using a period-luminosity relation. This value is within the uncertainties of the parallax value from the Gaia Early Data Release 3 \citep{Gaia-2016,Gaia-EDR3}, which gives a distance of $374 \pm 22$~pc after applying the corrections of \cite{Lindegren2021}.}

\begin{table}
\caption{Stellar and circumstellar parameters of W~Aql used in radiative transfer modelling.}             
\label{tab:stellarparams}      
\centering                          
\begin{tabular}{l l}        
\hline\hline                 
Distance & 395~pc\\
Effective temperature, $T_\star$ & 2300~K\\
Dust optical depth\tablefootmark{a} at $10\um$ & 0.6\\
Luminosity, $L_\star$ & $7500~\lsol$\\
Stellar LSR velocity, $\upsilon_\mathrm{LSR}$ & $-23~\kms$\\
Expansion velocity, $\upsilon_\infty$ & $16.5~\kms$\\
Minimum velocity, $\upsilon_0$ & $3~\kms$\\
Velocity index, $\beta$ & 2\\
Dust condensation radius, $R_\mathrm{in}$ & $2\e{14}$~cm\\
Mass-loss rate, $\dot{M}$ & $3\e{-6}\spy$\\
\hline                                   
\end{tabular}
\tablefoot{Parameters taken from \cite{Danilovich2014,Ramstedt2017}; (\tablefootmark{a}) See text for details of dust treatment for AlCl.}
\end{table}

For AlCl and AlF, we use a combination of spectral lines and azimuthally averaged radial profiles {(centred on the continuum peak)} to determine the molecular abundances and extents of the molecular envelopes. From earlier models of spatially resolved ALMA observations \citep[such as][]{Danilovich2019}, we found that this method was best for finding the abundance distribution of the inner wind --- most clearly seen in the ALMA azimuthal profile --- while still constraining the outer wind. In general, fainter outer emission is often below the noise in the azimuthal profiles, but its signatures are more clearly seen in spectral lines, especially those with emitting regions further out in the CSE, such as the $\varv=0$ AlF line.

Since the emitting regions for AlCl and AlF are quite different (see Fig. \ref{alfvsalcl}), there are some differences in how we treat the two molecules, {particularly with respect to how dust is implemented in the models}. AlF is treated similarly to other molecules in \cite{Brunner2018}, using the dust characteristics obtained by \cite{Danilovich2014}, {since a significant portion of the AlF emission is detected outside the dust condensation radius} \citep[$R_\mathrm{in}=2\times 10^{14}$~cm,][]{Danilovich2014}. 

From the ALMA observations, AlCl is predominantly found within 0\farcs05 ($\approx 3\times 10^{14}$~cm) of the star --- i.e. a significant portion of the AlCl is located within the dust condensation radius. One limitation of our ALI code is that it is not possible to simultaneously consider a dust-free inner region and a dusty outer region. To overcome this limitation, we run a dust-free model for AlCl but include an additional infrared radiation field, based on a blackbody with a temperature of 700~K, which is close to the dust temperature at $3\e{14}$~cm in the \cite{Danilovich2014} model. 
This method is only used for AlCl, since it is the most compact emission we see. For HCl and HF, we start our model at $2\times 10^{14}$~cm (i.e. the dust condensation radius), following the method in \cite{Danilovich2014}, since we have no observational information on the behaviour of these molecules in the inner wind.

Although AlF is more extended than AlCl, it is still relatively compact, compared with, for example, the molecules studied towards W~Aql by \cite{Brunner2018}, notably CS, SiS, and (quasi-thermal) SiO, 
{which extend out to $\sim2$\arcsec{} from the continuum peak.}
The predicted distributions of HCl and HF {(see Sect. \ref{chemmod})} are more similar {in spatial extent} to the molecules modelled by \cite{Brunner2018}, so we include the overdensity derived in that study, increasing the density of the wind by a factor of five between $8\e{15}$~cm and $1.5\e{16}$~cm. This overdensity, intended to represent the denser region of a spiral arm \citep[see][for plots of the CO distribution]{Ramstedt2017}, is outside of the region modelled for AlF or AlCl, but may contribute to the flux of the HCl and HF lines, {which extend past the overdense region}.

\subsection{Molecular data}

A radiative transfer analysis of molecular observations requires a comprehensive list of molecular energy levels and transitions, including Einstein A coefficients and collisional (de-)excitation rates. We prefer to use collisional (de-)excitation rates measured or calculated for collisions between the target molecule and \h2, but these are not always available --- or are not available for our domain of interest --- so some substitutions must be made. For most of the molecules studied here, we obtained the level and radiative information from the ExoMol database\footnote{\url{https://exomol.com}} \citep{Tennyson2016} and the collisional rates from a variety of sources detailed for each molecule {in Appendix \ref{colrates}. In particular, we use newly updated collisional rates for AlF, which cover the high temperatures seen for AGB CSEs and are described in detail in Sect. \ref{alfrates}.}

Generally, collisional rates have only been calculated for the ground vibrational states of various molecules. Hence, we ensure that the rates used cover, at minimum, all levels that participate in our observed transitions in the ground vibrational state. Often, the available collisional rates are calculated for lower temperatures than seen in our models, for example, rates calculated for maximum kinetic temperatures of 300~K are common. In these cases, collisional rates for higher kinetic temperatures are {linearly extrapolated in log-log space} from the given rates.

\subsection{Radial distribution of HCl and HF from chemical modelling}\label{chemmod}

In the absence of spatially resolved observations of HCl and HF, and having only a small number of spectrally unresolved lines, we turned to predictions from chemical models to determine the HCl and HF abundance distributions.
We use the one-dimensional chemical kinetics model of \cite{Van-de-Sande2018b}, and the publicly available gas-phase only \textsc{Rate12} reaction network\footnote{\url{http://udfa.ajmarkwick.net/index.php?mode=downloads}} \citep{McElroy2013}. The chemical model assumes the same mass-loss rate and stellar radius as the radiative transfer model, {and a constant expansion velocity of $16.5~\kms$. 
The retrieved gas temperature profile (Fig. \ref{kintemp}) was reproduced in the chemical model using two power-laws ($r < 2\e{14}$~cm, $T =  2100 (r/6.8\e{13})^{-0.16}$~K; $r \geq 2\e{14}$~cm, $T = 3350 (r/8.6\e{13})^{-0.7}$~K). }

To estimate the shape of the abundance profiles of HF and HCl, we assumed an initial abundance of F and Cl corresponding to their solar abundance minus the inner abundances calculated for AlCl and AlF from early models (see Sect. \ref{sec:alclresults} and \ref{sec:alfmod}). 
Both halogens are efficiently hydrogenised {into HF and HCl} at the start of the model. The other Cl- and F-bearing species included in the reaction network (which does not include AlCl or AlF) play a negligible role.
{Our chemical model is expanded to include Al chemistry in Sect. \ref{sec:chemistry}, where we discuss the reactions important in the syntheses of AlCl and AlF.}

The shape of the predicted radial abundance distribution does not change with changes in initial Cl or F abundances. Hence, {in our radiative transfer analysis,} we find the best fitting model by scaling the radial abundance distribution until the model lines best reproduce the observed data. When finding upper limits on the abundance, we scale the radial abundance profile such that the model lines do not exceed the observed lines.


%
%

\section{Model results}\label{sec:modresults}

\subsection{AlCl}\label{sec:alclresults}

\begin{figure*}
   \centering
   \includegraphics[width=\hsize]{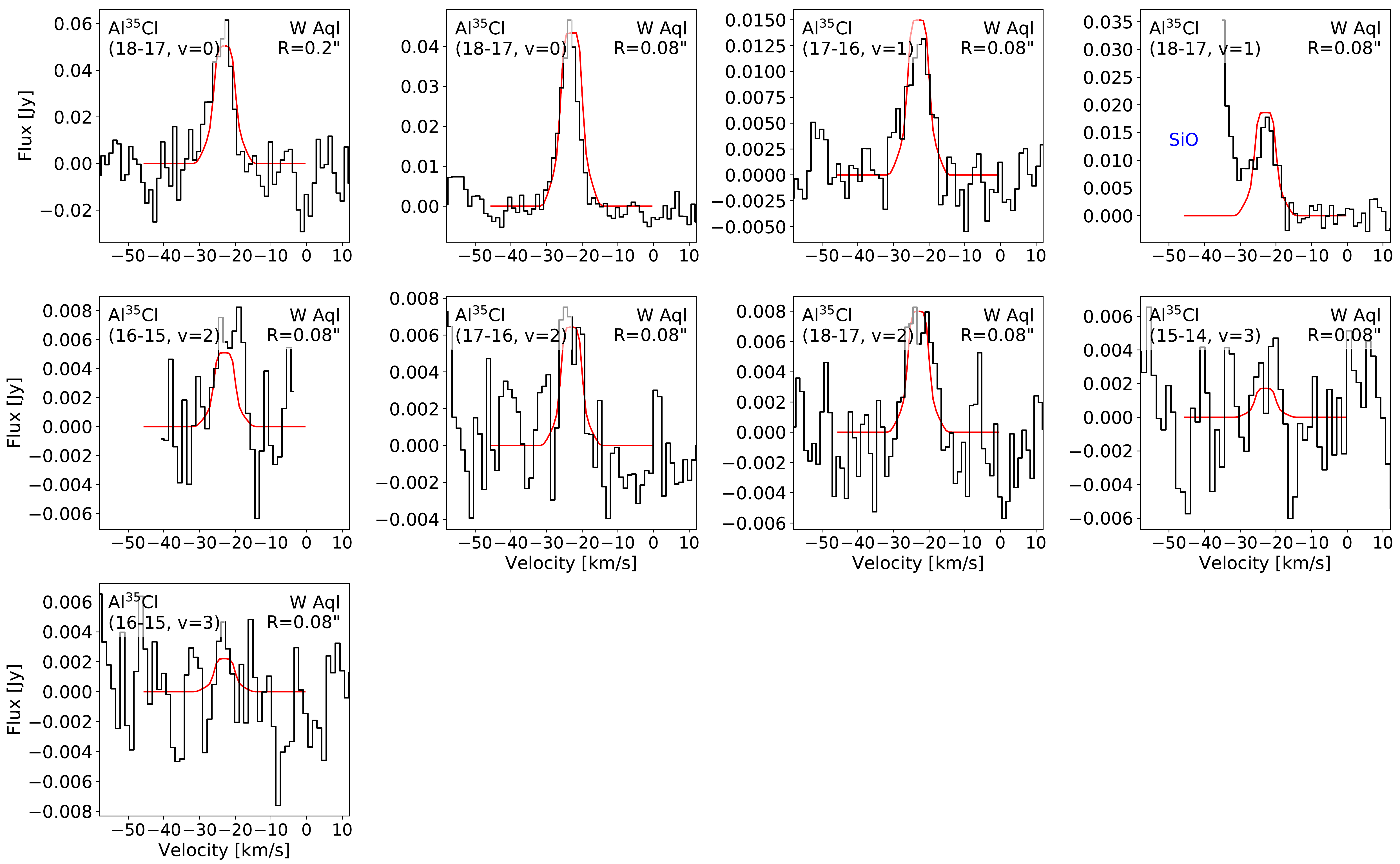}
      \caption{Spectra of Al$^{35}$Cl observed towards W Aql with ALMA (black histograms), with calculated line profiles (red curves) superposed on the observed profiles. {Nearby lines of other molecules are labelled in blue.}}
         \label{alclmod}
   \end{figure*}

\begin{figure}
   \centering
   \includegraphics[width=\hsize]{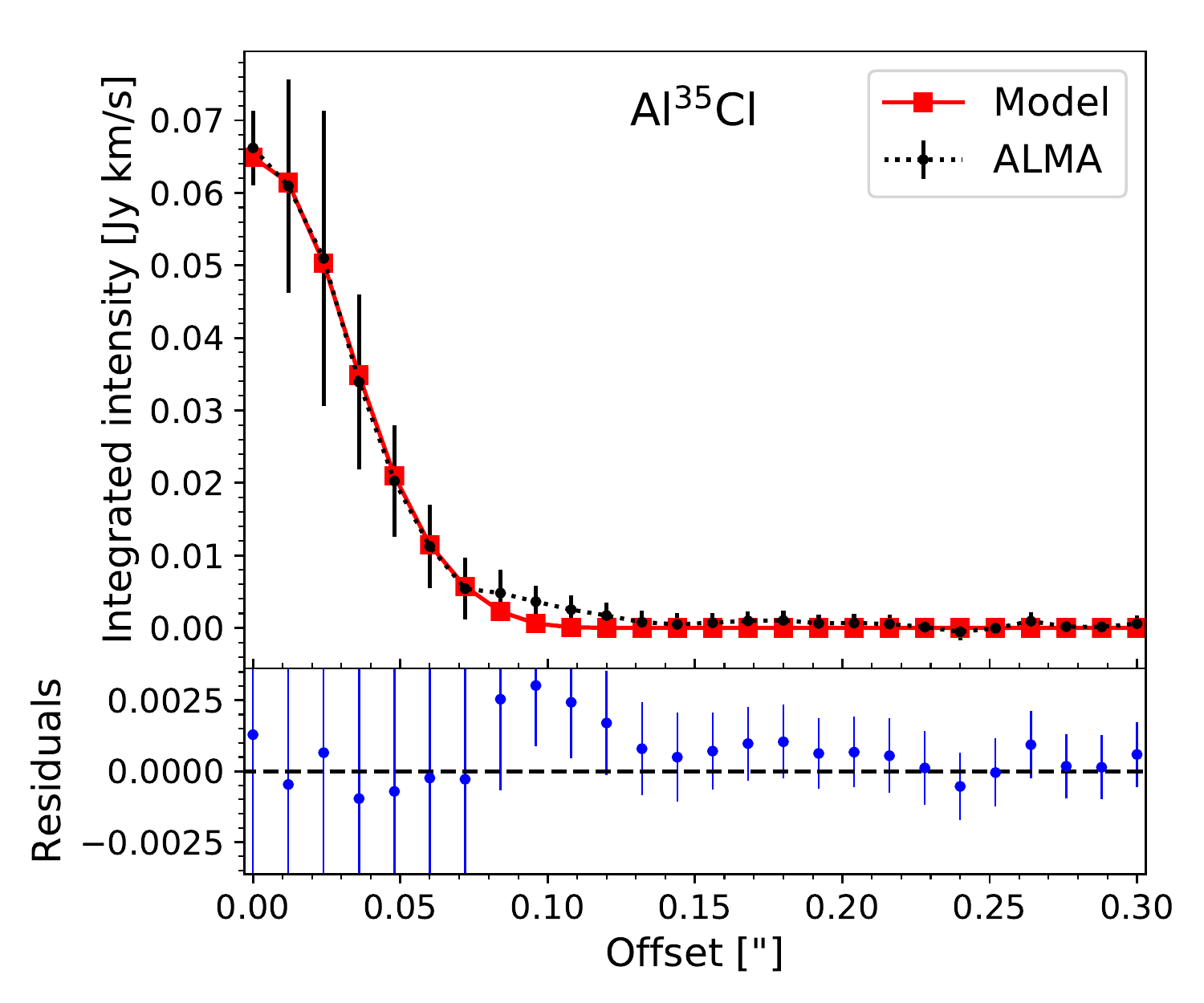}
   \includegraphics[width=\hsize]{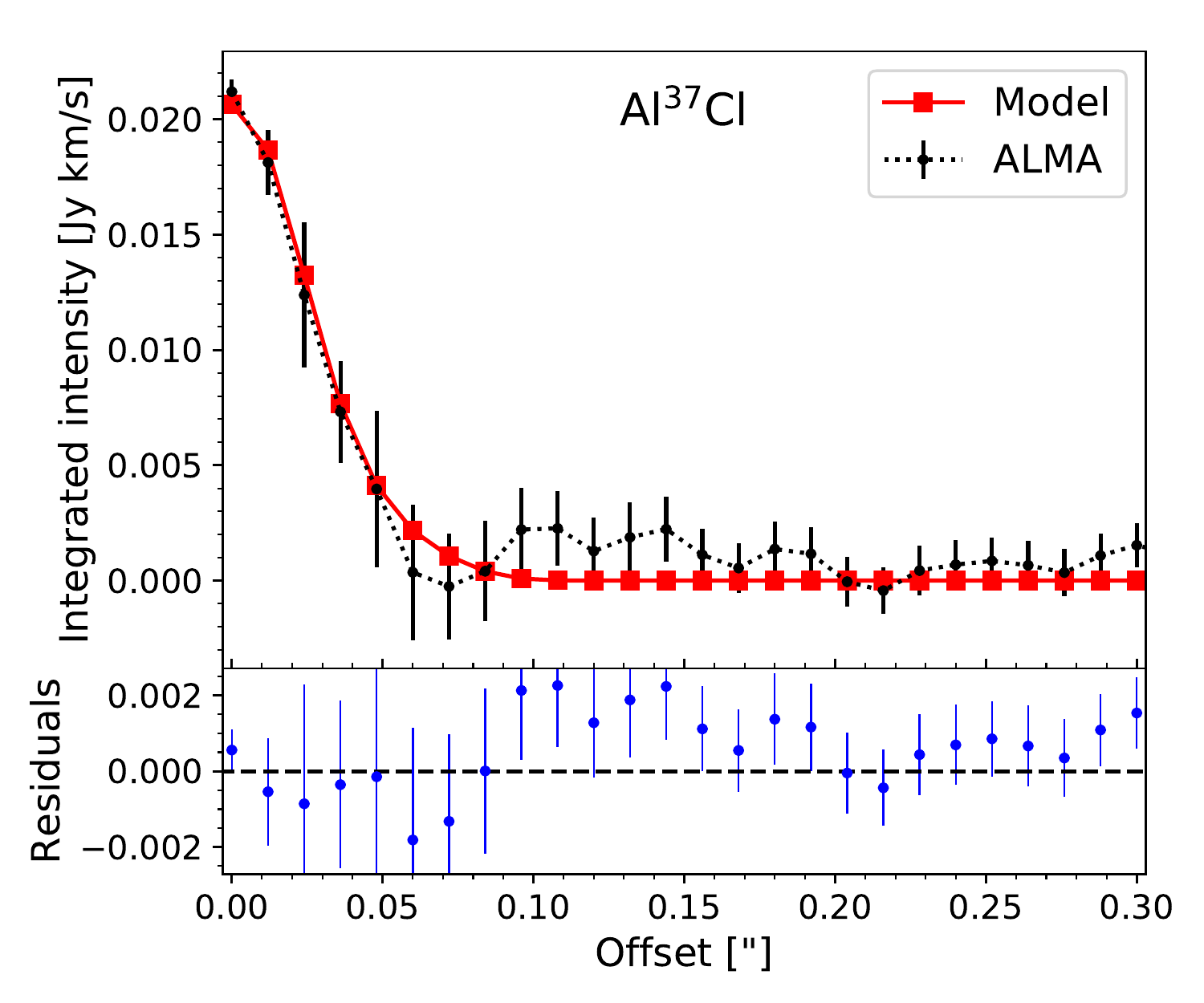}
      \caption{Azimuthally averaged radial profiles extracted from ALMA $\varv=0$ lines of AlCl (black dotted lines and points with error bars) plotted with the corresponding modelled radial profiles (red lines and squares). {Residuals are plotted in the lower panels, showing the difference between the observed and modelled lines, with error bars from the observations included.}}
         \label{alclradprofs}
   \end{figure}

In attempting to fit the observed spectral lines of Al$^{35}$Cl, we started with a constant radial abundance profile to first determine the outer extent of the molecular envelope. To also fit the inner part of the azimuthally averaged radial profile, we needed to include a step down in abundance in the inner region. Our final best-fitting model, which agreed well with the radial profile and the spectral lines, had an inner abundance $f_0=8.5\e{-8}$, relative to \h2, from the stellar surface out to $R_\mathrm{step}=1.4\e{14}$~cm ($\approx 3.6~R_\star$). Then between $1.4\e{14}$~cm and an outer radius $R_\mathrm{max}=5\e{14}$~cm ($\approx 13~R_\star \approx 0\farcs08$), we find a relative abundance of $1.7\e{-7}$. The model lines are plotted with the observed spectra in Fig. \ref{alclmod}. We also plot the observed and modelled azimuthally averaged radial profiles in Fig. \ref{alclradprofs}, {with a residual plot showing the difference between the modelled and observed radial profile points}. As can be seen in those plots, our model reproduces the observed data well. The main deviation between our model and the observations is the small ``tail'' in the radial profile seen around 0\farcs1. We were not able to reproduce this with our model, {possibly because it is caused by asymmetry in the detected emission (see Figs. \ref{alclchanmaps} and \ref{alclmom0}). When} we increased the outer radius of the model we reproduced the ``tail'' but failed to reproduce the shapes of the spectral line profiles. Using a larger outer radius in our model tended to increase the amount of emission coming from outside of $2\e{14}$~cm, which is the region in which the gas accelerates. Emission from outside of $2\e{14}$~cm is generally broader and a model that reproduced the ``tail'' caused the calculated line profile to have a broad base with wings approximately 30\% of the peak line flux. This is contrary to what is seen in the observations, especially for the $\varv=0$ ($18\to17$) line, which has the highest signal-to-noise ratio. Our best model includes part of the acceleration region, resulting in small wings on the model lines which are consistent with the observations.

\begin{figure}
   \centering
   \includegraphics[width=\hsize]{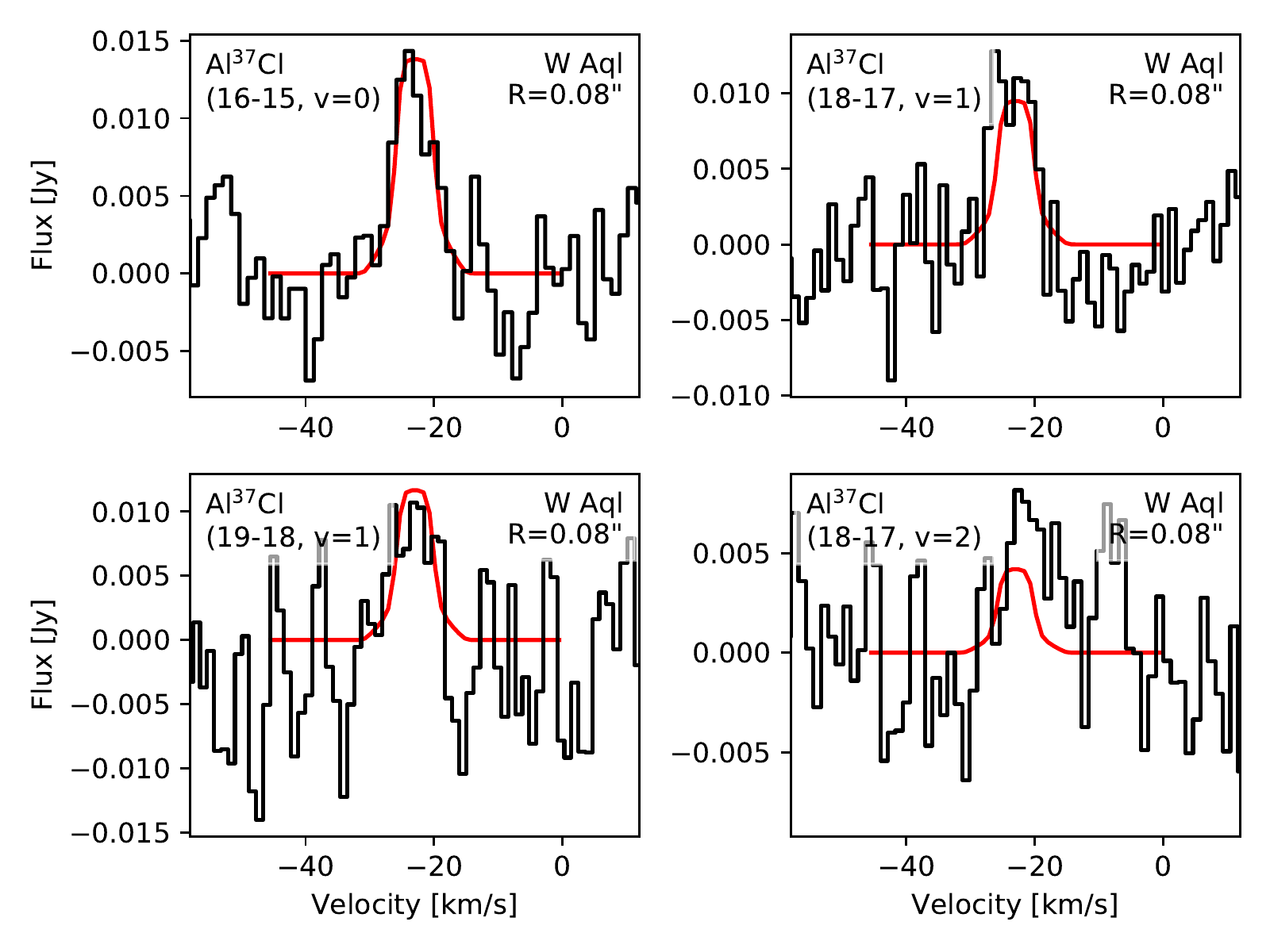}
      \caption{Spectra of Al$^{37}$Cl observed towards W Aql with ALMA (black histograms), with calculated line profiles (red curves) superposed on the observed profiles.}
         \label{al37clmod}
   \end{figure}

For Al$^{37}$Cl we used a similar modelling strategy. The lower abundance of the $^{37}$Cl isotopologue leads to fainter lines and a noisier azimuthally averaged radial profile. Therefore, it is more difficult to fit a model to the radial profile. We find that a model with the same outer radius as for Al$^{35}$Cl fits the data well. However, we do not require a step function to fit the radial profile of Al$^{37}$Cl. {In fact, a test using a step function with the same $f_0/f_1$ as for Al$^{35}$Cl did not {reproduce} the Al$^{37}$Cl data {as well}}. Instead we find the best model to have a constant abundance of $7\e{-8}$ relative to \h2. 
If we require the abundance profile to have the same $f_0/f_1$ as for Al$^{35}$Cl, then we can only obtain models with higher $\chi^2$ values and, although we can find a model with a radial profile within most of the uncertainties of the observed radial profile, we are unable to reproduce the central point in this way.
This is most likely {a result of noise in the observations or} could be because Al$^{37}$Cl is present or excited in different clumps to Al$^{35}$Cl, leading to a difference in the azimuthally averaged profile (see Fig. \ref{alclchanmaps} and \ref{alclmom0}). 
Since it best reproduces our data, we use the constant abundance model as our best fit model.
The difference in the Al$^{35}$Cl and Al$^{37}$Cl distributions can be seen in the left and centre panels of Fig. \ref{alclmom0}, where we plot the zeroth moment maps of the $\varv=0$ lines for both isotopologues. The calculated line profiles from our best model are plotted with the observed spectra in Fig. \ref{al37clmod} and the model and observed azimuthal profiles are plotted together in Fig. \ref{alclradprofs}, {along with a residual plot showing the difference between the modelled and observed radial profile points. Our results are tabulated in Table \ref{tab:results}.}

\begin{table*}
\caption{Detailed model results for W~Aql}             
\label{tab:results}      
\centering                          
\begin{tabular}{c c c c c c}        
\hline\hline                 
Molecule	&	$R_\mathrm{in}$	&	$f_0$	&	$R_\mathrm{step}$	&	$f_1$	&	$R_\mathrm{max}$	\\
	&		&	Rel. to \h2	&	[cm]	&	Rel. to \h2	&	[cm]	\\
\hline
Al$^{35}$Cl	&	$R_\star$	&$	8.5\e{-8}	$&$	1.4\e{14}	$&$	1.7\e{-7}	$&$	5\e{14}	$\\
Al$^{37}$Cl	&	$R_\star$	&$	7\e{-8}	$&	. . .	&$	7\e{-8}	$&$	5\e{14}	$\\
AlF	&	$R_\star$	&$	1\e{-7}	$&$	6\e{14}	$&$	4\e{-8}	$&$	3.5\e{15}	$\\
H$^{35}$Cl	&	$2\e{14}$~cm	&$	6.8\e{-8}	$&	. . .	&	. . .	&$	1.6\e{16}$*\\
H$^{37}$Cl	&	$2\e{14}$~cm	&$	2.9\e{-8}	$&	. . .	&	. . .	&$	1.6\e{16}	$*\\
HF	&	$2 R_\star$	&$	1\e{-8}	$&	. . .	&	. . .	&$	4\e{16}	$*\\
\hline                                   
\end{tabular}
\tablefoot{(*) For HCl and HF we give $R_{10}$, the radius at which the abundance has dropped to 10\% of $f_0$, in place of the model outer radius.}
\end{table*}

Since we do not find the same shape for the Al$^{35}$Cl and Al$^{37}$Cl abundance distributions, we are unable to determine a single $^{35}$Cl/$^{37}$Cl ratio across the entire emitting region. In the inner region of our model (inside $1.7\e{14}$~cm) we find Al$^{35}$Cl/Al$^{37}$Cl~$=1.2$ and in the outer region we find it is twice as high: Al$^{35}$Cl/Al$^{37}$Cl~$=2.4$. {However, given the uncertainties of our observations (see Fig. \ref{alclradprofs} and Sect. \ref{errors}), it is possible that the difference in these ratios may actually be less than a factor of 2.} Since the outer region represents a larger {total volume of our circumstellar model}, we adopt this value as the ratio for $^{35}$Cl/$^{37}$Cl.


\subsection{AlF}\label{sec:alfmod}

  \begin{figure}
   \centering
   \includegraphics[width=\hsize]{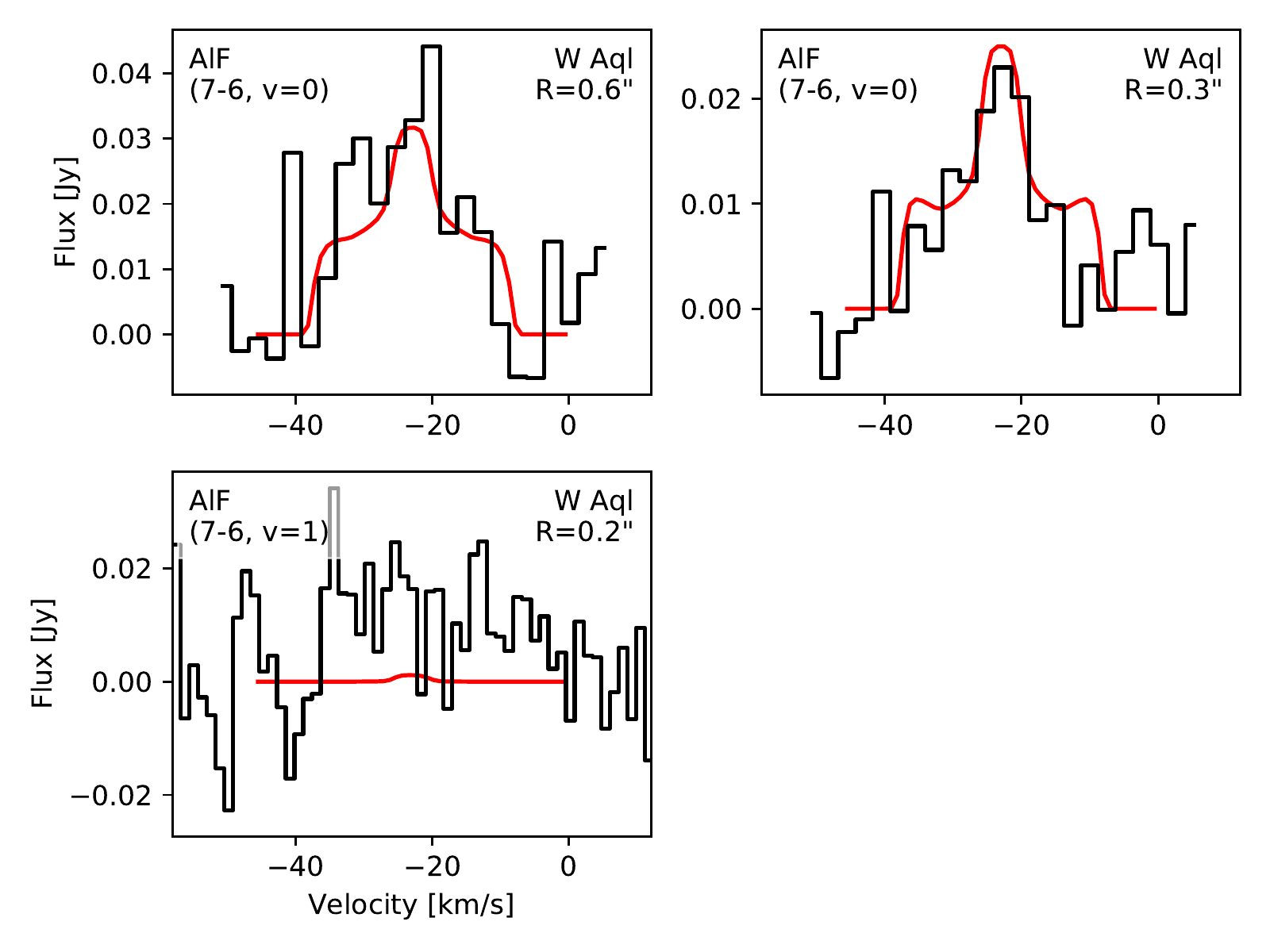}
      \caption{Spectra of AlF observed towards W Aql with ALMA (black histograms), with calculated line profiles (red curves) superposed on the observed profiles.}
         \label{alfmod}
   \end{figure}
   
      \begin{figure}
   \centering
   \includegraphics[width=\hsize]{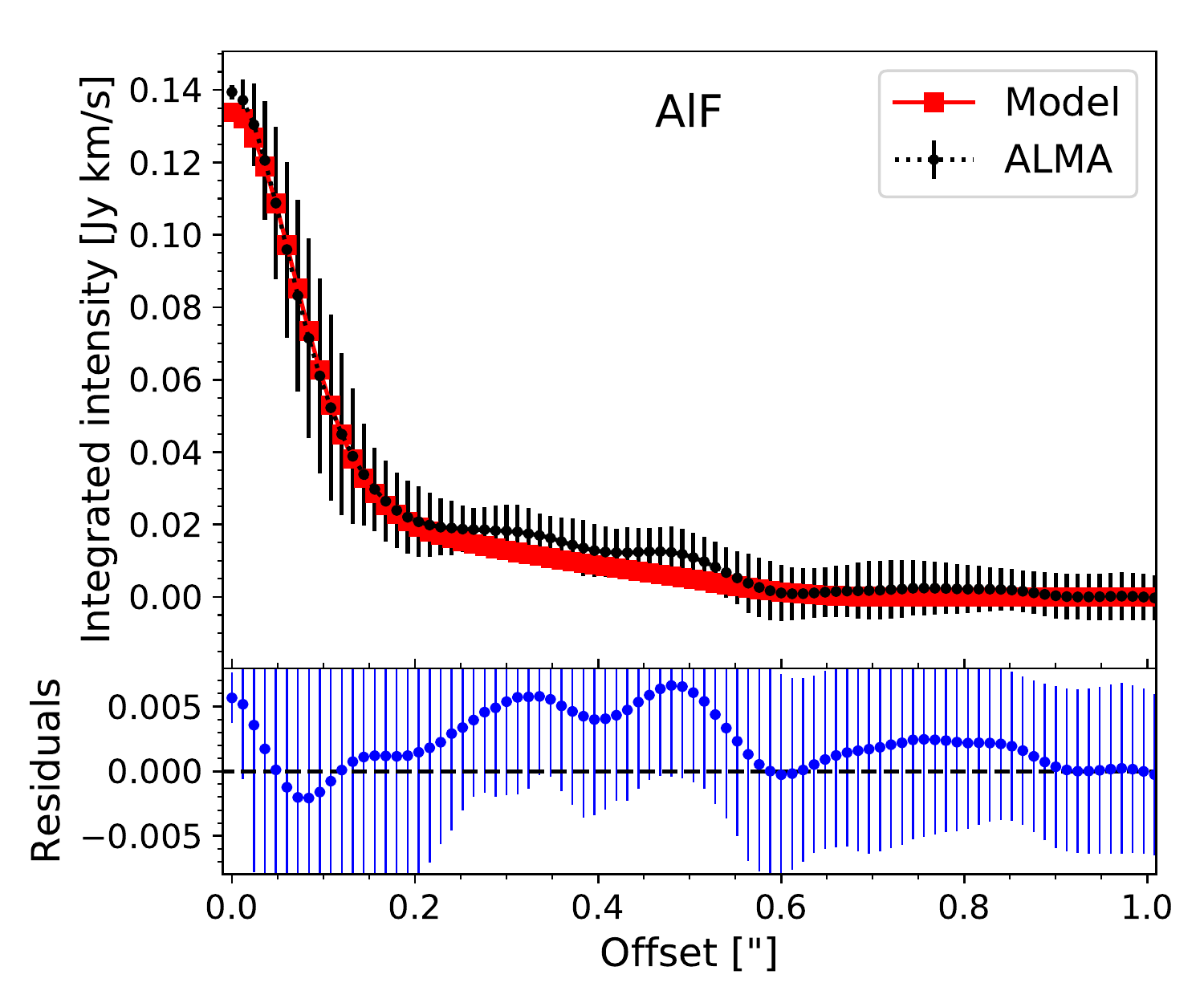}
      \caption{The azimuthally averaged radial profile of AlF, extracted from the ALMA ($7\to6$) $\varv=0$ line (black dotted lines and points with error bars) plotted with the corresponding modelled radial profiles (red lines and squares). {Residuals are plotted in the lower panels, showing the difference between the observed and modelled lines, with error bars from the observations included.}}
         \label{alfradprofs}
   \end{figure} 

For AlF we also started with a constant abundance model and then introduced a step function to fit the observed data. The azimuthally averaged radial profile (Fig. \ref{alfradprofs}) has a bright inner component out to $\sim0\farcs2$, then what looks like an extended plateau from $\sim0\farcs2$ to $\sim0\farcs5$. These features roughly correspond to the central inner region and the more diffuse emission seen in the zeroth moment map in Fig. \ref{alfvsalcl}. Since the emission seen in Fig. \ref{alfvsalcl} is not centred on the star, our spherically symmetric model is unable to reproduce it perfectly. Also, as noted in Sect. \ref{sec:rtmodov}, we include dust in the AlF model as described in \cite{Danilovich2014}, since the majority of the AlF emission comes from beyond the dust condensation radius. {Although we assume silicate dust opacities in our model \citep[based on the results of][]{Danilovich2014}, W~Aql does not have strong silicate features in its infrared spectrum \citep[see][and discussion in Sect \ref{sec:carbonhalides}]{Hony2009}. Hence, we also tested an AlF model with amorphous carbon dust \citep{Suh2000} in place of silicate dust. Using silicate dust,}
we find the best model with an abundance of $f_0=1\e{-7}$ relative to \h2 in the inner region, with a step down at $R_\mathrm{step}=6\e{14}$~cm ($\approx0\farcs1\approx15.6~R_\star$) to $f_1=4\e{-8}$ relative to \h2, and the outer radius of the model at $R_\mathrm{max}=3.5\e{15}$~cm ($\approx0\farcs6\approx90~R_\star$). {Using a model with amorphous carbon dust instead, we find a slightly lower inner abundance of $f_0=7.2\e{-8}$ but the same values for $R_\mathrm{step}$, $f_1$, and $R_\mathrm{max}$. This difference, of less than 30\%, is most likely owing to radiative pumping, since the term energies of the AlF vibrational levels ($12.6~\um$ for $\varv=1$ up to $2.2~\um$ for $\varv=6$) overlap with the wavelengths for which dust (re)radiation contributes significantly to the radiation field. In the absence of a detailed characterisation of S-type AGB dust and for consistency with our other results, and previous studies of W~Aql \citep{Danilovich2014,Ramstedt2017,Brunner2018}, we preferentially refer to the model results for silicate dust, unless otherwise specified.}

The calculated line profiles {for the silicate dust model} are plotted with the observed lines in Fig. \ref{alfmod}. The narrow central component and broader wings of the line profiles are the result of the velocity profile (Eqn. \ref{eq:vel}) and are a good fit to the observed spectral lines. As can also be seen in Fig. \ref{alfmod}, the model predicts a very faint $\varv=1$ line for AlF, with flux lower than the noise of the observed spectrum\footnote{This general result is unchanged for a model using amorphous carbon dust.}. The model and observed radial profiles are plotted together in Fig. \ref{alfradprofs}, {with a residual plot showing the difference between the modelled and observed radial profile points}. The plateau part of the model radial profile mainly fits within the error bars of the observed radial profile, the discrepancy arising from the lack of spherical symmetry in the data. The innermost few points are underpredicted by 7\% or less and were not notably improved by adding a third step up in abundance in innermost regions.

We note that the relative abundance in the inner region of our AlF model exceeds the solar abundance of F \citep[$7.2\e{-8}$][]{Asplund2009} by almost 40\%. This is discussed in more detail in Sect. \ref{sec:Fab}.

%
%


\subsection{Models of PACS observations}

The PACS spectrum of W~Aql (Fig. \ref{hclmod} and \ref{hfmod}) is noisiest in the $80\um$ region \citep[the B2B band, see][for band ranges]{Poglitsch2010}, therefore the HCl ($6\to5$) and the HF ($3\to2$) lines are not tightly constrained by the data. Although the HCl and HF lines in the $60\um$ region (the B2A band) are not formally detected above the noise, there are suggestive tentative detections of both of the HCl ($7\to6$) lines and the HF ($4\to3$) line at the expected wavelengths. We use these features as upper limits for our models.

\subsubsection{HCl}\label{sec:hclmod}

Taking the abundance distribution calculated from the chemical model described in Sect. \ref{chemmod}, we scaled the abundance by a constant factor until we found a model that best fit our data without overpredicting any of our undetected lines (Table \ref{tab:PACSlines}). We also fixed the H$^{35}$Cl/H$^{37}$Cl abundance ratio to 2.4, based on the outer Al$^{35}$Cl/Al$^{37}$Cl ratio (which is found in a region overlapping with the assumed inner part of the HCl distribution).

\begin{figure}
   \centering
   \includegraphics[width=0.9\hsize]{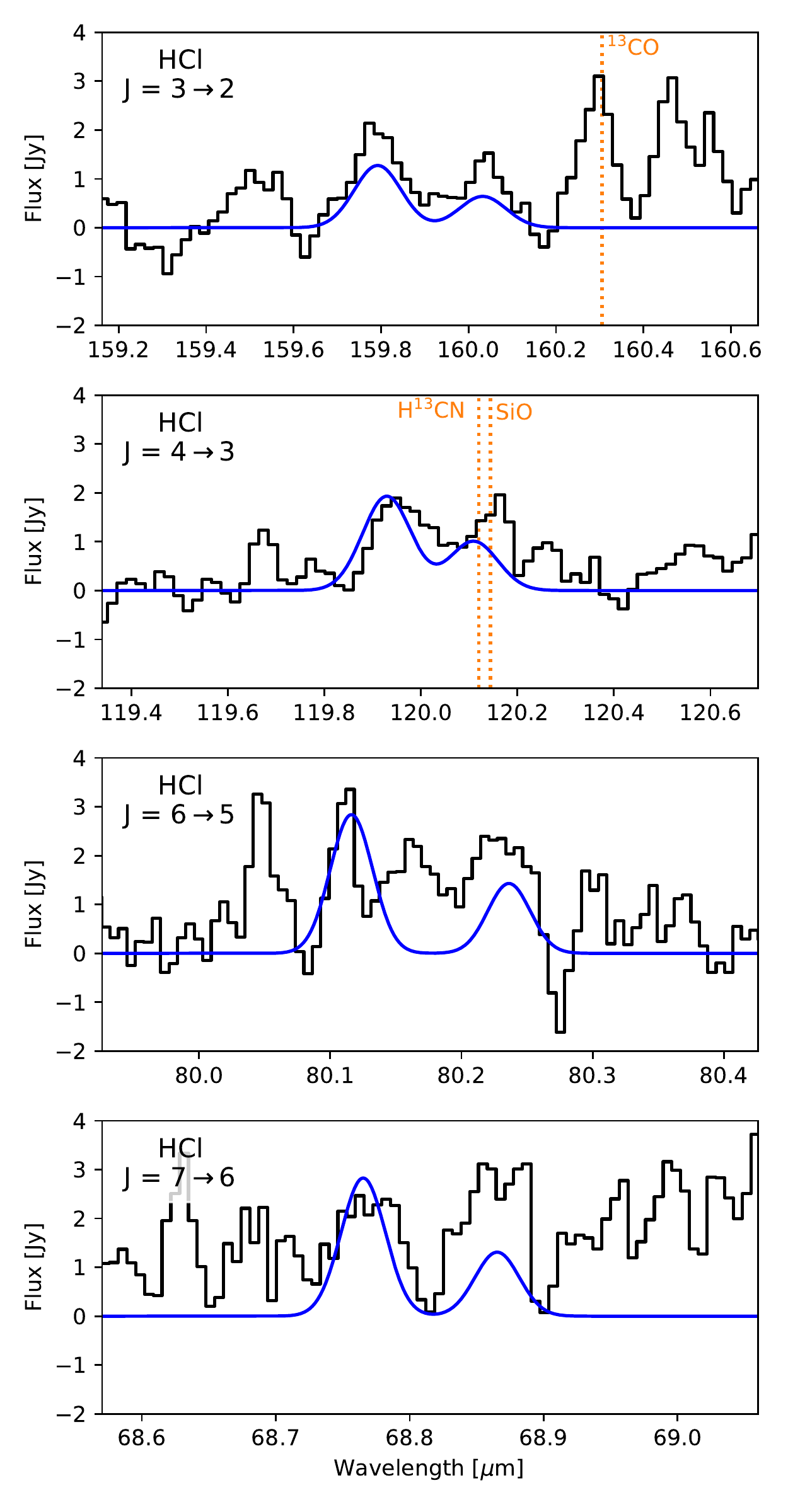}
      \caption{PACS spectra (black histograms) and model results (blue curves) for HCl towards W Aql. For each pair of lines, H$^{35}$Cl is shown on the left since it has the shorter wavelength and H$^{37}$Cl is on the right, with the longer wavelength. Some known nearby and blended lines are indicated in orange (but not all nearby lines have been identified).}
         \label{hclmod}
   \end{figure}

Our best fitting model had an inner H$^{35}$Cl abundance of $6.8\e{-8}$, relative to \h2, and an inner H$^{37}$Cl abundance of $2.9\e{-8}$. 
{Although the model was fit to the detected lines (see Table \ref{tab:PACSlines}), the uncertainties inherent in the PACS data, particularly with regards to possible line blends, mean that these results should be considered upper limits.}
The model results, convolved with the PACS spectral resolution, are shown with the observed spectra in Fig. \ref{hclmod}.

\subsubsection{HF}\label{sec:hfmod}

      \begin{figure*}
   \centering
   \includegraphics[width=0.8\hsize]{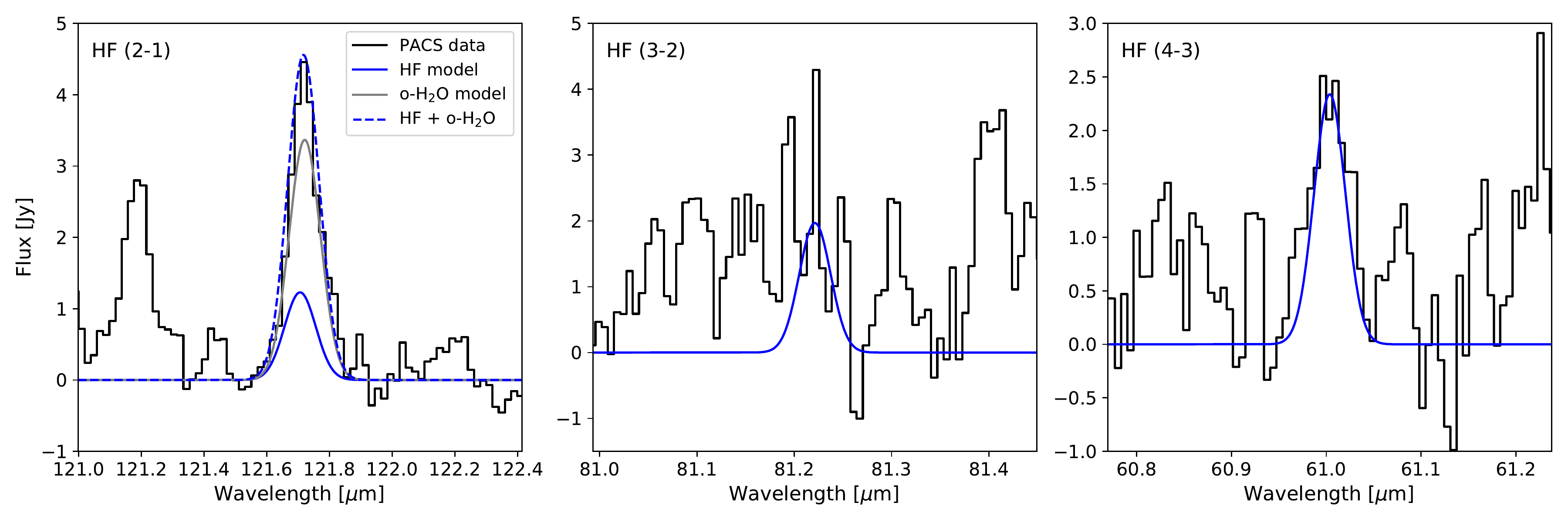}
      \caption{PACS spectra (black histograms) and model results (blue curves) for HF towards W Aql. The HF ($2\to1$) line is dominated by a blend with the o-\h2O line at $121.721\um$, which is plotted in grey. See text for details.}
         \label{hfmod}
   \end{figure*}

No lines of HF were clearly detected in the PACS spectrum of W~Aql. 
Nevertheless, we use the same method as for HCl, scaling the abundance distribution derived from the chemical model described in Sect. \ref{chemmod}. {For consistency with our $\chi$~Cyg results (see Sect. \ref{sec:stypehalides} and Appendix \ref{chicygmod}), we set the inner radius for HF as $R_\mathrm{in}=2R_\star$.} We find an upper limit on the HF abundance of $\leq1\e{-8}$ relative to \h2.

A plot of our HF model, convolved to the PACS spectral resolution, is shown in Fig. \ref{hfmod}, with the observed spectra. For the HF ($2\to1$) line, which is blended with the \h2O line at $121.721\um$, we use the \h2O model intensity from \cite{Danilovich2014} as a proxy for the \h2O contribution to the observed PACS line (shown in grey in Fig. \ref{hfmod}). The sum of our HF model line and the \h2O line is in good agreement with the line seen in the PACS spectrum.

   \begin{figure}
   \centering
   \includegraphics[width=\hsize]{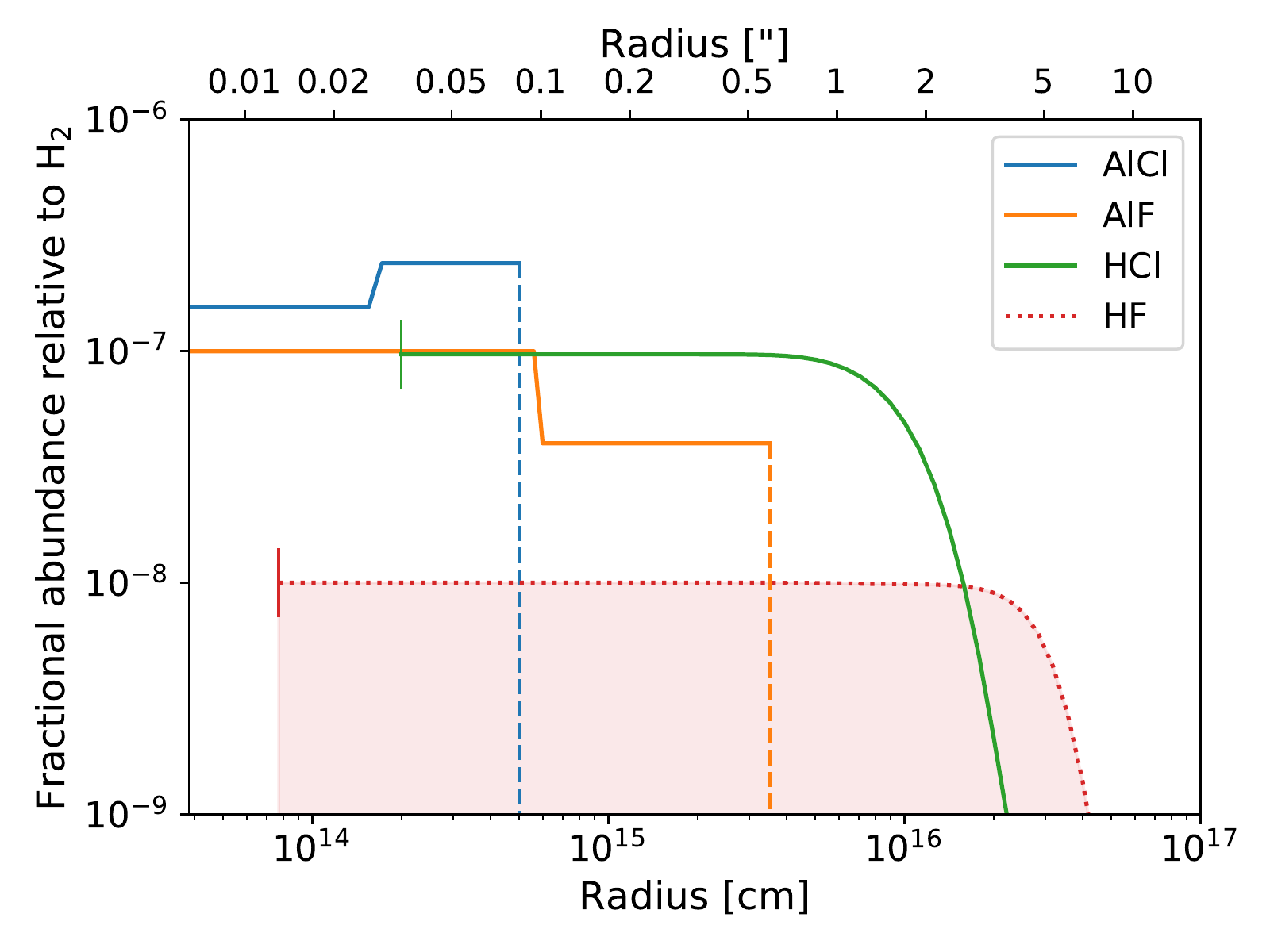}
      \caption{Abundance profiles for the halide molecules modelled here. Abundances for $^{35}$Cl and $^{37}$Cl isotopologues have been combined. Vertical dashed lines indicate the outer radius of the corresponding model and short vertical lines indicate the inner radii of the HCl and HF models. The dotted curve and corresponding shading indicates an upper limit abundance for HF. See text for details.}
         \label{abdist}
   \end{figure}

\subsection{Model uncertainties}\label{errors}

{The formal uncertainties on our models for a 90\% confidence interval are around 20\% for AlCl and 5\% for AlF, when $\chi^2$ statistics are calculated primarily from the radial profiles. However, our 1D model cannot account for the 3D effects that produce deviations from spherical symmetry, especially in the case of AlF (see Fig. \ref{alfvsalcl}). This means that our actual uncertainties are much larger than the formal uncertainties and cannot easily be quantified. This holds even though the error bars on the azimuthally averaged radial profiles (shown in Figs. \ref{alclradprofs} and \ref{alfradprofs}) take into account deviations from azimuthal symmetry in the distribution of emission, as well as stochastic noise.}

{The uncertainties on the HCl and HF model abundances are also larger than can be easily quantified from the formal errors. Owing to the low spectral resolution of PACS, there is substantial uncertainty as to whether the lines of interest are blended with nearby lines (for example the H$^{35}$Cl and H$^{37}$Cl ($3\to2$) lines are separated by almost 3~GHz, {despite overlapping in their wings at the PACS resolution}). This uncertainty can only be alleviated with spectrally resolved observations.}

{An additional source of uncertainty in our models, particularly for the case of AlF (see Sect. \ref{sec:alfmod}), is the choice of dust type. We primarily use silicate dust for consistency with \cite{Danilovich2014}, who also performed SED modelling to derive the dust optical depths. When we tested amorphous carbon dust with our AlF model, we used the same optical depth for the dust since running new SED models is beyond the scope of the present study. However, \cite{Hony2009} found that dust around S-type stars bears some similarities to M-type dust with some variation in features (see Sect. \ref{sec:carbonhalides}).}

{Another source of uncertainty comes from our extrapolation of the CSE model of \cite{Danilovich2014} inwards towards the star. Our model does not consider gas infall or stellar pulsations, which are likely to have an effect on material in the innermost regions of the CSE. The gas number
density of our model in these inner regions is extrapolated inwards following the power law $n= \dot{M}/(4\pi r^2 \upsilon(r))$, {which assumes an expanding CSE with a constant mass-loss rate}. However, recent models of the warm molecular layer close to AGB stars \cite[including W~Aql,][]{Khouri2016a} have yielded higher number densities (by around an order of magnitude) in this region. Since part of the AlCl and AlF emission is thought to come from this innermost warm molecular layer, this adds further uncertainty to the inner 2--$3R_\star$ ($\sim15$~mas) of our models.}

\section{Discussion}\label{sec:discussion}

\subsection{Halogens towards other AGB stars}\label{sec:AGBhals}

We searched the literature and in Table \ref{tab:halroundup} we have compiled the measured abundances of the aluminium and hydrogen halides for AGB stars. In the following subsections, we discuss the abundances of these molecules for individual sources, grouped into S-type AGB stars (Sect. \ref{sec:stypehalides}), carbon stars (Sect. \ref{sec:carbonhalides}), and oxygen-rich AGB stars (Sect. \ref{sec:mtypehalides}). We also touch on mid-infrared observations of HCl and HF in Sect. \ref{sec:rovibobs}.

\begin{table*}
\caption{Abundances relative to \h2 of halide molecules found towards AGB stars}             
\label{tab:halroundup}      
\centering                          
\begin{tabular*}{\textwidth}{@{\extracolsep{\fill}}*{7}{c}}
\hline\hline                 
& \multicolumn{3}{c}{S-type stars} & Carbon star & \multicolumn{2}{c}{Oxygen-rich stars}\\
\cline{2-4} \cline{5-5} \cline{6-7}
\rule{0pt}{2ex}	Molecule&	\object{W Aql}	&	\object{$\chi$ Cyg}	&	\object{R And}	&	\object{CW Leo}	&	\object{R Dor}	&	\object{IK Tau}	\\
\hline
AlCl	&	$1.6\e{-7}$--$2.4\e{-7}$	&	. . .	&	. . .	&	$7\e{-8}$	&	$2.5\e{-8}$	&	$9\e{-10}$	\\
AlF	&	$1\e{-7}$--$4\e{-8}$	&	. . .	&	. . .	&	$1\e{-8}$	&	. . .	&	. . .	\\
HCl	&	$9.7\e{-8}$	&	$6.5\e{-8}$	&	$1.5\e{-8}$	&	$1\e{-7}$	&	. . .	&	. . .	\\
HF	&	$\leq 1\e{-8}$	&	$1.2\e{-8}$	&	. . .	&	$8\e{-9}$	&	. . .	&	. . .	\\
\rule{0pt}{2.5ex}
Ref.	&	This work	&	This work$^{a}$	&	1	&	2	&	3	&	3	\\
\hline                                   
\end{tabular*}
\tablefoot{For comparison, the solar abundances of F and Cl, relative to \h2, are $7.2\e{-8}$ and $6.3\e{-7}$, respectively \citep{Asplund2009}. ($^{a}$) See Sect. \ref{sec:stypehalides} and Appendix \ref{chicygmod}.
\textit{References:} (1) \cite{Yamamura2000}; (2) \cite{Agundez2011,Agundez2012}; (3) \cite{Decin2017}.}
\end{table*}

\subsubsection{S-type AGB stars}\label{sec:stypehalides}

No halide molecules (i.e. AlF, AlCl, NaCl, KCl) were detected towards $\pi^1$~Gru, the only other S-type star in the ATOMIUM survey. This may be partly because the unusual torus + bipolar structure of the CSE of $\pi^1$~Gru \citep{Doan2017,Homan2020} could make it harder to detect less abundant molecules or might even interfere with molecule formation. We are not aware of any other detections of halide molecules towards S-type stars with ALMA. 

PACS spectra were taken for only three S-type AGB stars: W~Aql, $\pi^1$~Gru and $\chi$~Cyg \citep{Groenewegen2011,Nicolaes2018}. We checked the PACS spectra of $\pi^1$~Gru and $\chi$~Cyg for the signatures of HCl and HF. We found evidence of HCl and HF towards $\chi$~Cyg and, tentatively, HF towards $\pi^1$~Gru. The aforementioned complex circumstellar structure of $\pi^1$~Gru is such that it cannot be modelled under the assumption of spherical symmetry \citep[see also the unusual CO line structure presented in][]{Danilovich2015a}. 
However, we are able to run radiative transfer models models for $\chi$~Cyg --- see details in Appendix \ref{chicygmod}.
We find inner relative abundances of $4.6\e{-8}$ and $1.9\e{-8}$ for H$^{35}$Cl and H$^{37}$Cl, respectively (assuming, in the absence of other data, the same $^{35}$Cl/$^{37}$Cl ratio as for W~Aql), and $1.2\e{-8}$ for HF.

The HCl abundances are around 50\% higher for W~Aql than for $\chi$~Cyg, and the HF abundance is 20\% higher for $\chi$~Cyg than the upper limit found for W~Aql. For both molecules, these differences are within the observational uncertainties of the PACS data, especially for the very weak lines detected towards W~Aql. The relative proximity of $\chi$~Cyg \citep[150 pc,][]{Schoier2011} results in lines with higher signal to noise ratios in the PACS spectrum (see Fig. \ref{chicyghclmod}), despite the similar abundances between the two stars, and the lower mass-loss rate of $\chi$~Cyg \citep[$7\e{-7}\spy$,][]{Schoier2011}. 
From the available data, we are unable to conclude whether the abundances of HCl and HF have a mass-loss rate dependence, as has been seen for some other molecules \citep[e.g. SiO][]{Gonzalez-Delgado2003,Ramstedt2009}.
Following similar arguments, we are also unable to conclude whether the abundances of HCl and HF depend on the C/O ratio of the star, despite W~Aql (S6/6e) and $\chi$~Cyg (S8/1) being categorised as one grade away from the SC and MS classifications, respectively \citep{Turnshek1985,Gray2009,Danilovich2015}.
A larger sample is required to draw firmer conclusions.

\subsubsection{Carbon stars}\label{sec:carbonhalides}

CW~Leo (IRC+10216) is the closest carbon star and all the halide molecules mentioned thus far have been detected in its circumstellar envelope. It is also the only other AGB star for which abundances of HCl and HF have been calculated from radiative transfer modelling. \cite{Agundez2011} found an inner abundance of $8\e{-8}$ relative to \h2 for H$^{35}$Cl (and an H$^{35}$Cl/H$^{37}$Cl ratio of 3.3) and of $8\e{-9}$ for HF. The HCl abundance for CW~Leo is comparable to that found for W~Aql but 50\% higher than that found for $\chi$~Cyg. Conversely, the HF abundance is comparable to the upper limit found for W~Aql but 33\% lower than that found for $\chi$~Cyg. {The observed abundances towards CW~Leo of these two molecules are a factor of $\sim5$ lower for HCl and  an order of magnitude higher for HF than the predictions made by the chemical kinetics model of \cite{Cherchneff2012}, which considers shocks for models of the inner $5R_\star$ of the CSE.}

Based on observations with the IRAM 30~m telescope {(with beam sizes ranging from 7\arcsec{} to 30\arcsec)}, abundances of the halide molecules AlCl, AlF, NaCl, and KCl, were modelled by \cite{Agundez2012} for CW~Leo. In the data presented by \cite{Agundez2012}, there are no detections of any vibrationally excited lines for any of these halide molecules \citep[though NaCl in the $\varv=1$ state was subsequently seen by][using ALMA]{Quintana-Lacaci2016a}. Despite AlCl and AlF having significantly lower dipole moments than NaCl and KCl, \cite{Agundez2012} saw similar line strengths for all four halogen-bearing species. They conclude that this is due to the higher abundances of AlCl and AlF compared with NaCl and KCl, which is borne out by their model results ($7\e{-8}$ and $1\e{-8}$ for AlCl and AlF compared with $1.8\e{-9}$ and $5\e{-10}$ for NaCl and KCl). The most striking difference between their model results for CW~Leo and ours for W~Aql is the much larger molecular envelope sizes of AlCl and AlF towards CW~Leo. Although \cite{Agundez2012} do not list specific parameters for the sizes of the envelopes, from their Fig. 13 it can be seen that AlCl is present at an appreciable abundance ($\sim1\e{-9}$ relative to \h2) out to $\sim2\e{16}$~cm and AlF out to $\sim4\e{16}$~cm. {This is much larger than our outer radii of $5\e{14}$~cm and $3.5\e{15}$~cm for AlCl and AlF.
They also observe AlCl and AlF lines with expansion velocities around $14.5~\kms$, equal to the terminal gas expansion velocity used in their models. This is in contrast with what we see for W~Aql, where AlCl is found in a region very close to the star (see Fig. \ref{alclchanmaps} and \ref{abdist}) and the expansion velocity of AlCl, as derived from the vibrational ground state, is around $4~\kms$ -- i.e. much lower than the terminal expansion velocity of $16.5~\kms$ \citep{Danilovich2014} {or the maximum velocity found by \cite{Gottlieb2021}} (see Fig. \ref{alclmod} and \ref{al37clmod}).
}

\cite{Quintana-Lacaci2016a} observed Al$^{35}$Cl, Al$^{37}$Cl, and AlF with ALMA towards CW~Leo. In all cases, they found there was resolved out flux in the ALMA observations, when compared with spectra obtained from the IRAM 30~m telescope. Al$^{37}$Cl was less affected than the other two species and, from the plots presented \citep[Figs. 12 and 13 of][]{Quintana-Lacaci2016a}, it can be seen that the Al$^{37}$Cl extends out to $\sim2\arcsec$ from the continuum peak. Assuming a distance of 130~pc \citep{Agundez2012}, this corresponds to $\sim4\e{15}$~cm, which is almost an order of magnitude larger than what we see for W~Aql (see Fig. \ref{abdist}). Another difference we see from comparison to the \cite{Quintana-Lacaci2016a} data is a lack of $v>0$ AlCl detections towards CW~Leo, despite $\varv=1,2$ lines being present in the covered frequency range, and even though the $\varv=0$ AlCl lines in the CW~Leo data set {have flux densities almost 18 times higher}
than our W~Aql data, {when measured in an equivalent aperture}\footnote{The ALMA beam size and rms are $\sim$0\farcs7, 4 mJy and 0\farcs023, 1 mJy for the CW Leo data and our data respectively. The $\varv=2$ line covered by both observations (considered since the $\varv=1$ is close to a brighter SiO line), has a total flux density of 13 mJy over a few beam areas at the line peak for W~Aql. 
The proportional flux density for the CW~Leo data, when compared with the $v=0$ line, would be ~23 mJy/beam. This is just over 5$\sigma$ but it is possible for weak emission to be missed in masking for cleaning, especially in the Cycle 0 data with fewer antennas and less refined calibration.}.
The absence of vibrationally excited lines could indicate different formation pathways if AlCl is not present as close to the star in the CSE of CW~Leo compared with W~Aql.

The AlF emission observed by \cite{Quintana-Lacaci2016a} for the ($8\to7$) transition is significantly resolved out by ALMA. Based on their observing setup, the authors estimate that smooth emission larger than 13--14\arcsec{} was filtered out. From their plot of the recovered AlF emission, it seems that the emission around the central $\sim15~\kms$ was the most resolved out. Based on their estimate of the maximum resolvable scale, we can expect that AlF emission extends out to at least 6.5--7\arcsec{} from the star, assuming relatively symmetric emission. At a distance of 130~pc, this corresponds to $\sim1.3\e{16}$~cm or about 5 times larger than our AlF model for W~Aql.

{Since the temperatures (agreement within 30~K) and luminosities (CW~Leo 17\% more luminous) of the two stars are similar, we can conclude that the differences between AlCl and AlF distributions should mainly be because of the different chemical compositions and wind densities of the two CSEs (the latter dependent mainly on the mass-loss rates: $\dot{M}=2\e{-5}\spy$ for CW~Leo compared with $\dot{M}=3\e{-6}\spy$ for W~Aql). This in itself is an interesting result since the equilibrium models of \cite{Agundez2020} predict very similar abundances between carbon stars and S-type stars for both Cl- and Al-bearing molecules.}

{Another possibility is that} dust composition and interactions between gas and dust species could also contribute to the differences seen between W~Aql and CW~Leo.
As a carbon star, CW~Leo is surrounded by carbon-rich dust. As an S-type star, the dust around W~Aql is less well understood. \cite{Hony2009} analyse infrared spectra from ISO/SWS for several AGB stars, with a focus on S-type stars. They note that there are weak features in the spectrum of W~Aql (and other S stars) which are similar but broader and less structured than the silicate and aluminium oxide features present in M-type AGB stars. {\citet{Smolders2012} studied a large sample of S-type stars (not including W~Aql) observed with Spitzer, and found that around half of them did not exhibit any alumina features. Although rotational lines of AlO were not detected towards W~Aql, despite two lines in the ground vibrational state being covered by the ATOMIUM programme, and despite a lack of clear aluminium features in the infrared spectrum}, it is possible that AlCl is incorporated into dust, hence explaining why it is seen only in regions close to the star. AlF may also be partly incorporated into dust, hence explaining the step down in abundance at approximately the same distance from the star as the outer edge of our AlCl model. {Both the step down in AlF abundance and the outer radius of AlCl are within a factor of $\sim 3$ of the dust condensation radius.} The persistence of AlF further out in the envelope, after AlCl has been destroyed, is likely to be at least partially due to the higher binding energy of AlF compared with AlCl \citep[681 kJ mol$^{-1}$ for AlF and 507 kJ mol$^{-1}$ for AlCl,][]{Curtiss2007}. {See also the more detailed discussion of chemistry in Sect. \ref{sec:chemistry}.}

\subsubsection{Oxygen-rich AGB stars}\label{sec:mtypehalides}

Aside from W~Aql, in this study, and CW~Leo, we have found no other published observations of AlF towards AGB stars. However, we note that AlF has been detected around several oxygen-rich ATOMIUM sources (Wallstr\"om et al, \textit{in prep.}), which will be modelled in a future study.
{In the ATOMIUM sample, AlCl was also detected towards the oxygen-rich AGB star GY~Aql, which will also be studied in a future publication.}

Al$^{35}$Cl has also been tentatively detected at relatively low abundances towards the oxygen-rich stars R~Dor (low mass-loss rate, $\dot{M}=1.6\e{-7}\spy$) and IK~Tau (higher mass-loss rate, $\dot{M}=5\e{-6}\spy$), as reported by \cite{Decin2017}. For both stars, Al$^{35}$Cl was found to be confined to the region close to the central star, similar to our results for W~Aql (although our ALMA observations are at a higher spatial resolution and can put more stringent constraints on the AlCl emission region). Based on models which only considered rotational levels in the ground vibrational state of Al$^{35}$Cl, \cite{Decin2017} derived abundances are $2.5\e{-8}$ and $9\e{-10}$ relative to \h2 for R~Dor and IK~Tau, respectively. These abundances are around one to two orders of magnitude lower than the Al$^{35}$Cl abundance we found for W~Aql. {Gaseous AlO and AlOH have also been detected around the same two oxygen-rich stars \citep{Decin2017,Danilovich2020a}, but at low enough abundances that we would not expect their presence to inhibit the production of AlCl.}

NaCl and KCl have been detected towards several oxygen-rich stars, including some in the ATOMIUM sample, to be presented in a future study. Spectral scans of IK~Tau, carried out using single-dish telescopes, detected several NaCl lines but no KCl lines \citep{Milam2007,Velilla-Prieto2017}. A spectral scan of R~Dor and IK~Tau using ALMA found clumpy NaCl emission towards IK~Tau \citep{Decin2018}, no NaCl emission towards R~Dor, and no KCl towards either star, {although \cite{De-Beck2018} tentatively find NaCl towards R~Dor in an APEX spectral scan}.
NaCl and KCl in the ground and several vibrationally excited states have been seen towards extreme OH/IR stars, oxygen-rich stars with high mass-loss rates ($\gtrsim10^{-5}\spy$), including OH~26+0.6 \citep{Justtanont2019} and OH~30.1-0.7 (Danilovich et al. \textit{in prep.}). Tentatively, it seems that there is a positive correlation between higher abundances of NaCl and KCl and higher mass-loss rates for oxygen-rich AGB stars, though this will be explored in more detail in a future study.

{The chemical kinetics model of \cite{Gobrecht2016} considered shocks in the inner wind and included chloride species for a model based on IK~Tau. They predict an average abundance for HCl of $3.8\e{-7}$ relative to H$_2$, around 4 times higher than our S-star results. They also predict a very low abundance of AlCl, with $3.8\e{-12}$ close to the stellar surface and $2.2\e{-10}$ at $6R_\star$. This is lower than the \cite{Decin2018} observational result for IK~Tau of $9\e{-10}$, by a factor of at 4 at $6R_\star$ and by two orders of magnitude close to the stellar surface. This corresponds to 3 to 5 orders of magnitude lower than our W~Aql results. In the \cite{Gobrecht2016} model, NaCl was a more significant carrier of Cl than AlCl but only reached $\sim3\%$ of the HCl abundance in the outer part of their model (at $9R_\star$).}

\subsubsection{Rovibrational observations of halides}\label{sec:rovibobs}

Thus far we have discussed observations of rotational transitions of halide molecules, but observations of rovibrational bands in the mid-infrared are also possible. \cite{Yamamura2000} report HCl lines in the spectrum of R~And, an S-type star with a moderately low mass-loss rate \citep[$5.3\e{-7}\spy$,][]{Danilovich2015a}. A rough comparison between their reported column density and CO gives an approximate relative abundance of HCl of $1.5\e{-8}$, which is within an order of magnitude of our values for W~Aql and $\chi$~Cyg.

\cite{Jorissen1992} observed a sample of AGB stars of different chemical types (and some non-AGB stars) and used mid infrared lines of HF as a proxy for the fluorine abundance. They find photospheric F abundances consistently higher in AGB stars than the solar abundance of fluorine. \cite{Abia2015} find similarly enhanced F abundances for their sample mostly of carbon and SC-type AGB stars, albeit to a lesser extent.
The ramifications of their results will be discussed in Sect. \ref{sec:Fab}, but the purely observational implication is that HF should be detectable for many AGB stars, if only the rotational lines were not so difficult to access from the ground.

%
%
%
%
%
%
%
%
%

\subsection{Chlorine abundance and isotopic ratio}


Both stable isotopes of chlorine, \ce{^35Cl} and \ce{^37Cl}, are believed to be primarily formed during the hydrostatic and explosive oxygen burning stages of supernova explosions \citep{Woosley1995}, through different nuclear reactions \citep[and predominantly through core-collapse supernovae][]{Kobayashi2020}. 
\cite{Esteban2015} and \cite{Henry2004} found a weak trend towards decreasing Cl abundance with galactic radius, based on data from H\textsc{ii} regions and planetary nebulae (PNe). 
They also found very similar relations between Cl and O gradients with galactic radius, suggesting that the production of Cl and O are correlated \citep{Maas2016,Maas2021} {and hence that Cl can be used as a tracer of metallicity}. 
From a study of PNe, \cite{Delgado-Inglada2015} found that Cl is a good indicator of metallicity in the progenitors of PNe (i.e., AGB stars). Hence, the comparison of an AGB Cl abundance with the solar abundance might indicate a higher or lower metallicity of the natal environment of the AGB star, without obfuscation from elements actively synthesised by AGB stars \cite[which $^{35}$Cl is not, based on the investigation of][]{Maas2016}.
However, the solar abundance of chlorine is very difficult to measure, for reasons discussed in detail by \cite{Maas2016}.
For the purposes of our study, we assume a solar abundance of Cl, relative to \h2, of $6.3\e{-7}$, based on the value given by \cite{Asplund2009}, when referring to solar chlorine abundance.

The highest total chlorine abundance we find for W~Aql, by summing the abundance of HCl and the higher abundance of AlCl, is just over half that of the solar abundance. While we can determine from non-detections in our ALMA observations that the abundances of NaCl and KCl are less significant than those of HCl and AlCl, we cannot be certain whether other molecules do not contribute to the total Cl abundance. For example, \cite{Agundez2020} predict that more unusual and as yet undetected molecules (such as CaCl$_2$, SiCl, ZrCl$_2$) may contribute a few percent to the overall Cl abundance, although they predict AlCl, HCl and atomic Cl to be the dominant species for S-stars. Furthermore, their models only look at the inner regions of AGB CSEs (out to $10R_\star$) under chemical equilibrium conditions and make no predictions for abundances outside of these regions. Hence, although we find a lower total Cl abundance for W~Aql than solar, we cannot with certainty say that W~Aql must have a lower metallicity than solar. The uncertainty of our HCl abundance is also significant and improved models based on spectrally resolved HCl lines would give rise to firmer conclusions.
{Although there are likely chemical differences affecting Cl-bearing molecules between carbon and S-type stars (see Sect. \ref{sec:carbonhalides}), we can make a first order approximation of the difference in total Cl abundance by considering the sum of abundances of the observed Cl-bearing molecules for W~Aql and CW~Leo \citep[see Table \ref{tab:halroundup} and][]{Agundez2012}. Considering similar inner regions of the CSEs (where our AlCl and HCl models overlap for W~Aql), we find a factor of 2 more Cl detected around W~Aql than CW~Leo. This could be because different chemical processes are in play for the different types of stars, or, if we assume the chemical processes are similar, this difference could indicate that W~Aql formed from a natal cloud with higher metallicity than CW~Leo.}

The solar system $^{35}$Cl/$^{37}$Cl ratio of 3.1 is well established \citep[see for example][]{Asplund2009}. In close agreement with the solar value, \cite{Agundez2011} found $^{35}$Cl/$^{37}$Cl $= 3.3\pm0.3$ towards CW~Leo from the modelling of HIFI observations of HCl, which is also in agreement with studies of metal chlorides in the same star \citep[e.g. $2.9 \pm 0.3$ from the][study of NaCl, KCl and AlCl]{Agundez2012}. Some variation has been found for this ratio in other astronomical sources, however. For example, \cite{Peng2010} observed the ($1\to0$) transition of both isotopologues of HCl towards several different galactic sources, including star-forming regions, molecular clouds and carbon stars\footnote{CW Leo was the only carbon star for which \cite{Peng2010} observed both isotopologues of HCl. Their ratio is lower than that found by \cite{Agundez2011} but still agrees within uncertainties. We consider only the \cite{Agundez2011} value here since they use more HCl transitions and more detailed radiative transfer models to obtain their result.}. For sources towards which both isotopologues were detected, they found a spread of $^{35}$Cl/$^{37}$Cl ratios, mostly in the 2.0--2.6 range, with values below 1 for two locations in the W3 star forming region and $\sim5$ for DR21(OH), a region of massive star formation. \citet{Maas2018} found $^{35}$Cl/$^{37}$Cl ratios ranging from 1.76 to 3.42 for a sample of six M giants.
All of these results point to significant variation in  $^{35}$Cl/$^{37}$Cl across the galaxy. 
{The stellar evolution models of \cite{Cristallo2015} and \cite{Karakas2016} predict modest decreases in $^{35}$Cl/$^{37}$Cl during the AGB phase, depending on metallicity and initial mass. For example, the most significant decrease of $^{35}$Cl/$^{37}$Cl (to 2.17 at the end of the AGB phase) in the \cite{Karakas2016} models is seen for a low metallicity star ($Z=0.007$) with initial mass $2.75~\msol$.}

The $^{35}$Cl/$^{37}$Cl values we find for W~Aql from AlCl are 1.2 in the innermost region and 2.4 in the outer region of the AlCl emission. Since the AlCl emission is relatively faint, especially in the case of Al$^{37}$Cl, it is unclear to what extent the different isotopic ratios are real or a product of observational uncertainty and noise, especially since chemical fractionation is not expected to play a significant role. 
This uncertainty could be reduced if we had sensitive observations of additional Al$^{35}$Cl and Al$^{37}$Cl lines in the ground vibrational state, rather than just one line for each isotopologue. Alternatively, checking for a similar discrepancy in another molecule could confirm it more strongly if it were found. For example, spectrally and spatially resolved observations of H$^{35}$Cl and H$^{37}$Cl ($1\to0$) are possible with ALMA and could give us more information about the spatial dependence of the $^{35}$Cl/$^{37}$Cl ratio. Additionally, spectrally (but not spatially) resolved observations of H$^{35}$Cl and H$^{37}$Cl up to ($4\to3$) are possible with SOFIA\footnote{Stratospheric Observatory for Infrared Astronomy}, and would allow us to independently determine the H$^{35}$Cl and H$^{37}$Cl abundances and constrain our HCl models better than the PACS data alone.

\subsection{Abundance of fluorine}\label{sec:Fab}

The cosmic origin of fluorine has not yet been fully constrained, with nucleosynthesis models under-predicting observed fluorine abundances \citep{Lugaro2004,Kobayashi2020}. A significant portion of the local fluorine abundance is thought to have been produced by AGB stars, around 51\%, according to the models of \cite{Kobayashi2020}. However, there are at present still several uncertainties in the calculations of nucleosynthesis yields, particularly when it comes to the treatment of convection and mass loss \citep{Kobayashi2020}. In a detailed study of the uncertainties in the nuclear reaction rates for $^{19}$F, \cite{Lugaro2004} concluded that these cause uncertainties in theoretical models of fluorine production of a factor of 2--7, depending on the initial stellar mass. 
A recent study by \cite{Ryde2020} argues for multiple sites of fluorine production, in particular at different metallicities.
Nevertheless, it has been clear for some time that AGB stars play a significant role in the production of fluorine. \cite{Jorissen1992} first determined fluorine abundances for sources outside of the solar system. From their observations of atmospheric HF, they concluded that, not only is F more abundant in M, S and carbon stars than in the Sun, it is generally further enhanced in carbon and S-type stars compared with M-type stars. Further evidence of the synthesis of F in AGB stars is provided by \cite{Zhang2005}, who find abundances of F in PNe higher than the solar abundance, hence surmising that F is produced in the AGB progenitors of PNe.

The solar abundance of elemental fluorine is still somewhat uncertain, not least because fluorine is the least abundant element in the range of atomic numbers from 6 to 20 \citep[carbon to calcium,][]{Asplund2009}, and is more difficult to measure. Nevertheless, the recent solar fluorine abundance found by \cite{Maiorca2014} is in agreement within the uncertainties with earlier determinations \citep{Lodders2003,Asplund2009}. Converting their values to abundances relative to \h2 for comparison with our results, we find solar abundances of F in the range 5.0 to $7.2\e{-8}$, with uncertainties at, or close to, a factor of two. Whichever value we adopt for the solar abundance of F, our abundance of AlF for W~Aql is higher than the solar F abundance, and increases if we include the upper limit we found for HF. This indicates that F synthesised in the AGB star has already been dredged up to the surface and injected into the wind.

As discussed in Sec. \ref{sec:AGBhals}, fluorine-bearing molecules have not been extensively studied towards AGB stars, with the carbon star CW~Leo being the only example of AlF and rotational lines of HF predating the present study. The total abundance of F found from AlF and HF for CW~Leo (see Table \ref{tab:halroundup}) is almost an order of magnitude less than what we find for W~Aql and lower than (any determination of) the solar abundance. {This could a result of different chemical processes in the winds of the two stars, or could be explained by CW~Leo having formed from a lower-metallicity natal cloud than W~Aql, which would affect the initial (and hence total) F abundance.}
Although the chemical equilibrium predictions of \cite{Agundez2020} predict that AlF and HF will be the dominant carriers of F in the inner winds of carbon stars as well as S-type stars (with very similar molecular abundances of F-bearing species between the two AGB types), it is possible there are some additional non-equilibrium processes at play that more strongly affect carbon stars and were not taken into account for that study. {For example, the presence of \h2O in the inner regions of CW Leo is a clear indicator of non-equilibrium processes in those regions \citep{Decin2010c}.}

An interesting direction of future study would be a more extensive look at AlF and HF towards a larger sample of AGB stars, especially S-type stars. As previously mentioned, AlF was detected towards some of the oxygen-rich AGB stars in the ATOMIUM sample, and will be analysed in future work. Additional observations of AlF towards S-type stars do not currently exist but could be be obtained with ALMA, potentially of multiple rotational lines. Observations of HF are more difficult to obtain, with rotational HF (found in the THz range) not presently accessible\footnote{At 1232.476 GHz \citep{Nolt1987,Pickett1998}, the \mbox{$J=1\to 0$} line falls close to a water line and just outside of the feasible observing ranges SOFIA is currently equiped for \citep{Duran2021}.}. Nevertheless, photospheric abundance determinations from infrared observations of HF towards AGB stars with known mass-loss rates and well-studied CSEs (including abundances of other molecules) would help fill out our understanding of F towards these stars. In particular, comparing the abundances of AlF + HF with physical parameters, such as mass-loss rates, pulsation periods, luminosity, expansion velocity, etc, could tell us about the evolutionary history of the star, especially in light of the synthesis of F during the AGB phase.

\subsection{Constraints on aluminium abundance}

AlO and AlOH were not detected towards W~Aql, despite being detected towards some of the oxygen-rich ATOMIUM sources (to be presented in a future study). This is notable since AlO and especially AlOH are expected to be the dominant carriers of Al under thermochemical equilibrium \citep{Agundez2020} and steady-state chemical models, even in the case of S-type stars. The absence or very low abundance (quantified below) of AlO and AlOH is indicative of other processes, such as dust formation or growth, limiting the gas-phase abundance of these molecules.
In Sect. \ref{alnds} we calculate rms values as detection limits for AlO and AlOH. Additionally, we ran some radiative transfer models to obtain abundance upper limits for AlO and AlOH.

We tested two abundance distributions for both AlO and AlOH: 1) a model with a constant abundance of the molecule from the stellar surface out to $4\e{14}$~cm, based partly on the results found for the oxygen-rich stars in \cite{Decin2017}; and 2) a model with an abundance profile based on the predictions of the chemical model described in Sect. \ref{chemmod} and expanded on in Sect. \ref{sec:chemistry}. In both cases the abundance was scaled until the lines predicted by the model were equal to and/or did not exceed the rms values given in Table \ref{tab:saltrms} (using the same extraction apertures). The molecular data used here is taken from \cite{Decin2017} and \cite{Danilovich2020a} for the AlOH and AlO models respectively. For the constant abundance models, we found upper limits of $f_\mathrm{AlO} \leq 6\e{-9}$ and $f_\mathrm{AlOH} \leq 3\e{-8}$, relative to \h2. For the abundance distributions predicted by the chemical model, we found upper limits of $f_\mathrm{AlO, peak} \leq 6.6\e{-9}$ and $f_\mathrm{AlOH, peak} \leq 6.5\e{-8}$, relative to \h2. Both sets of upper limits are plotted in Fig. \ref{aluplims}. {The AlOH upper limit exceeds the abundances found for the M-type stars R Dor and IK Tau by around an order of magnitude, but the AlO upper limit is around an order of magnitude smaller than the abundances found for the same stars \citep{Decin2017}.}

\subsection{Chemistry of AlCl and AlF}\label{sec:chemistry}

\begin{table*}
\caption{Calculated rate coefficients for reactions between Al and halogen species (see text in Appendix \ref{chemsupinfo} for further details).}             
\label{tab:reactions}      
\centering                          
\begin{tabular}{c l}        
\hline\hline                 
Reaction & \multicolumn{1}{c}{Rate coefficient\tablefootmark{a}} \\
\hline
\multicolumn{2}{l}{Fluorine reactions:}\\
\ce{Al +HF \to AlF + H} & $k_1 = 1.1\e{-10}\exp(-125/T)+9.3\e{-10}\exp(-2750/T)$\\
\ce{AlOH + HF  \to  AlF + H2O}	& $k_2 = 7.3\e{-12}\exp(-1800/T)+5.4\e{-12}(T/300)^{-1.8}$\\
\ce{AlO + HF  \to  AlF + OH}		& $k_3 = 7.1\e{-10}\exp(145/T)$\\
\ce{AlF + H \to Al +HF} & $k_{-1} = 1.5\e{-10}\exp(-13816/T)$\\
\ce{AlF + H2O \to AlOH + HF}	& $k_{-2} = 3.9\e{-12}\exp(-6125/T)$\\
\ce{AlF + OH \to AlO + HF}		& $k_{-3} = 4.1\e{-10}\exp(-4105/T)$\\
\multicolumn{2}{l}{Chlorine reactions:}\\
\ce{Al + HCl  \to  AlCl + H	}	& $k_4=1.4\e{-10}\exp(-890/T)+2.0\e{-9}\exp(-4036/T)$\\
\ce{AlOH + HCl  \to  AlCl + H2O}	& $k_5=1.9\e{-14} (T/300)^{1.98}\exp(-630/T)$\\
\ce{AlO + HCl  \to  AlCl + OH}		& $k_{6a}=2.0\e{-10} \exp(-793/T)$\\
\phantom{\ce{AlO + HCl }.} $\to$ AlOH + Cl		& $k_{6b}=8.8\e{-10} \exp(-27/T)$\\
\ce{AlCl + H \to Al + HCl	}	& $k_{-4}=2.0\e{-10} \exp(-10670/T)$\\
\ce{AlCl + H2O \to AlOH + HCl}	& $k_{-5}=3.6\e{-14} (T/300)^{2.10}\exp(-3010/T)$\\
\ce{AlCl + OH \to AlO + HCl}		& $k_{-6a}=1.7\e{-10}\exp(-1266/T)$\\
\phantom{\ce{AlCl + OH}.} $\to$ AlOH + Cl		& $k_{-6b}=8.0\e{-10}\exp(-18.9/T)$\\
\hline
\end{tabular}
\tablefoot{(\tablefootmark{a}) Units for the bimolecular reaction rate coefficients are cm$^{3}$ molecule$^{-1}$ s$^{-1}$.}
\end{table*}

{After the initial chemical model results were used as input for the radiative transfer modelling of HCl and HF (i.e. see Sect. \ref{chemmod} and the derived abundance results in Fig. \ref{abdist}), we extended the {\sc Rate12} chemical model {by including reactions describing the chemistry of aluminium,} in an attempt to reproduce the observed abundance distributions of the halide species.}
The aluminium halides studied here can be produced by the exothermic reactions of Al, AlOH and AlO with the corresponding hydrogen halides:
\begin{align}
&\ce{Al +HF \to AlF + H} &\Delta H^\ominus =-113~\mathrm{kJ~mol}^{-1}\tag{R1}\\
&\ce{AlOH + HF  \to  AlF + H2O}	&\Delta H^\ominus  = -52~\mathrm{kJ~mol}^{-1}	\tag{R2}\\
&\ce{AlO + HF  \to  AlF + OH}		&\Delta H^\ominus  = -37~\mathrm{kJ~mol}^{-1}	\tag{R3}\\
&\ce{Al + HCl  \to  AlCl + H	}	&\Delta H^\ominus  = -81~\mathrm{kJ~mol}^{-1}	\tag{R4}\\
&\ce{AlOH + HCl  \to  AlCl + H2O}	&\Delta H^\ominus  = -19~\mathrm{kJ~mol}^{-1}	\tag{R5}\\
&\ce{AlO + HCl  \to  AlCl + OH}		&\Delta H^\ominus  = -5~\mathrm{kJ~mol}^{-1}\tag{R6}
\end{align}
The standard reaction enthalpies, $\Delta H^\ominus$, (at 0 K) are calculated using the very accurate G4 method within the Gaussian suite of programs \citep{Frisch2016}. Theoretical calculations of the rate coefficients for reactions R1 to R6 (i.e. $k_{1}$ to $k_6$) and the reverse reactions R-1 to R-6 ($k_{-1}$ to $k_{-6}$) are described in Appendix \ref{chemsupinfo} and the rate coefficients are listed in Table \ref{tab:reactions}. Note that the theoretical estimate of $k_5$ is in good agreement with a measurement between 475 and 1275 K \citep{Rogowski1989}. The {aluminium} halides can be removed by reaction with H (reactions R-1 and R-4), with \ce{H2O} (R-2 and R-5) or with OH (R-3 and R-6). In addition, they can undergo photolysis:
\begin{align}
&\ce{AlF + h\nu  \to  Al + F	}\tag{$J_1$}\\
&\ce{AlCl + h\nu \to  Al + Cl	}\tag{$J_2$}
\end{align}

The chemical outflow model used here is based on \cite{McElroy2013} and adapted by \cite{Van-de-Sande2018b}, as described in Sect. \ref{chemmod}. 
The complete list of reactions added to the network is given in Table \ref{tab:addrates} and discussed in Appendix \ref{chemsupinfo}. 

\begin{figure}
   \centering
   \includegraphics[width=\hsize]{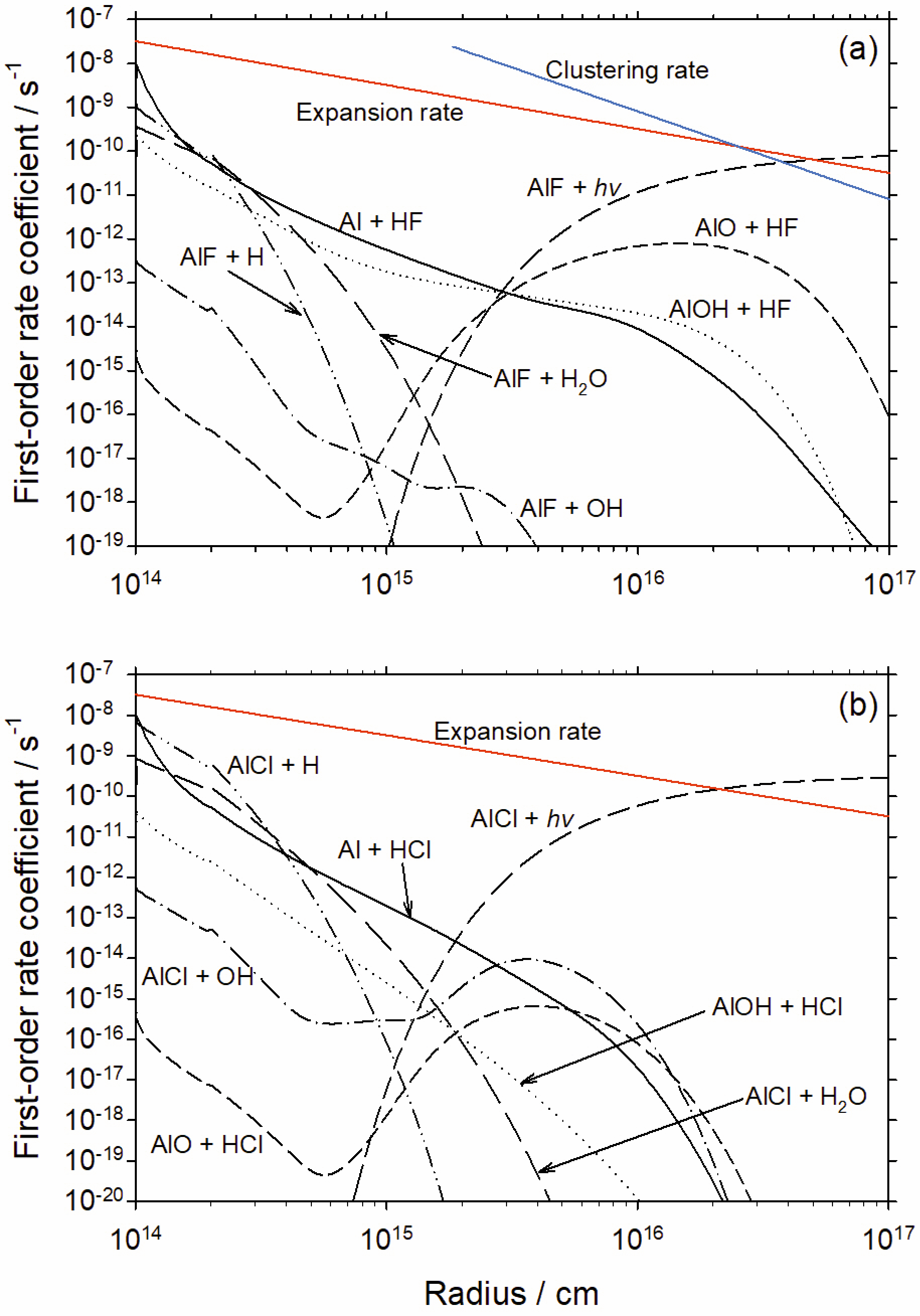}
      \caption{(a) Calculated first-order rates for conversion of HF to AlF by reaction with Al, AlOH and AlO; and loss of AlF by photolysis and reaction with H, \ce{H2O} and OH. The clustering rate with metallic compounds (blue line, see text for further details) is shown in the region where the temperature is below 500~K. (b) Calculated first-order rates for conversion of HCl to AlCl by reaction with Al, AlOH and AlO; and loss of AlCl by photolysis and reaction with H, \ce{H2O} and OH. The red lines show the molecular expansion rate of the outflow at a constant velocity of $16.5~\kms$.}
         \label{ratesvsrad}
   \end{figure}

The first-order rates (s$^{-1}$) for the formation and destruction of AlF and AlCl (R1 to R6, R-1 to R-6, and $J_1$ and $J_2$) can now be calculated as a function of radius in the outflow, using the temperature and concentrations of \h2O, OH, H, HF and HCl from the 1D  model of W~Aql \citep{Van-de-Sande2018b}.  These rates are plotted in Fig. \ref{ratesvsrad}(a) and (b) for AlF and AlCl, respectively, along with the molecular expansion rate (expressed as $2\upsilon_\infty/R$, where $\upsilon_\infty = 16.5~\kms$). This shows that the production and loss rates of both AlF and AlCl are slower than the expansion rate out to $2 \times 10^{16}$~cm. In the case of AlF, Fig. \ref{ratesvsrad}(a) shows that AlF is mostly produced in the inner region of the model by the reaction of Al with HF (R1), although production by AlO + HF (R3) becomes more important beyond $2 \times 10^{15}$~cm. Removal of AlF by reaction with \ce{H2O} (R-2) is most important in the inner region between $2 \times 10^{14}$ and $1.5 \times 10^{15}$~cm, because R-1 has a substantial activation energy so that reactions with H are not 
a competitive loss term for AlF, and the abundance of OH is relatively low. At distances $> 2 \times 10^{15}$~cm, photolysis by interstellar radiation becomes the dominant removal process. 

\begin{figure*}
   \centering
   \includegraphics[width=\hsize]{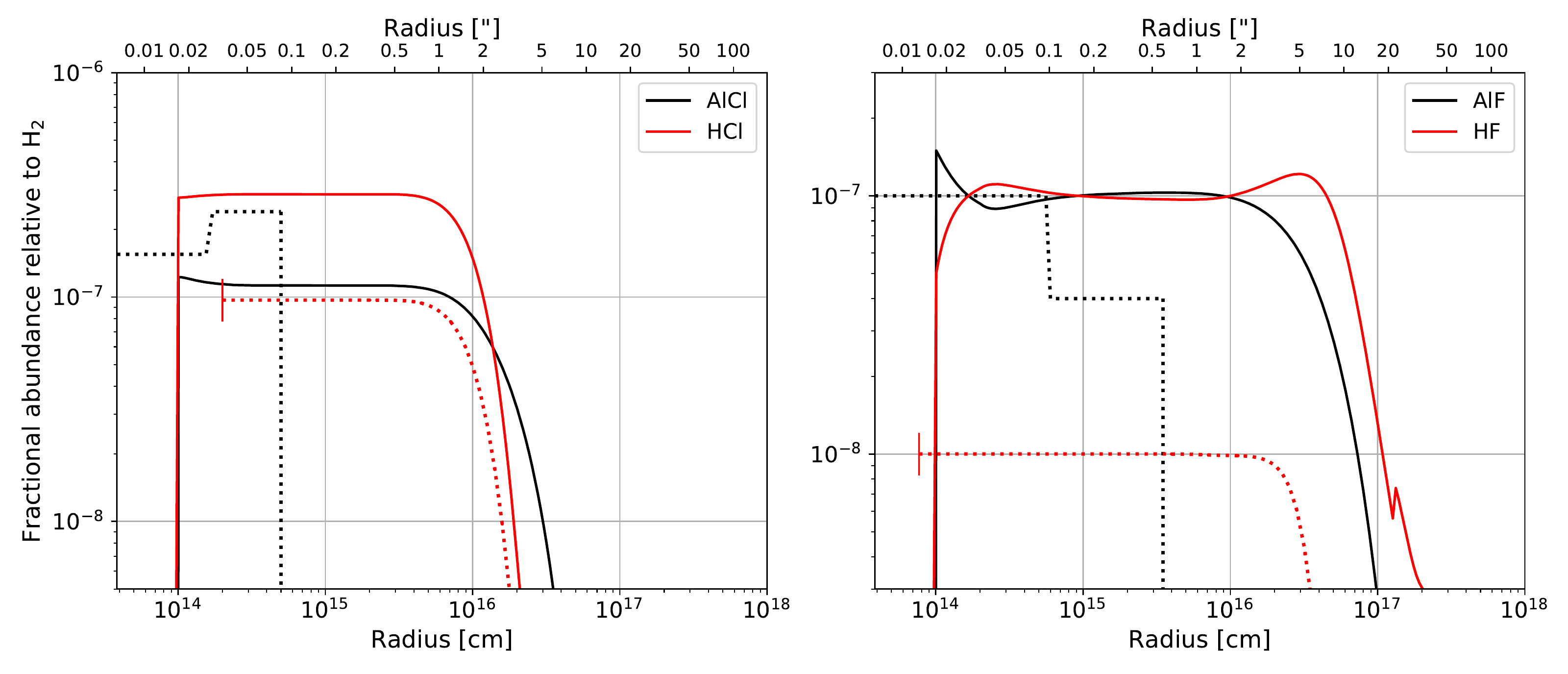}
      \caption{Fractional abundance relative to \h2 of AlCl and HCl (left panel), and AlF and HF (right panel), obtained from chemical modelling {(solid lines), plotted with the results of the radiative transfer models (dotted lines)}. The assumed physical conditions and parent species are given in the text and Appendix \ref{chemsupinfo}.}
         \label{fracabun}
   \end{figure*}

The {newly} modelled concentration profiles of the halogen species
are shown as a function of radius in Fig. \ref{fracabun}, {plotted with the results of the radiative transfer models (Sect. \ref{sec:modresults})}.  The model successfully simulates the observed absolute relative abundance of AlF at radii $< 3 \times 10^{15}$~cm (c.f. Fig. \ref{abdist}). However, the AlF photolysis rate does not start to approach the expansion rate until a radial distance of $5 \times 10^{16}$~cm. Hence, AlF is nearly unchanged until $> 10^{16}$~cm (Fig. \ref{fracabun}), in comparison with the observed disappearance of AlF around $3 \times 10^{15}$~cm (Fig. \ref{abdist}). This discrepancy probably indicates that AlF is efficiently removed in the cooler part of the outflow ($>10^{15}$~cm, where the temperature is below 400 K) by clustering with other metallic molecules {such as oxides (e.g. FeO and MgO) and hydroxides (FeOH and MgOH), as well as small dust particles (e.g. \ce{(Al2O3)_n}, \ce{(FeMgSiO4)_n}, $n > 1$)}. The total concentration of the major metals (Mg, Fe, Al and Na) relative to \ce{H2} is $7.6 \times 10^{-5}$ \citep{Asplund2009}. Clustering is likely to be fast because metal-containing molecules have large dipole moments. Assuming a clustering rate coefficient of $5 \times 10^{-10}$~cm$^3$~molecule$^{-1}$~s$^{-1}$ {\citep[i.e. a typical dipole-dipole capture frequency][]{Saunders2006}}, 
then the blue line in Fig. \ref{ratesvsrad}(a) shows that the first-order clustering rate in this cooler region is faster than the expansion rate out to $2 \times 10^{16}$~cm. {This means that the observed disappearance of AlF by $4\e{15}$~cm (Fig. \ref{abdist}) could be explained by cluster formation or uptake on dust particles.}

In the case of AlCl, the production and loss rates as a function of distance are illustrated in Fig. \ref{ratesvsrad}(b). The reaction of Al with HCl (R4) is the most important AlCl production term over the entire outflow, and photolysis dominates AlCl removal beyond $2 \times 10^{14}$~cm. The {newly} simulated HCl density (Fig. \ref{fracabun}) is in good accord with observations, including its disappearance beyond $10^{16}$~cm. {However, the location of the drop-off in HCl abundance is not tightly constrained by our current radiative transfer model, in the absence of higher quality data (see Sect. \ref{sec:hclmod}).} Although the model simulates well the absolute AlCl density out to $5 \times 10^{14}$~cm, it fails to reproduce the rapid decrease in AlCl further out. The reason(s) for this are unclear. Removal of AlCl on dust or molecular clustering might be an explanation, but this would require that AlCl was removed much more efficiently than AlF, {which seems unlikely because AlF is much less volatile than AlCl. For example, the heat of vaporisation of AlF from \ce{AlF3} is 1227~kJ~mol$^{-1}$ at 1000~K, whereas that of AlCl from \ce{AlCl3} is only 620~kJ~mol$^{-1}$ \citep{Chase1985}.}
So this remains an interesting challenge for future study.

\section{Conclusions}\label{conclusion}


We presented observations of AlCl and AlF towards an S-type AGB star for the first time. We detected rotational lines of AlCl up to the second vibrationally excited state and one rotational line of AlF in the ground vibrational state. AlCl was found in regions very close to the star, within 0\farcs1, while AlF was found in a larger region of the envelope, out to 0\farcs2 -- 0\farcs6. The distribution of both molecules was slightly asymmetric, probably due to regions of higher and lower density around the star.

The observations of AlCl and AlF were azimuthally averaged and analysed using 1D non-LTE radiative transfer models. We found step-function abundance profiles best reproduce the ALMA observations, with Al$^{35}$Cl increasing at $\sim 3.6R_\star$ from an abundance of $8.5\e{-8}$ to $1.7\e{-7}$, relative to \h2, and no longer present from $\sim 13R_\star$. For Al$^{37}$Cl we did not find a step-function in abundance (possibly due to the fainter data) and instead found a constant abundance of $7\e{-8}$ relative to \h2, also out to $\sim 13R_\star$. For AlF we found a higher abundance in the inner region close to the star, with abundance $1\e{-7}$ relative to \h2 out to $\sim16R_\star$, after which it dropped down to $4\e{-8}$ until $\sim 90R_\star$, beyond which it was not detected.
The AlCl and AlF abundances found for W~Aql are higher than those seen for the carbon star CW~Leo, and distributed differently in the CSE. This points to different chemical processes taking part in the creation and destruction of these molecules in the S-type W~Aql and the carbon-rich CW~Leo.

In addition to the ALMA observations of AlCl and AlF, we used radiative transfer models and unresolved PACS spectra of HCl and HF towards W~Aql to constrain the abundances of those molecules, using predictions from chemical models to determine the size of the corresponding molecular envelopes. We found an HCl abundance of $9.7\e{-8}$, relative to \h2, and were able to put an upper limit on HF of $\leq 1\e{-8}$. We also modelled HCl and HF for another S-type AGB star: $\chi$~Cyg and found a slightly lower abundance of HCl ($6.5\e{-8}$ relative to \h2)  and a higher abundance of HF ($1.2\e{-8}$).

The total abundance of F calculated for W~Aql (even if we exclude the upper limit found for HF) is higher than the solar abundance of F, indicating that not only has F been synthesised in W~Aql, as is expected for AGB stars, but has also been dredged up to the surface and ejected into the circumstellar envelope.

{From an analysis of chemical reactions in the wind, we find that gas-phase reactions alone cannot explain the abundance distributions of AlCl and AlF found from the observations and radiative transfer modelling. We conclude that AlF is most likely removed from the gas phase due to clustering (i.e. as part of the dust formation process). However, the very rapid removal of AlCl may be due to an additional factor that cannot yet be fully explained.}

\begin{acknowledgements}
We thank the anonymous referee for their thoughtful feedback on the manuscript.
      TD, MVdS and SHJW acknowledge support from the Research Foundation Flanders (FWO) through grants 12N9920N, 12X6419N, and 1285221N respectively. TJM is grateful to the STFC for support through grant ST/P000312/1. JMCP was supported by STFC grant number ST/T000287/1. LD, JMCP, WH, SHJW, DG acknowledge support from ERC consolidator grant 646758 AEROSOL. EDB acknowledges support from the Swedish National Space Agency. EC acknowledges funding from the KU Leuven C1 grant MAESTRO C16/17/007.
      This paper makes use of the following ALMA data: ADS/JAO.ALMA\#2018.1.00659.L. ALMA is a partnership of ESO (representing its member states), NSF (USA) and NINS (Japan), together with NRC (Canada), MOST and ASIAA (Taiwan), and KASI (Republic of Korea), in cooperation with the Republic of Chile. The Joint ALMA Observatory is operated by ESO, AUI/NRAO and NAOJ.
      We acknowledge excellent support from the UK ALMA Regional Centre (UK ARC), which is hosted by the Jodrell Bank Centre for Astrophysics (JBCA) at the University of Manchester. The UK ARC Node is supported by STFC Grant ST/P000827/1.
      PACS has been developed by a consortium of institutes led by MPE (Germany) and including UVIE (Austria); KU Leuven, CSL, IMEC (Belgium); CEA, LAM (France); MPIA (Germany); INAF-IFSI/OAA/OAP/OAT, LENS, SISSA (Italy); IAC (Spain). This development has been supported by the funding agencies BMVIT (Austria), ESA-PRODEX (Belgium), CEA/CNES (France), DLR (Germany), ASI/INAF (Italy), and CICYT/MCYT (Spain).
      This work has made use of data from the European Space Agency (ESA) mission {\it Gaia} (\url{https://www.cosmos.esa.int/gaia}), processed by the {\it Gaia} Data Processing and Analysis Consortium (DPAC, \url{https://www.cosmos.esa.int/web/gaia/dpac/consortium}). Funding for the DPAC has been provided by national institutions, in particular the institutions participating in the {\it Gaia} Multilateral Agreement.
\end{acknowledgements}

%
%
\bibliographystyle{aa} 
\bibliography{WAql_halides} 

\begin{appendix}

\section{Non-detections of other halide and aluminium-bearing molecules}

\begin{table*}
\caption{The rms values for undetected lines of NaCl, KCl, AlO, and AlOH towards W Aql.}             
\label{tab:saltrms}      
\centering                          
\begin{tabular}{c c c c c c}        
\hline\hline                 
Molecule & Transition & Frequency & Extended ($R=0\farcs1$) & Mid ($R=0\farcs6$)& Vel res\\
 & $(\varv=0)$ & [GHz] & [mJy] & [mJy] & [$\kms$]\\
\hline
NaCl	&	($17 \to 16$)	&	221.260	&	4.1	&	4.9	&	1.3	\\
NaCl	&	($19\to18$)	&	\phantom{.$^{a}$}247.240\tablefootmark{a}	&	4.5	&	6.0	&	1.2	\\
NaCl	&	($20 \to 19$)	&	260.223	&	\multicolumn{2}{c}{Overlap with H$^{13}$CN (3-2) $\nu_2 = 1$}		& ...			\\
KCl	&	($28 \to 27$)	&	215.008	&	4.0	&	4.4	&	1.4	\\
KCl	&	($30 \to 29$)	&	230.321	&	3.8	&	4.3	&	1.3	\\
KCl	&	($32 \to 31$)	&	\phantom{.$^{b}$}245.624\tablefootmark{b}	&	5.6	&	5.6	&	1.2	\\
KCl	&	($33 \to 32$)	&	\phantom{.$^{b}$}253.271\tablefootmark{b}	&	3.8	&	7.1	&	1.2	\\
KCl	&	($35 \to 34$)	&	268.559	&	6.5	&	12.3	&	1.1	\\
AlO	&	($6\to5$)	&	229.670	&	3.3	&	4.9	&	1.3	\\
AlO	&	($7\to6$)	&	267.937	&	6.4	&	11.7	&	1.1	\\
AlOH	&	($7\to6$)	&	220.330	&	3.4	&	3.7	&	1.3	\\
AlOH	&	($8\to7$)	&	251.794	&	4.1	&	8.3	&	1.2	\\
\hline                                   
\end{tabular}
\tablefoot{Wavelengths and energies taken from CDMS for NaCl: \citep{Clouser1964,Uehara1989,Caris2002,Timp2012,Cabezas2016}; KCl: \citep{Clouser1964,Caris2004,Barton2014}; AlO: \cite{Torring1989,Yamada1990}; and from the JPL Molecular Spectroscopy Database for AlOH: \cite{Apponi1993}. (\tablefootmark{a}) Line falls on the edge of a band, and rms is measured nearby, rather than at the line frequency, so as to exclude artefacts on the band edge. (\tablefootmark{b}) Due to nearby lines, rms is measured only to one side of the listed line, to avoid contamination.}
\end{table*}

\subsection{NaCl and KCl}\label{saltnds}

No NaCl or KCl lines were detected towards W~Aql in the ATOMIUM survey. To give an indication of the detection limits of our data, we measured the rms values of the spectra near the vibrational ground state NaCl and KCl lines covered by ATOMIUM. For each line listed in Table \ref{tab:saltrms} we calculated the rms over a velocity range of 100~$\kms$, centred on the line frequency. The calculation was done for the spectrum extracted from the extended array with a 0\farcs1 radius aperture and for the spectrum from the mid array extracted with a 0\farcs6 radius aperture. These were chosen because we do not know a priori the extent of the possible NaCl or KCl lines and by checking for compact and more extended emission we can be sure of the non-detection of the salt lines. Where a line fell close to the edge of a band or close to a detected line, we measured the rms only for the unaffected half of the spectrum (over a velocity range of 50~$\kms$). The NaCl ($20\to19$) line at 260.223~GHz was excluded from the measurement since it is dominated by an overlap with a $\nu_2=1$ H$^{13}$CN line. The rms values are given in Table \ref{tab:saltrms}.

\subsection{AlO and AlOH}\label{alnds}

\begin{figure}
   \centering
   \includegraphics[width=\hsize]{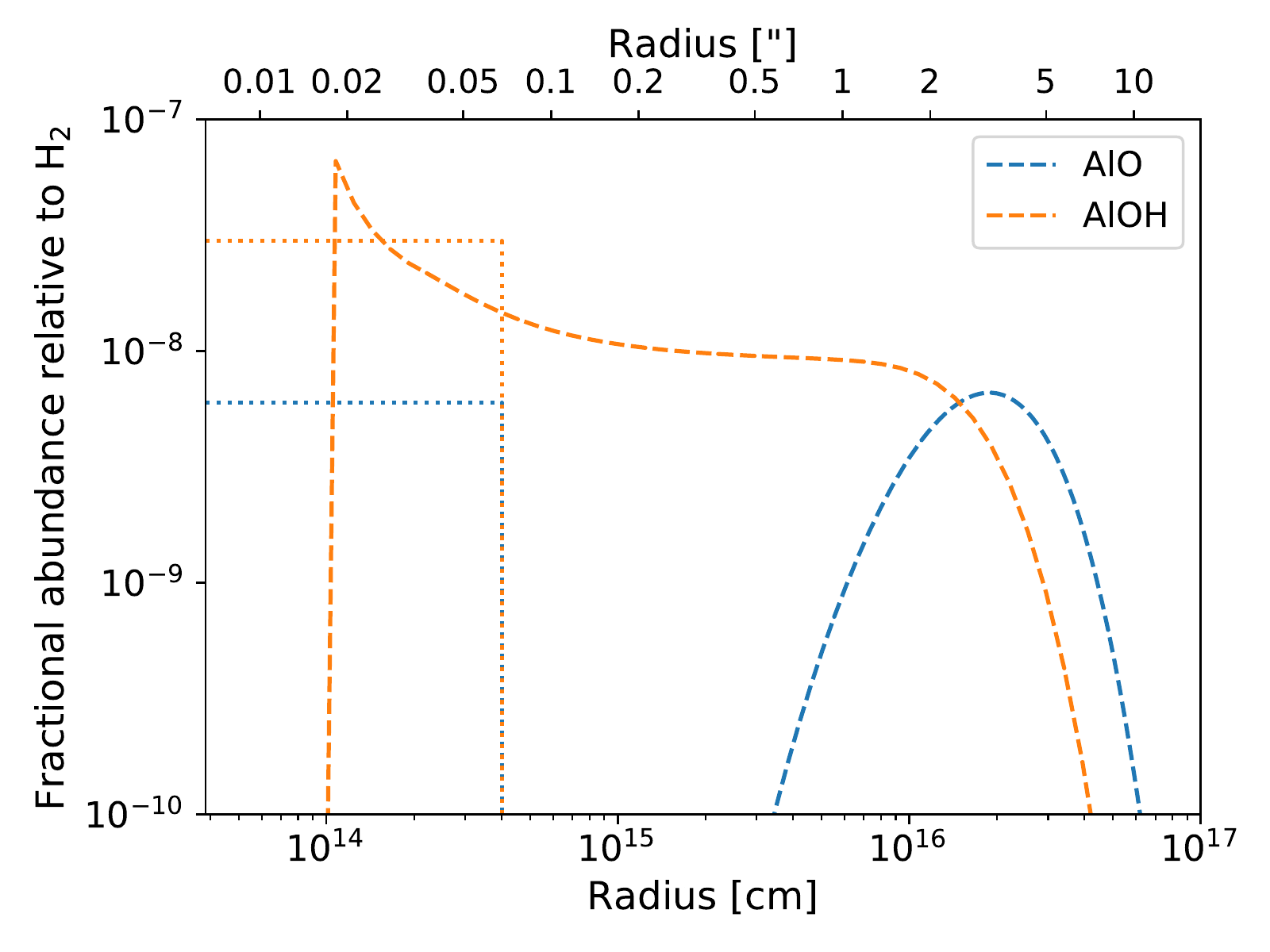}
      \caption{Upper limits for AlO (blue) and AlOH (orange) abundance distributions. Dotted lines show the constant abundance models and dashed lines show abundance distributions predicted by chemical models. See text for details.}
         \label{aluplims}
   \end{figure}

We performed similar measurements for AlO and AlOH, with the rms values also listed in Table \ref{tab:saltrms}. For these two molecules we additionally ran some radiative transfer models to ascertain the upper limits on their abundances. {The results for two sets of models are shown in Fig. \ref{aluplims}.} This was done for the aluminium-bearing molecules because they are expected to play a key role in circumstellar chemistry (see Sect. \ref{sec:chemistry}).

\section{Molecular data and collisional rates}\label{colrates}

\subsection{AlCl}

For both AlCl isotopologues, we included molecular data for levels with $J\leq40$ and $v\leq 10$. The maximum vibrationally excited level of $\varv=10$ was chosen because the term energy of this level is $2.2\um$, close to the wavelength of peak flux for W~Aql, as seen from its SED \citep{Danilovich2014}.
The radiative information used in our AlCl models comes from \cite{Yousefi2018}, accessed via the ExoMol database. The only AlCl collisional rates we found were for AlCl-He collisions by \cite{Pamboundom2016}. However, the rates calculated in that study only go up to the $J=17,\, \varv=0$ rotational level of AlCl, whereas our ALMA observations cover the $J=18\to17,\, \varv=0$ transition. Hence, using the rates of \cite{Pamboundom2016} would not give us the most accurate model results, even if they were scaled to account for the different mass of the AlCl-\h2 system. 

We searched for available rates for a molecule with similar characteristics to AlCl. Via the BASECOL\footnote{\url{https://basecol.vamdc.eu}} database \citep{Dubernet2013}, we found rates calculated by \cite{Klos2008} for collisions between SiS-\h2, considering both ortho- and para-\h2 and going up to the $J=40$ rotational level, in the ground vibrational state. SiS has a dipole moment of 1.74~D \citep{Hoeft1969,Murty1969}, very close to the dipole moment of AlCl: 1.63~D \citep{Yousefi2018}. SiS and AlCl also have very similar molecular masses and comparable level energies. These similar properties suggest that the collisional rates of SiS with \h2 are an adequate stand-in for the collisional rates of AlCl with \h2. For our model, we weighted the rates assuming an \h2 ortho-to-para ratio of 3, as is typical for warm environments.

\subsection{AlF}\label{alfrates}

For AlF we include molecular data for levels with $J\leq30$ and $v\leq 6$. As for AlCl, the maximum vibrationally excited level was chosen for its term energy, which is also $2.2\um$, close to the peak flux of W~Aql. 
We neglect the hyperfine structure of AlF because the separation of the hyperfine components is smaller than the spectral resolution of our observations \citep[based on parameters calculated by][]{Wyse1970}.
The radiative information used in our AlF models comes from \cite{Yousefi2018}, accessed via the ExoMol database. 

For the collisional rates, a new calculation was performed to obtain rates for sufficiently high temperatures as seen in our circumstellar model.
This calculation is an extension of the earlier work by \cite{Gotoum2012}, in which rates were given for temperatures up to 70~K. 
Using the potential energy surface (PES) for AlF-\h2 given by \cite{Gotoum2012}, we broadened the calculations of the integral cross sections for kinetic energy up to 10000~cm$^{-1}$.  Indeed the earlier work was limited for total energy $E_c \leq 350$~cm$^{-1}$ where cross sections were calculated in the quantum mechanical close coupling formalism \citep[CC,][]{Arthurs1960}. We broadened the calculation up to 500~cm$^{-1}$ using the CC method with $J_\mathrm{max}$ = 20, then up to 10000~cm$^{-1}$ using the coupled-state (CS) approach \citep{McGuire1974} with $J_\mathrm{max}=50$. The following energy steps were considered, where all values are in cm$^{-1}$: 1 for $350 \leq E \leq 500$, 5 for $500 \leq E \leq 1000$, 20 for $1000 \leq E \leq 2000$ and finally with uniform steps of 50 up to 10000. All these computations are done using the MOLSCAT package \citep{Hutson2012}.

\begin{figure}
   \centering
   \includegraphics[width=0.85\hsize]{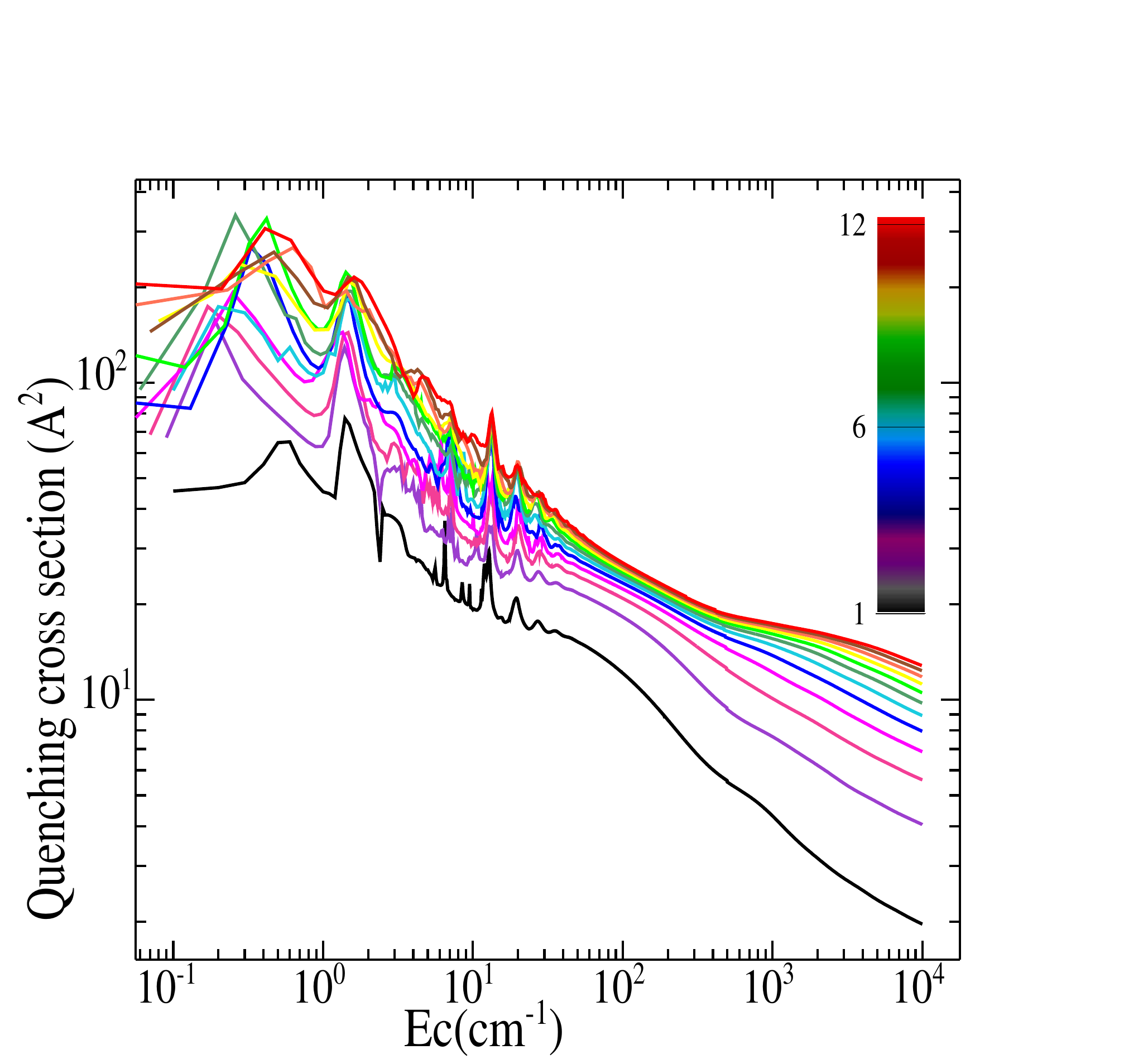}
      \caption{Rotational quenching cross sections of AlF induced by collision with \ce{H2} as a function of the kinetic energy, presented with color diagram, with $J=1$ in black to $J=12$ in red.}
         \label{alfh2qxs}
   \end{figure}

In Fig. \ref{alfh2qxs}, we present the rotational quenching cross sections of AlF in collision with \ce{H2} for the $J = 1$ to 12 rotational levels. This figure shows that for low kinetic energies cross sections present shape and Feshbach resonances and a large overlap between them. The quenching cross sections for $J=1$ and $J=2$ are separated from the others from $E_c=20$~cm$^{-1}$, while the others explode at around 500 to 1000~cm$^{-1}$.

Generally, as the rotational level increases, the energy gap of the rotational transitions increases, and the efficiency of the rotational quenching decreases. That is why the quenching cross sections vary only slightly as a function of rotational levels for $J>2$.

\begin{figure}
   \centering
   \includegraphics[width=0.85\hsize]{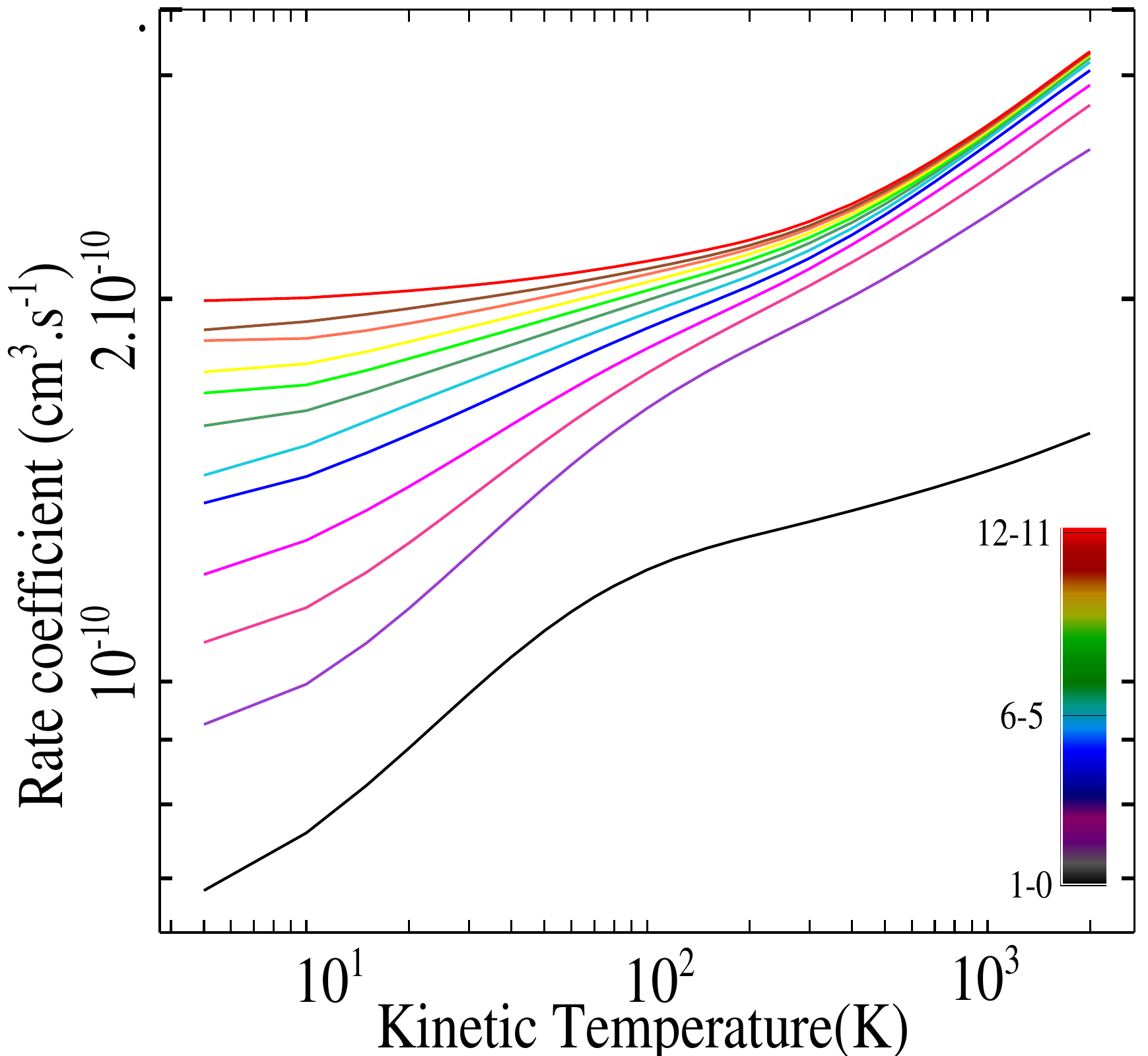}
      \caption{Log-log scale variation of downward rate coefficients for transitions with $\Delta J=-1$, as a function of the kinetic temperature.}
         \label{alfh2rates}
   \end{figure}

These cross sections were averaged over the Boltzmann distribution of velocities to determine the downward rate coefficients of AlF in collision with \ce{H2} for kinetic temperature up to 2000~K. We present in Fig. \ref{alfh2rates} these rates as a function of temperature for $\Delta J=-1$, with $J$ from 1 to 12. The rate curves exhibit the same trends and increase with increasing $J$, and the gap between the plots narrow considerably. In addition, the collision rate coefficients reflect the same behaviour as the general trends observed previously for its valence isomers: AlOH-\ce{H2} \citep{Naouai2019} and HCP-\ce{H2} \citep{Hammami2008}. We should note here that the reason this latter behaviour is also observed for the quenching cross sections for high energy values is probably because of the small rotational constant of AlF ($\sim 0.5$). In this case the kinetic energy is very large compared to the rotational energies and hence the statistical approach can be valid.

\subsection{HCl}

For HCl we include molecular data for levels with $J\leq16$ and $v\leq 2$, neglecting hyperfine structure since our PACS observations are spectrally unresolved. The spectroscopic information used in our models of HCl comes from \cite{Gordon2017}. We used HCl-\h2 collisional rates from \cite{Lanza2014a} accessed from BASECOL \citep{Dubernet2013}. \cite{Lanza2014a} calculate rates for both ortho- and para-\h2 and we again weighted these assuming an \h2 ortho-to-para ratio of 3.

\subsection{HF}

For HF we use a molecular data file obtained from the LAMDA database\footnote{\url{https://home.strw.leidenuniv.nl/~moldata/}} \citep{Schoier2005,van-der-Tak2020}, which includes levels up to $J\leq8$ for $\varv=0$ and up to $J\leq6$ for $\varv=1$. The data for the ground vibrational state was taken from \cite{Nolt1987} via the JPL Molecular Spectroscopy Database\footnote{\url{https://spec.jpl.nasa.gov/home.html}} \citep{Pickett1998}.
Rovibrational transition frequencies came from \cite{Goddon1991} and Einstein A values were computed from \cite{Pine1985}. We used the collisional rates for HF-\h2 calculated by \cite{Guillon2012}, again assuming an \h2 ortho-to-para ratio of 3. 

\section{$\chi$ Cyg HCl and HF}\label{chicygmod}

As noted in Sect. \ref{sec:stypehalides}, we found evidence of HCl and HF emission in the PACS spectrum of \object{$\chi$ Cyg}. To be able to directly compare the abundances of HCl and HF between $\chi$~Cyg and W~Aql, we also ran radiative transfer models for $\chi$~Cyg, using the same methods we used for W~Aql in Sections \ref{sec:hclmod} and \ref{sec:hfmod}. For the HCl and HF abundance distributions, we re-ran the chemical model described in Sect. \ref{chemmod}, adjusted for the mass-loss rate and expansion velocity of $\chi$~Cyg \citep[$\dot{M}=7\e{-7}\spy$ and $\upsilon_\infty=8.5~\kms$,][]{Schoier2011}. The stellar and circumstellar parameters for $\chi$~Cyg were taken from \cite{Schoier2011}. As for W~Aql, we used the non-detected lines to constrain the upper limits of the models.

\begin{figure}
   \centering
   \includegraphics[width=0.9\hsize]{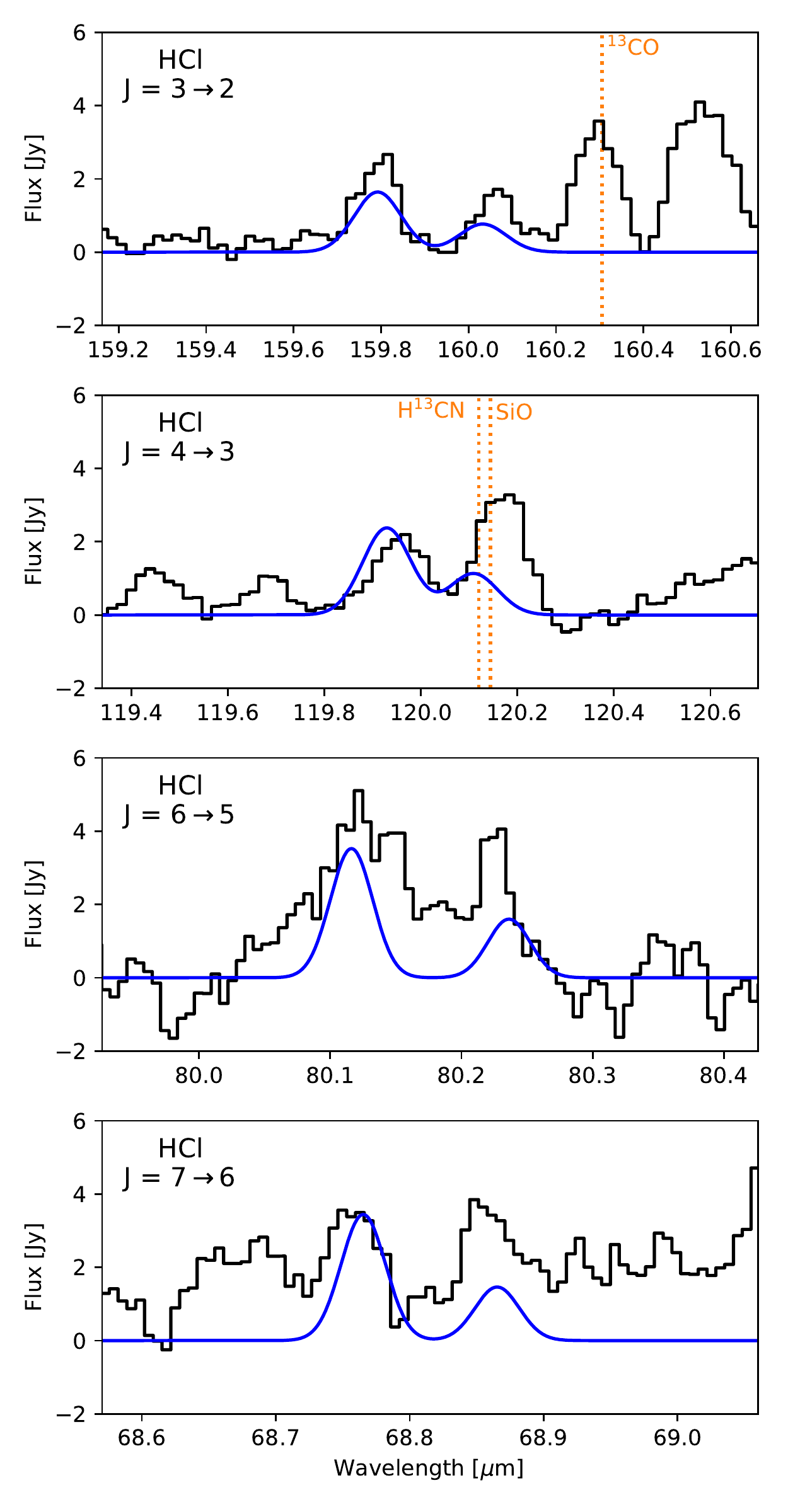}
      \caption{PACS spectra (black histograms) and model results (blue curves) for HCl towards $\chi$~Cyg. For each pair of lines, H$^{35}$Cl is shown on the left since it has the shorter wavelength and H$^{37}$Cl is on the right, with the longer wavelength. Some known nearby and blended lines are indicated in orange (but not all nearby lines have been identified).}
         \label{chicyghclmod}
   \end{figure}

\begin{figure}
   \centering
   \includegraphics[width=\hsize]{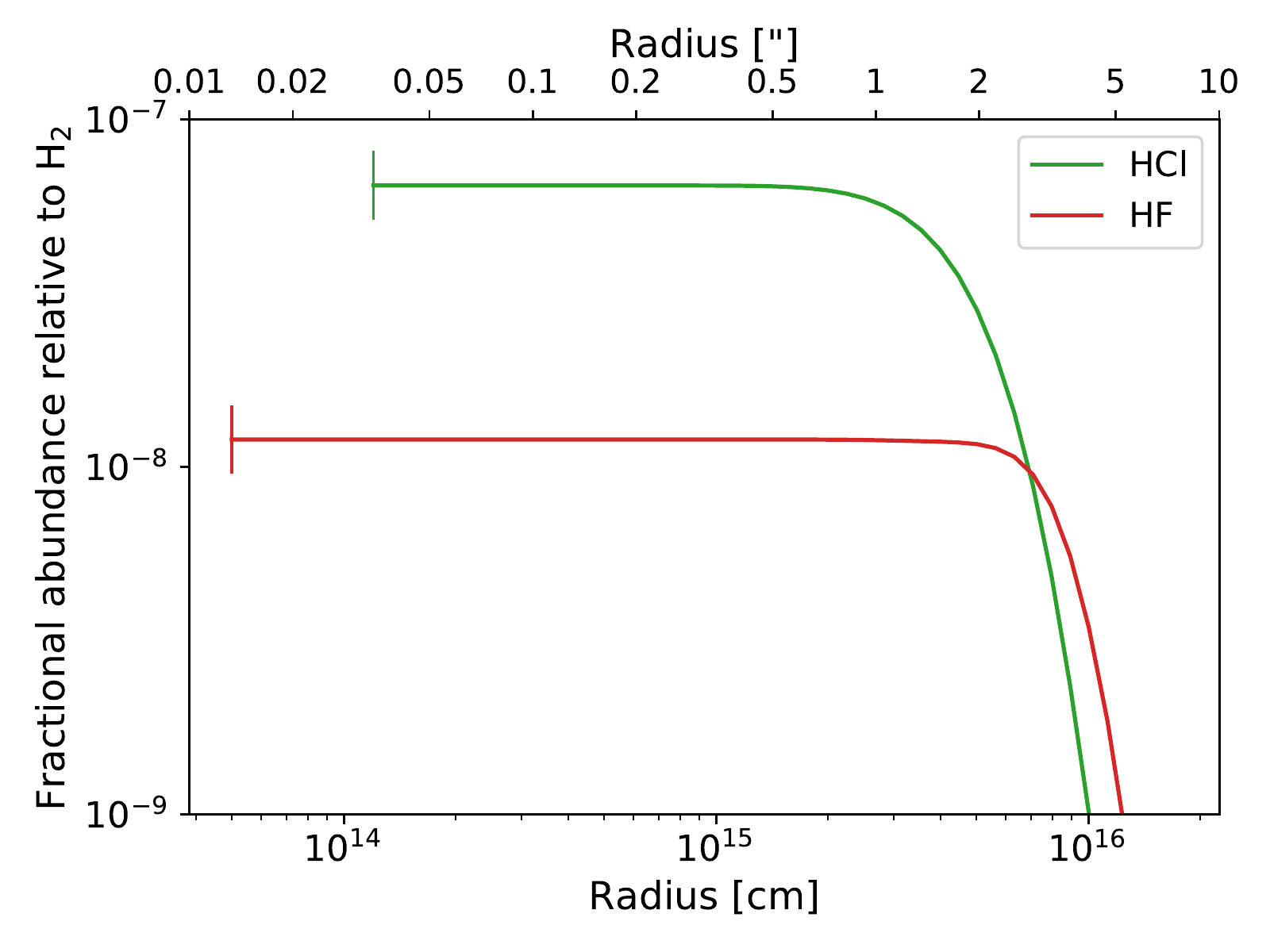}
      \caption{Radial abundance profiles for HCl (green) and HF (red) as derived for $\chi$ Cyg. The short vertical lines indicate the inner radius of the corresponding model.}
         \label{chicygabs}
   \end{figure}

For the HCl model, we found that the results were more strongly dependent on the choice of inner radius for $\chi$~Cyg than they were for W~Aql. The inner radius used by \cite{Schoier2011} $R_\mathrm{in}=2\e{14}$~cm resulted in models that significantly under-predited the higher-$J$ HCl lines. When we treated the inner radius as a free parameter, we found the best fitting model to have $R_\mathrm{in}=1.2\e{14}$~cm and abundances of $4.6\e{-8}$ for H$^{35}$Cl and $1.9\e{-8}$ for H$^{37}$Cl. The $^{35}$Cl/$^{37}$Cl ratio was fixed based on our W~Aql AlCl results, since the PACS data is not of sufficiently high quality to independently determine the isotopic ratio.
The PACS spectra of HCl are plotted with the model HCl lines in Fig. \ref{chicyghclmod} and the radial abundance profile is shown in Fig. \ref{chicygabs}. The fact that changing the $R_\mathrm{in}$ from $2\e{14}$~cm to $1.2\e{14}$~cm had little effect on the intensity of the ($3\to2$) model line but an increasingly significant effect on the higher-$J$ lines suggests that these lines are probably emitted from the 1.2--2~$\e{14}$~cm region. However, higher quality observations, preferably spectrally and spatially resolved, are needed to confirm this.

\begin{figure*}
   \centering
   \includegraphics[width=0.8\hsize]{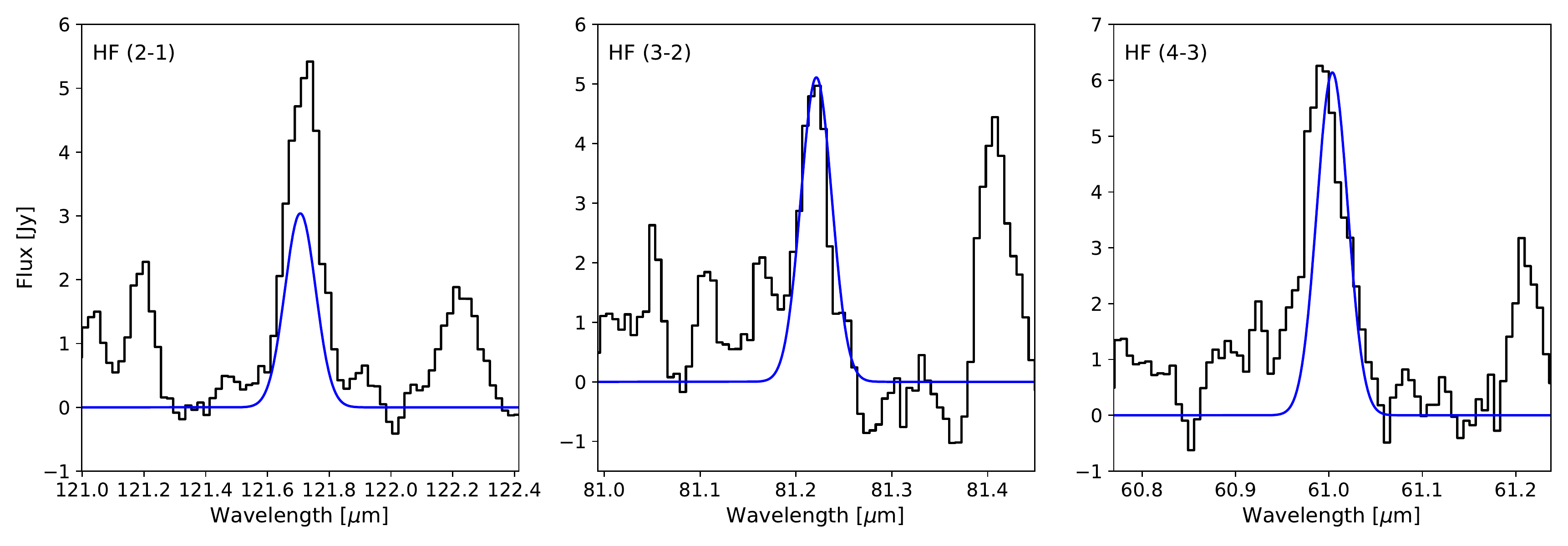}
      \caption{PACS spectra (black histograms) and model results (blue curves) for HF towards $\chi$~Cyg. The HF ($2\to1$) line is blended with the o-\h2O line at $121.721\um$, which is not shown here. See text for details.}
         \label{chicyghfmod}
   \end{figure*}

For HF, we found that the ($3\to2$) and ($4\to3$) lines were more clearly seen towards $\chi$~Cyg than W~Aql, possibly due to the closer proximity of that star (150~pc for $\chi$~Cyg compared with 395~pc for W~Aql). As for W~Aql, the ($2\to1$) line towards $\chi$~Cyg is blended with \h2O. However, although \cite{Schoier2011} modelled \h2O towards $\chi$~Cyg (to find an abundance about 80\% of the W~Aql \h2O abundance), their model was based only on HIFI observations and did not include any PACS data. Hence we are unable to estimate the contribution of \h2O to the \h2O+HF line blend, as was done for W~Aql. Instead, we focus our HF modelling on the other two HF lines, which are not known to be blended. Similar to HCl, we found that the line ratios of these two HF lines were very sensitive to the choice of inner radius. Leaving $R_\mathrm{in}$ as a free parameter again, we found the best fitting model had $R_\mathrm{in}=6\e{13}~\mathrm{cm}\approx 2R_\star$ and an inner HF abundance of $1.2\e{-8}$ relative to \h2. These best fitting model lines are plotted with the PACS spectra of HF in Fig. \ref{chicyghfmod} and the radial abundance profile is shown in Fig. \ref{chicygabs}.

\section{Supplementary information concerning the chemistry of AlF and AlCl}\label{chemsupinfo}

\begin{figure}
   \centering
   \includegraphics[width=\hsize]{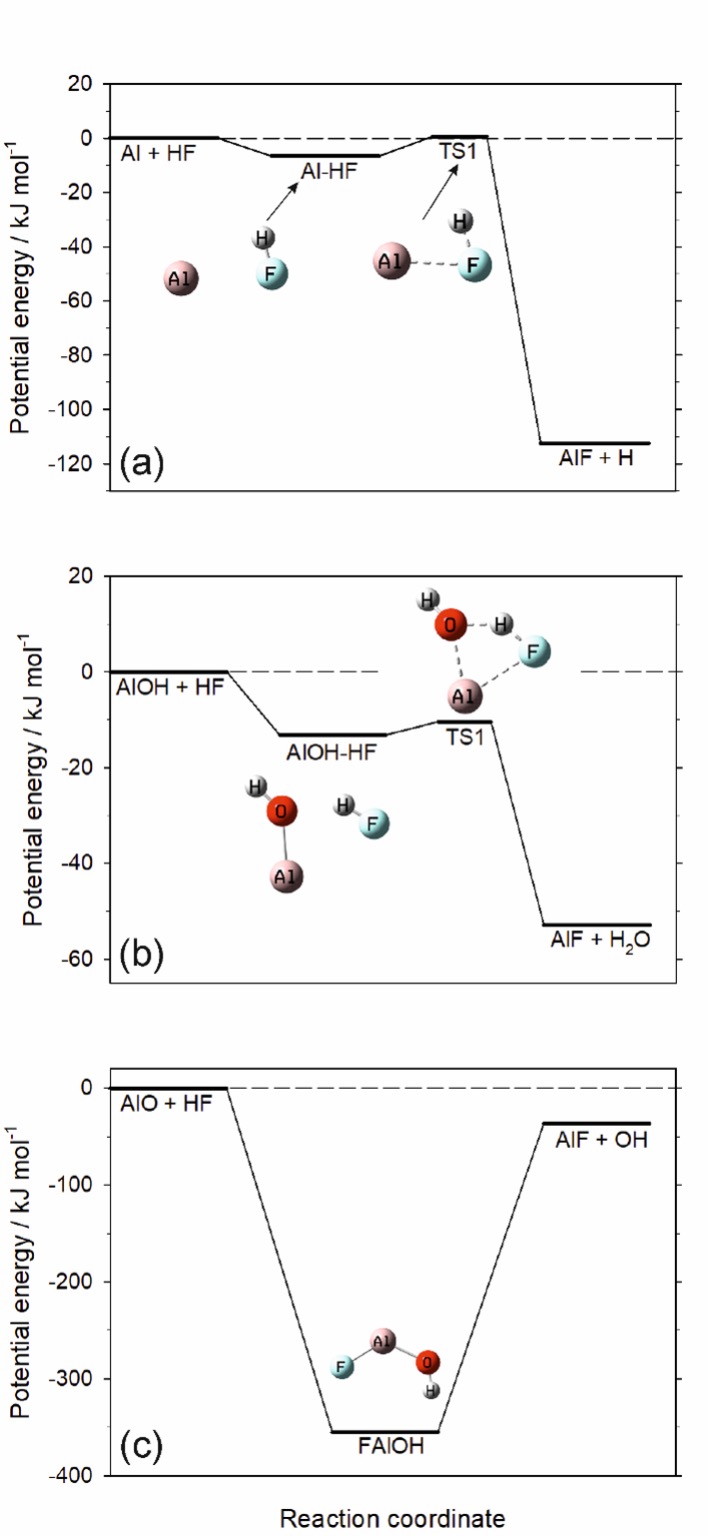}
      \caption{Potential energy surfaces calculated at the G4 level of theory for: (a) Al + HF (R1); (b) AlOH + HF (R2); (c) AlO + HF (R3). Note that the very endothermic channels are not shown.}
         \label{fig:A1}
   \end{figure}

\begin{figure}
   \centering
   \includegraphics[width=\hsize]{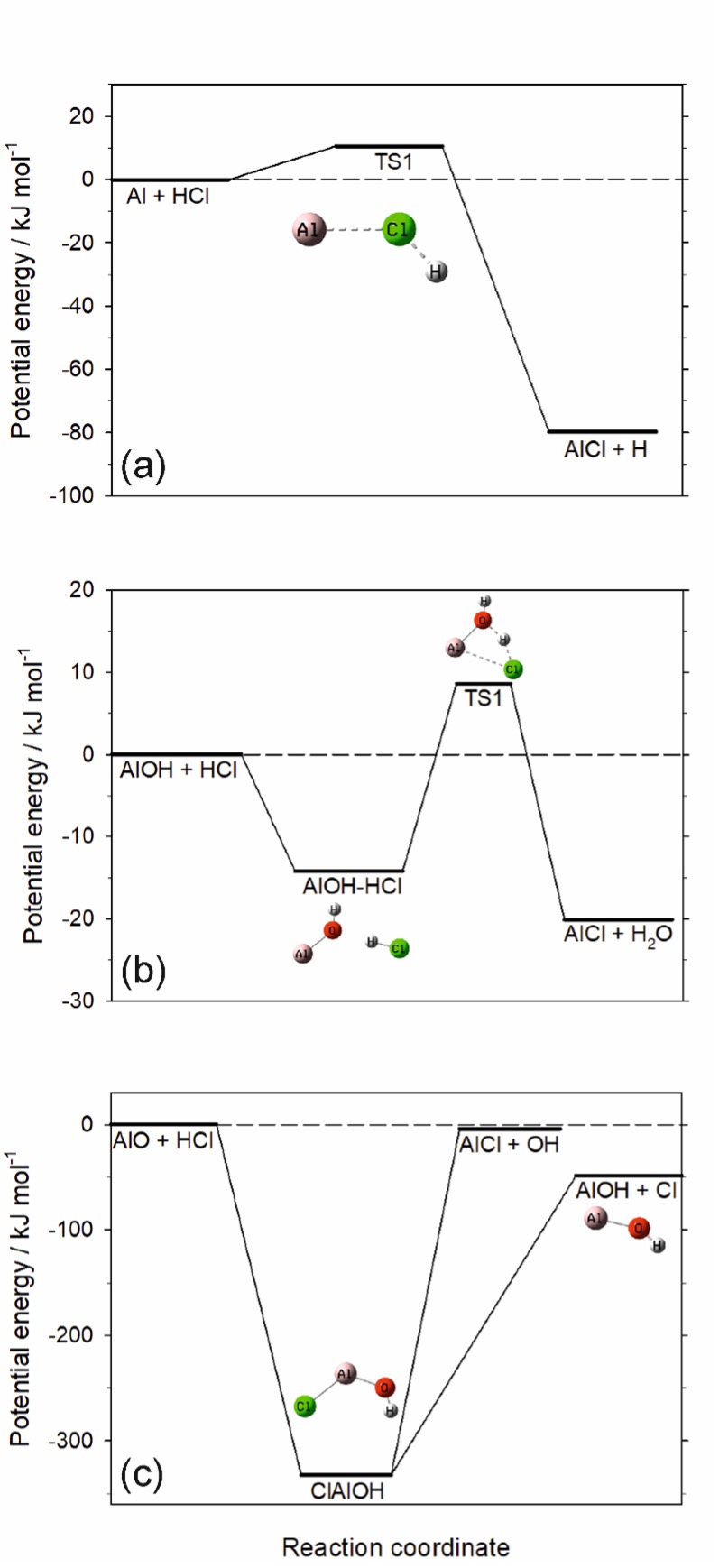}
      \caption{Potential energy surfaces calculated at the G4 level of theory for: (a) Al + HCl (R1); (b) AlOH + HCl (R2); (c) AlO + HCl (R3). Note that the very endothermic channels are not shown.}
         \label{fig:A2}
   \end{figure}

The rate coefficients for the reactions of Al, AlO and AlOH with HF and HCl to produce AlF and AlCl, respectively, were estimated by combining electronic structure calculations with Rice-Ramsperger-Kassel-Markus (RRKM) statistical rate theory. Accurate energies, geometries and vibrational frequencies were determined at the G4 level of theory \citep{Curtiss2007} within the Gaussian 16 suite of programs \citep{Frisch2016}. The Cartesian coordinates, rotational constants, vibrational frequencies and heats of formation of the relevant molecules are listed in Table \ref{tab:A1} for reactions producing AlF (R1 -- R3), and Table \ref{tab:A2} for the AlCl-forming reactions (R4 -- R6). The potential energy surfaces for the six reactions are illustrated in Figs. \ref{fig:A1} and \ref{fig:A2}, which also show the geometries of the stationary points. Note that the relative energies include zero-point energy corrections.

The Master Equation Solver for Multi-Energy well Reactions (MESMER) program \citep{Glowacki2012} was then used to estimate rate coefficients. Apart from R4, these reactions proceed via the formation of an energised adduct. This adduct can either dissociate back to the reactants, dissociate to bimolecular products sometimes involving a barrier (R1, R2 and R5, but not R3 or R6), or be stabilised by collision with a third body (which in the case of the outflow is \ce{H2}). The internal energies of the stationary points on the surface (i.e. reactants, intermediates, transition states and products) were divided into a contiguous set of grains (width = 150~cm$^{-1}$) containing a bundle of rovibrational states. The density of states of each stationary point was calculated using the vibrational frequencies and rotational constants listed in Tables \ref{tab:A1} and \ref{tab:A2}. The vibrations were treated as harmonic oscillators, and a classical densities of states treatment was used for the rotational modes. 

Each adduct grain was then assigned a set of microcanonical rate coefficients for dissociation to the reactants and products. These were determined using inverse Laplace transformation to link them directly to the relevant capture rates. These capture rates were calculated using long-range transition state theory \citep{Georgievskii2005}, and are listed in Table \ref{tab:A3}. 

\begin{table}
\caption{Capture rate coefficients calculated using long range transition state theory \citep{Georgievskii2005}.}             
\label{tab:A3}      
\centering                          
\begin{tabular}{l c c}        
\hline\hline                 
& Reaction & Rate coefficient \\
& & [cm$^3$~molecule$^{-1}$~s$^{-1}$]\\
\hline
R1 & \ce{Al +HF \to AlF + H} & $7.7\e{-10}\exp(-86/T)$\\
R2 & \ce{AlOH + HF  \to  AlF + H2O}	& $7.8\e{-10}\exp(-86/T)$\\
R3 & \ce{AlO + HF  \to  AlF + OH}		& $1.0\e{-9}(T/298)^{-0.17}$\\
R4 & \ce{Al + HCl  \to  AlCl + H	}	& $1.0\e{-9}\exp(-86/T)$\\
R5 & \ce{AlOH + HCl  \to  AlCl + H2O}	& $1.0\e{-9}\exp(-86/T)$\\
R6 & \ce{AlO + HCl  \to  AlCl + OH}		& $1.0\e{-9} \exp(-86/T)$\\
\hline
\end{tabular}
\end{table}

The probability of collisional transfer between grains was estimated using the exponential down model \citep{Gilbert1990}: the average energy for downward transitions, \mbox{$\langle\Delta E\rangle_\mathrm{down}$}, was set to 200~cm$^{-1}$ with no temperature dependence, and the probabilities for upward transitions are determined by detailed balance. The collision rate of \ce{H2} with the adduct as a function of temperature was calculated using Lennard-Jones parameters ($\sigma = 3.0~\AA$ and $\varepsilon/k = 200$~K) to characterise the intermolecular potential. The ME, which describes the evolution with time of the adduct grain populations, was then expressed in matrix form and solved to yield the rate coefficients for recombination and bimolecular reaction at a specified pressure and temperature. In fact, at the low pressures in the outflow (\ce{[H2]} < $10^{8}$~cm$^{-3}$ beyond a radius of $2 \times 10^{14}$~cm), it is only the bimolecular channels that matter. The rate coefficients for the reverse reactions were calculated by detailed balance, using the molecular parameters in Tables \ref{tab:A1} and \ref{tab:A2} to calculate the relevant equilibrium constants. In the case of reaction R6 between AlO and HCl, the AlOH + Cl channel is 44~kJ~mol$^{-1}$ more exothermic than AlCl + OH, and so AlOH is the major product (91\% at 900~K, 79\% at 1900~K).

\begin{table*}
\caption{Molecular properties and heats of formation (at 0 K) of the stationary points on the Al, AlOH and AlO + HF potential energy surfaces. } 
\label{tab:A1}
\centering
\begin{tabular}{llllr}
\hline
\multicolumn{1}{c}{Molecule} & \multicolumn{1}{c}{Geometry$^{a}$} & \multicolumn{1}{c}{Rotational constants$^{a}$} & \multicolumn{1}{c}{Vibrational  freq.$^{a}$} & \multicolumn{1}{c}{$\delta$$_{f}$H$^{o}$(0 K)$^{b}$} \\
\multicolumn{1}{c}{(electronic state)} & \multicolumn{1}{c}{[Cartesian co-ords in $\AA$]} & \multicolumn{1}{c}{[GHz]} & \multicolumn{1}{c}{[cm$^{-1}$]} & \multicolumn{1}{c}{[kJ mol$^{-1}$]}\\
\hline\hline
\multicolumn{5}{c}{Al + HF  $\to$  AlF + H}					\\
HF&	F, 0., 0., 0.090&	624.62&	4117&	-273.6	\\
&	 H, 0., 0., -0.829	\smallskip\\
Al-HF complex&	Al, 0.151, -1.228, 0.&	607.47 8.1347 8.0272&	159, 467, 3541&	47.3	\\
& H, -0.944, 0.847, 0. \\
& F, -0.032, 1.098, 0. \smallskip\\
TS from Al-HF complex&	Al, 0.048, -0.843, 0. & 	485.83 11.330 11.072&	894.2$i$, 519, 1418&	56.2	\\
to AlF + H (TS1)&	H, -0.991, 0.757, 0.	\\
&	F, 0.042, 1.143, 0.	\smallskip\\
AlF&	F, 0., 0., -0.983 &	16.4524&	828&	-275.0	\\
&	Al, 0., 0., 0.677	\\
\hline
\multicolumn{5}{c}{AlOH + HF  $\to$  AlF + H$_{2}$O}					\\
AlOH&	Al, 0.026, 0., 0.018 & 	2590.6 15.754 15.658&	215, 849, 3960&	-186.6	\\
&	O, -0.063, 0., 1.702	\\
&	H, 0.364, 0., 2.555	\smallskip\\
AlOH-HF complex&	Al, -0.226, -0.705, 0.001&	14.301 8.2297 5.2238&	131, 139, 338, 	&-471.3	\\
&	O, -0.202, 1.069, -0.003&&559, 705, 793, 	\\
&	 H, -0.888, 1.736, 0.002 &&1080, 3151, 3919	\\
&	F, 1.893, 0.063, 0.001	\\
&	H, 1.347, 0.865, -0.000	\smallskip\\
TS from AlOH-HF &	Al, 0.525, -0.801, 0.021 &	13.443 9.7664 5.6824&	-263$i$, 267, 399, &	-465.8	\\
complex to AlF + H$_{2}$O&	O, 0.640, 1.020, -0.069 &&586, 674, 1030, 	\\
&	H, -0.714, 0.845, 0.014	&& 1237, 2187, 3901\\
&	F, -1.368, 0.037, -0.001	\\
&	H, 1.217, 1.647, 0.371	\smallskip\\
H$_{2}$O&	O, 0.001, 0., 0.001 &	798.21 438.23 282.91&	1672, 3802, 3906&	-247.2	\\
&	H 0, 0.0123, 0., 0.963 	\\
&	H, 0.933, 0., -0.237	\\
\hline
\multicolumn{5}{c}{AlO + HF $\to$ AlF + OH}					\\
AlO&	Al, 0., 0., 0.002 &	19.0158&	967&	70.3	\\
&	O, 0., 0., 1.628	\smallskip\\
FAlOH&	Al, 0.704, 0.261, 0. &	43.502 6.5157 5.6669&	215,  308, 622, &	-556.4	\\
&	O, -0.923, -0.219, 0. &&786, 923, 3873	\\
&	H, -1.179, -1.144, 0.	\\
&	F, 1.949, -0.829, 0.	\smallskip\\
OH&	O, 0., 0., 0.002&	559.17&	3691&	34.8	\\
&	 H, 0., 0., 0.978&\\
\hline
\end{tabular}
\tablefoot{TS denotes a transition state. ($^{a}$) Calculated at the G4 level of theory \citep{Curtiss2007,Frisch2016}. ($^{b}$) Calculated at the G4 level of theory \citep{Curtiss2007,Frisch2016} with JANAF reference values for $\Delta_f H^o$(Al) = 327.3~kJ~mol$^{-1}$, $\Delta_f H^o$(O) = 246.8~kJ~mol$^{-1}$, $\Delta_f H^o$(F) = 77.3~kJ~mol$^{-1}$, and $\Delta_f H^o$(H) = 216.0~kJ~mol$^{-1}$ \citep{Chase1985}.}
\end{table*}

\begin{table*}
\caption{Molecular properties and heats of formation (at 0 K) of the stationary points on the Al, AlOH and AlO + HCl potential energy surfaces.} 
\label{tab:A2}
\centering
\begin{tabular}{llllr}
\hline
\multicolumn{1}{c}{Molecule} & \multicolumn{1}{c}{Geometry$^{a}$} & \multicolumn{1}{c}{Rotational constants$^{a}$} & \multicolumn{1}{c}{Vibrational  freq.$^{a}$} & \multicolumn{1}{c}{$\delta$$_{f}$H$^{o}$(0 K)$^{b}$} \\
\multicolumn{1}{c}{(electronic state)} & \multicolumn{1}{c}{[Cartesian co-ords in $\AA$]} & \multicolumn{1}{c}{[GHz]} & \multicolumn{1}{c}{[cm$^{-1}$]} & \multicolumn{1}{c}{[kJ mol$^{-1}$]}\\
\hline\hline
HCl&	H, 0.,0., -1.212&	313.07&	2958&	-90.3	\\
&	Cl, 0.,0., 0.072				\smallskip\\
TS from Al + HCl to &	Al, 0.036, 1.415, 0.&	424.11 5.6019 5.5289&	-801$i$, 353, 548&	248.6	\\
AlCl + H (TS1)&	H, -1.083, -1.946, 0.				\\
&	Cl, 0.038, -0.963, 0.				\smallskip\\
AlCl&	Al, 0., 0., -1.224&	7.1141&	463&	-59.4	\\
&	Cl, 0., 0., 0.935				\\
\hline
\multicolumn{5}{c}{AlOH + HCl   $\to$  AlCl + H$_{2}$O}				\smallskip\\
AlOH: see Table \ref{tab:A1} \smallskip\\
AlOH-HCl complex&	Al, -1.825, -0.734, -0.034&	34.350 1.6277 1.5541&	53, 145, 157, 437,&	-290.7	\\
&	O, -0.755, 0.619, -0.009	&& 466, 714, 772, 			\\
&	H, -0.901, 1.567, 0.0228	&&2601, 3880			\\
&	Cl, 2.382, 0.529, 0.018				\\
&	H, 1.072, 0.540, 0.003				\smallskip\\
TS from AlOH-HCl &	Al, 1.315, -0.704, 0.019 &	13.899 3.5906 2.8599&	-560$i$, 188, 232,6&	-264.7	\\
complex to AlCl + H$_{2}$O &	O, 0.929, 1.071, -0.088 	&& 517, 623, 785, 931, 			\\
&	H, -0.363, 0.838, -0.014 		&&1392, 385		\\
&	Cl, -1.494, -0.100, 0.011 				\\
&	H, 1.325, 1.837, 0.337				\smallskip\\
H$_{2}$O: see Table \ref{tab:A1}					\\
\hline
\multicolumn{5}{c}{AlO + HCl  $\to$ AlOH + Cl, AlCl + OH}					\\
AlO: see Table \ref{tab:A1} \smallskip\\
ClAlOH&	Al, 0.011, 0.817, 0.&	32.739 3.5743 3.2225 &	179, 309, 469, &	-351.8	\\
&	O, 1.697, 0.979, 0.		&&626, 873, 3880		\\
&	H, 2.296, 0.228, 0.				\\
&	Cl, -0.927, -1.096, 0.			\smallskip	\\
AlOH: see Table \ref{tab:A1}\smallskip\\
OH: see Table \ref{tab:A1} \smallskip\\
\hline
\end{tabular}
\tablefoot{TS denotes a transition state. ($^{a}$) Calculated at the G4 level of theory \citep{Curtiss2007,Frisch2016}. ($^{b}$) Calculated at the G4 level of theory \citep{Curtiss2007,Frisch2016} with JANAF reference values for $\Delta_f H^o$(Al) = 327.3~kJ~mol$^{-1}$, $\Delta_f H^o$(O) = 246.8~kJ~mol$^{-1}$, $\Delta_f H^o$(Cl) = 119.6~kJ~mol$^{-1}$, and $\Delta_f H^o$(H) = 216.0~kJ~mol$^{-1}$ \citep{Chase1985}.}
\end{table*}

\begin{figure}
   \centering
   \includegraphics[width=\hsize]{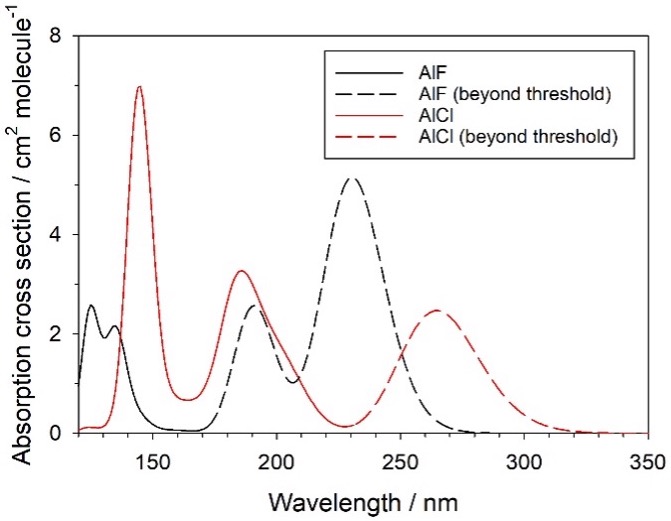}
      \caption{Absorption cross sections of AlF and AlCl calculated at the TD-B3LYP//6-311+g(2d,p) level of theory \citep{Frisch2016}. Photolysis is possible in the portions of the absorption curves shown with solid lines.}
         \label{abscross}
   \end{figure}

The absorption cross sections for AlF and AlCl were calculated by first optimising their geometries at the B3LYP/6-311+g(2d,p) level of theory \citep{Frisch2016}, before determining the vertical excitation energies and transition dipole moments for transitions from their ground electronic states to the first 50 electronically excited states, using the time-dependent density function theory (TD-DFT) method \citep{Bauernschmitt1996}. The resulting cross sections are illustrated in Fig. \ref{abscross}. Because the AlF bond strength is so large \citep[681 kJ mol$^{-1}$ at the G4 level of theory,][]{Curtiss2007} compared with that of AlCl (507 kJ mol$^{-1}$), the wavelength threshold for photolysis of AlF is 175~nm compared with 235~nm for AlCl. Interstellar radiation is attenuated by dust within the outflow, where the extinction is assumed to be equal to that of the ISM, i.e. $1.87 \times 10^{21}$ atoms cm$^{-2}$ mag$^{-1}$ \citep{Cardelli1989}.

The outflow model was initialised with the parent species listed in Table \ref{tab:parents}. Additional reactions, beyond those published in \cite{McElroy2013} as part of \textsc{Rate12} are listed in Table \ref{tab:addrates}. The tabulated constants give the rate coefficient via a temperature-dependent Arrhenius-type formula:
\begin{equation}
k=\alpha\left(\frac{T}{300}\right)^{\beta}\exp\left(\frac{-\gamma}{T} \right)~\mathrm{cm}^3~\mathrm{s}^{-1},
\end{equation}
except for reactions with (interstellar) photons (h$\nu$), which are parameterised as
\begin{equation}
k=\alpha \exp (-\gamma A_V)~\mathrm{s}^{-1},
\end{equation}
where $A_V$ is the dust extinction at visible wavelengths.

The rate coefficients for R1 to R6 were adjusted by varying the heights of the transition states within the expected uncertainty of the G4 level of theory (6 kJ mol$^{-1}$), to optimise agreement with the observations in Fig. \ref{abdist}. These rate coefficients are listed in Table \ref{tab:A3}.

\begin{table}
\caption{Parent species initial fractional abundances relative to \h2.} 
\label{tab:parents}
\centering
\begin{tabular}{cc}
\hline
Species & Abundance\\
\hline\hline
He & 0.17 \\
CO &$6.2 \times 10^{-4}$ \\
\ce{N2}& $4.0 \times 10^{-5}$ \\
\ce{H2O}& $1.5 \times 10^{-5}$\\
HCN& $3.3 \times 10^{-6}$\\
SiO &$3.2 \times 10^{-6}$\\
\ce{NH3}& $1.7 \times 10^{-5}$\\
SiS& $1.6 \times 10^{-6}$\\
CS &$1.2 \times 10^{-6}$\\
HS &$1.7 \times 10^{-5}$\\
Al &$5.6 \times 10^{-6}$\\
Cl &$4.0 \times 10^{-7}$\\
F &$2.0 \times 10^{-7}$\\
\hline
\end{tabular}
\end{table}


\begin{table*}
\caption{Additional reactions for halide and aluminium-bearing molecules included in our chemical modelling.} 
\label{tab:addrates}
\centering
\begin{tabular}{cccccl}
\hline
Reaction & $\alpha$ & $\beta$& $\gamma$ & Temp range$^{a}$ [K] & Source$^b$ \\
\hline\hline
\ce{	AlO	+	h\nu	\to	Al	+	O	} &	3.09e-10	&	...	&	1.7	&	10	--	3000	&	C	:	\cite{Plane2021}$^c$	\\
\ce{	AlCl	+	h\nu	\to	Al	+	Cl	} &	3.52e-10	&	...	&	1.7	&	10	--	3000	&	C	:	\cite{Plane2021}$^c$	\\
\ce{	AlOH	+	h\nu	\to	AlO	+	H	} &	4.57e-10	&	...	&	1.7	&	10	--	3000	&	C	:	\cite{Plane2021}	\\
\ce{	AlF	+	h\nu	\to	Al	+	F	} &	9.70e-11	&	...	&	2	&	10	--	3000	&	C	:	This study	\\
\ce{	HF	+	h\nu	\to	H	+	F	} &	1.38e-10	&	...	&	3	&	10	--	3000	&	M	:	\cite{Heays2017}	\\
\ce{	HCl	+	h\nu	\to	H	+	Cl	} &	1.73e-09	&	...	&	2.88	&	10	--	3000	&	M	:	\cite{Heays2017}	\\
\ce{	Al	+	H2O	\to	AlOH	+	H	} &	7.66e-14	&	3.59	&	-526	&	298	--	1174	&	C	:	\cite{Mangan2021}	\\
\ce{	AlOH	+	H	\to	AlO	+	H2	} &	8.89e-11	&	0.0	&	9092	&	10	--	3000	&	C	:	\cite{Mangan2021}	\\
\ce{	AlOH	+	H	\to	Al	+	H2O	} &	4.31e-11	&	0.0	&	9457	&	10	--	3000	&	C	:	\cite{Mangan2021}	\\
\ce{	Cl	+	NH3	\to	NH2	+	HCl	} &	1.08e-11	&	0.0	&	1370.0	&	290	--	566	&	M	:	\cite{Gao2006}	\\
\ce{	Cl	+	CH4	\to	CH3	+	HCl	} &	6.60e-12	&	0.0	&	1240.0	&	200	--	300	&	L	:	\cite{Atkinson2006}	\\
\ce{	Cl	+	H2CO	\to	HCO	+	HCl	} &	8.20e-11	&	0.0	&	34.0	&	200	--	500	&	L	:	\cite{Atkinson2006}	\\
\ce{	AlO	+	H2	\to	AlOH	+	H	} &	5.37e-13	&	2.77	&	2190	&	10	--	3000	&	C	:	\cite{Mangan2021}	\\
\ce{	AlO	+	H2O	\to	AlOH	+	OH	} &	3.89e-10	&	0.0	&	1295	&	10	--	3000	&	C	:	\cite{Mangan2021}	\\
\ce{	F	+	H2	\to	HF	+	H	} &	2.54e-11	&	1.848	&	-6.182	&	20 -- 295	&	M	:	\cite{Tizniti2014}	\\
&	1.20e-10	&	0.0	&	470	&	295	--	376	&	M	:	\cite{Stevens1989}	\\
\ce{	Cl	+	H2	\to	HCl	+	H	} &	5.27e-12	&	1.4	&	1760	&	199	--	2940	&	M	:	\cite{Kumaran1994}	\\
\ce{	AlOH	+	HCl	\to	AlCl	+	H2O	} &	6e-14	&	1.5	&	200.0	&	100	--	2000	&	C	:	This study$^d$	\\
\ce{	AlO	+	HCl	\to	AlCl	+	OH	} &	3.10e-10	&	0	&	1116	&	100	--	2000	&	C	:	This study$^d$	\\
\ce{	AlCl	+	H	\to	Al	+	HCl	} &	1.90e-10	&	0	&	11027	&	100	--	2000	&	C	:	This study$^d$	\\
\ce{	AlCl	+	OH	\to	AlOH	+	Cl	} &	3.95e-10	&	0	&	11.3	&	10	--	2000	&	C	:	This study$^d$	\\
\ce{	AlCl	+	OH	\to	AlO	+	HCl	} &	3.95e-10	&	0	&	11.3	&	10	--	2000	&	C	:	This study$^d$	\\
\ce{	AlCl	+	H2O	\to	AlOH	+	HCl	} &	7e-14	&	1.7	&	3200.0	&	100	--	2000	&	C	:	This study$^d$	\\
\ce{	Al	+	HF	\to	AlF	+	H	} &	7.7e-10	&	0.0	&	86.0	&	100	--	2000	&	C	:	This study$^d$	\\
\ce{	AlOH	+	HF	\to	AlF	+	H2O	} &	7.8e-10	&	0.0	&	86.0	&	100	--	2000	&	C	:	This study$^d$	\\
\ce{	AlO	+	HF	\to	AlF	+	OH	} &	1.03e-09	&	0.16	&	0	&	10	--	399	&	C	:	This study$^d$	\\
								&	6.20e-10	&	0	&	-224	&	400	--	2000	&	C	:	This study$^d$	\\
\ce{	AlF	+	H	\to	Al	+	HF	} &	1.00e-10	&	0	&	13756	&	100	--	2000	&	C	:	This study$^d$	\\
\ce{	AlF	+	OH	\to	AlO	+	HF	} &	4.50e-10	&	0	&	4147	&	100	--	2000	&	C	:	This study$^d$	\\
\ce{	AlF	+	H2O	\to	AlOH	+	HF	} &	8.10e-10	&	0	&	7242	&	100	--	2000	&	C	:	This study$^d$	\\
\hline
\end{tabular}
\tablefoot{($^a$): The range over which the fit parameters are valid. ($^b$): C, M, and L indicate whether a fit is calculated, measured, or a literature survey value (i.e. a recommended value based on a review of available data), respectively. 
($^c$): Calculated at the same theoretical level as the AlOH photolysis reaction in \cite{Plane2021}.
($^d$): The rate coefficient listed in Table 6 has been adjusted by varying the height of the transition state within the expected uncertainty of the G4 level of theory (6 kJ mol$^{-1}$), to optimise agreement with the observations in Fig. \ref{abdist}.} 
\end{table*}

\end{appendix}

\end{document}